\documentclass[aps,pra,twocolumn,showpacs]{revtex4-1}
\usepackage{amsmath}
\usepackage{amssymb}
\usepackage{upgreek}
\usepackage{color}
\usepackage{graphicx}
\usepackage{subfigure}
\usepackage{enumerate}

\graphicspath{{figures_semi/}}

\begin{document}

\title[]{Dissipative quantum phase transitions of light in a generalized Jaynes-Cummings-Rabi model}

\author{R.~Guti\'{e}rrez-J\'{a}uregui}
\email[Email:]{r.gutierrez@auckland.ac.nz}
\author{H.~J.~Carmichael}
\email[Email:]{h.carmichael@auckland.ac.nz}
\affiliation{The Dodd-Walls Centre for Photonic and Quantum Technologies, Department of Physics, University of Auckland, Private Bag 92019, Auckland, New Zealand}

\date{\today}
\pacs{PACS numbers go here}

\begin{abstract}
The mean-field steady states of a generalized model of $N$ two-state systems interacting with one mode of the radiation field in the presence of external driving and dissipation are surveyed as a function of three control parameters: one governs the interaction strength relative to the resonance frequency, thus accessing the Dicke quantum phase transition, a second the relative strength of counter-rotating to rotating-wave interactions, and a third the amplitude of an external field driving the cavity mode. We unify the dissipative extension of the Dicke quantum phase transition with the recently reported breakdown of photon blockade [H.~J.~Carmichael, Phys.\ Rev.\ X {\bf 5}, 031028 (2015)]; key to the unification is a previously unreported phase of the Dicke model and a renormalized critical drive strength in the breakdown of photon blockade. For the simplest case of one two-state system, we complement mean-field results with a full quantum treatment: we derive quasi-energies to recover the renormalized critical drive strength, extend the multi-photon resonances of photon blockade to a counter-rotating interaction, and explore quantum fluctuations through quantum trajectory simulations.
\end{abstract}

\maketitle

\section{Introduction}
The relationship between phase transitions away from thermal equilibrium and open systems in quantum optics was first addressed in early laser days \cite{degiorgio&scully_1970,graham&haken_1970,grossmann&richter_1971}. The theme was then carried forward by work on optical bistability \cite{bonifacio&lugiato_1976,bonifacio_etal_1978,drummond&walls_1980,rosenberger_etal_1991}, the degenerate parametric oscillator \cite{carmichael_2008}---with loss added to the quantum theory of parametric amplification \cite{mollow&glauber_1967}---and collective radiative phenomena like cooperative fluorescence \cite{drummond&carmichael_1978,carmichael_1980}, to name just a few of the examples. As counterpoint to these phase transitions of light away from equilibrium, and contemporary with the early laser work, Hepp and Lieb  introduced the celebrated Dicke-model phase transition \cite{hepp&lieb_1973a,dicke_1954}---a phase transition for photons in thermal equilibrium.

While the dissipative platforms provided by the laser, optical bistability, and parametric oscillator encouraged wide experimental activity, Hepp and Lieb's proposal lay
dormant on the experimental front. Its call for a dipole coupling strength between light and matter in excess of atomic transition frequencies posed an extreme technical challenge, and also undermined approximations adopted in the Dicke model \cite{rzazewski_etal_1975,bialynicki-birula&rzazewski_1979,gawedzki&rzazewski_1981,rzazewski&wodkiewicz_1991}.  The long wait ended in 2010, however, with the experimental work of Baumann \emph{et al.} \cite{baumann_etal_2010,baumann_etal_2011}, who realized the $T=0$ phase transition of Hepp and Lieb with a superfluid gas in an optical cavity.  The key to success was their engineering of the Dicke-model Hamiltonian as an effective Hamiltonian, by employing an external Raman drive to realize the phase transition in a dissipative setting \cite{dimer_etal_2007,baden_etal_2014}.

In a separate development rooted in research on open systems in quantum optics, cavity and circuit QED have shown that where many material particles and photons might traditionally be required for the strong  interaction of matter and light, it is now possible to achieve strong interactions, sufficient to access nonlinearities, with one particle (e.g.\ two-state system) and photon numbers that range from just one to the relatively low tens, hundreds, or thousands.  Thus, with regard to the laser
and optical bistability, there are cavity QED versions of both \cite{rice&carmichael_1994,savage&carmichael_1988}, and even realizations of the parametric oscillator where single photons are enough to access the nonlinearity \cite{leghtas_etal_2015}. Although the thermodynamic limit of equilibrium phase transitions does not apply under these conditions, it still remains that a mean-field treatment and phase transition perspective can guide much of the phenomenology; albeit with the caveat that fluctuations might add more than just minor corrections.

In this paper we unify the dissipative extension of the Dicke-model quantum phase transition \cite{baumann_etal_2010,baumann_etal_2011,dimer_etal_2007,baden_etal_2014}
with the recently reported  breakdown of photon blockade \cite{carmichael_2015,fink_etal_2017}. The former is addressed by Hepp and Lieb through the thermodynamic limit ($N\to\infty$ two-state systems), while the latter has been approached in a cavity or circuit QED setting \cite{carmichael_2015,fink_etal_2017,alsing&carmichael_1991,kilin&krinitskaya_1991,armen_etal_2009} (one two-state system). Both phenomena may be engineered, however, in either of the two ways, and we therefore first consider mean-field results for both (Sec.~\ref{sec:mean-field}) before turning to results specific to one two-state system (Sec.~\ref{sec:quantum_fluctuations}).

We achieve the proposed unification within the framework of a generalized Dicke-model Hamiltonian, where two extensions of the analysis in Ref.~\cite{dimer_etal_2007} are made: first, we allow for rotating and counter-rotating interactions of independently adjustable coupling strength (see Ref.~\cite{dimer_etal_2007}, Eq.~(12)); and, second, we add external coherent driving of the field mode. The first extension was made by Hepp and Lieb \cite{hepp&lieb_1973b}, in a quick followup to their original paper; the generalized interaction Hamiltonian is also featured in a number of recent publications \cite{stepanov_etal_2008,schiro_etal_2012,tomka_etal_2014,xie_etal_2014,tomka_etal_2015,wang_etal_2016,moroz_2016,kirton_etal_2018}. A key link in our unification is a phase that went unreported by Hepp and Lieb. Beyond this, though, the added coherent drive is also key, since the breakdown of photon blockade is organized around a critical drive strength, identified, to date, in the driven Jaynes-Cummings model (no counter-rotating interaction) alone \cite{alsing&carmichael_1991,alsing_etal_1992,carmichael_2015}. We show that the critical drive is a feature of the generalized Hamiltonian, rotating and counter-rotating interactions included, and thus links the Dicke model quantum phase transition to the breakdown of photon blockade.

We begin in Sec.~\ref{sec:background} with a detailed review, building up our generalized Jaynes-Cummings-Rabi model while making connections to prior work. We then survey the mean-field steady states of the model in Sec.~\ref{sec:mean-field} and show how a common critical drive strength links the dissipative extension of the Dicke-model quantum phase transition to the breakdown of photon blockade.  Finally, in Sec.~\ref{sec:quantum_fluctuations}, we turn from the mean-field treatment to full quantum mechanical calculations for the special case of one two-state system. We recover the critical drive strength from the quasi-energy spectrum of the model Hamiltonian and show how mean-field predictions can  still provide a guide to the physics in the presence of quantum fluctualtions. Conclusions are presented in Sec.~\ref{sec:conclusions}

\section{Background}
\label{sec:background}
\subsection{The Dicke quantum phase transition in the rotating-wave approximation}
\label{sec:Dicke_rotating_wave}
In their original paper \cite{hepp&lieb_1973a} ``On the Superradiant Phase Transition for Molecules in a Quantized Radiation Field: the Dicke Maser Model,'' Hepp and Lieb first introduce an ``interesting caricature$\ldots$invented by Dicke'' \cite{dicke_1954} of the interaction between quantized radiation in a box and a system of $N$ molecules. The caricature assumes single-mode radiation, two-state molecules, and the rotating-wave approximation; it generalizes the Tavis-Cummings model \cite{tavis&cummings_1968} to non-zero detuning, and, adopting natural units with $\hbar=1$, is defined by the Hamiltonian
\begin{equation}
H_0=\omega a^\dagger a+\omega_0J_z+\frac{\lambda}{\sqrt N}(aJ_++a^\dagger J_-),
\label{eqn:hamiltonian_rotating_wave}
\end{equation}
where $\omega$ is the frequency of the field, $\omega_0$ the resonance frequency of the two-state molecules, and $\lambda$ is a coupling strength; annihilation and creation operators for the field mode obey the boson commutation relation, $[a,a^\dagger]=1$, and the collective operators for $N$ two-state systems obey angular momentum commutation relations, $[J_-,J_+]=-2J_z$, $[J_\mp,J_z]=\pm J_\mp$. Hepp and Lieb exactly compute thermodynamic functions in the limit $N\to\infty$ and find a critical temperature, $T_c>0$, for any coupling strength above
\begin{equation}
\lambda_0=\sqrt{\omega\omega_0}.
\label{critical_point_eta=0}
\end{equation}
Considering zero temperature, as we do in this paper, $\lambda_0$ has the significance of a critical coupling strength, where for $\lambda\le\lambda_0$ the photon number is zero in the ground state, while it follows the formula
\begin{equation}
\frac{\langle a^\dagger a\rangle_0}N=\frac{\omega_0}{4\omega}\frac{\lambda^4-\lambda_0^4}{\lambda^2\lambda_0^2}
\label{eqn:photon_number_eta=0}
\end{equation}
when $\lambda>\lambda_0$. Soon after the rigorous calculation of Hepp and Lieb, the same result was derived by Wang and Hioe \cite{wang&hioe_1973} using a simpler method [see their Eq.~(40)].
\subsection{Counter-rotating terms}
\label{sec:Dicke_counter_rotating}
The method of Wang and Hioe readily generalizes to an interaction without the rotating-wave approximation: $aJ_++a^\dagger J_-\to (a+a^\dagger)(J_-+J_+)$. The calculation, made by Hepp and Lieb \cite{hepp&lieb_1973b} and Carmichael \emph{et al.} \cite{carmichael_etal_1973}, retains the phase transition and the form of Eq.~(\ref{eqn:photon_number_eta=0}), but unlike in the rotating-wave approximation, the state of nonzero photon number now assigns a definite phase to the field, and the critical coupling is changed to $\sqrt{\omega\omega_0}/2$. In fact Hepp and Lieb \cite{hepp&lieb_1973b} consider a Hamiltonian generalized in the form
\begin{eqnarray}
H_\eta&=&\omega a^\dagger a+\omega_0J_z\notag\\
&&+\frac{\lambda}{\sqrt N}(aJ_++a^\dagger J_-)+\eta\frac{\lambda}{\sqrt N}(a^\dagger J_++aJ_-),
\label{eqn:hamiltonian_counter_rotating}
\end{eqnarray}
with $\eta$ a parameter. We let $\eta$ vary from 0 to 1 and show (Sec.~\ref{sec:epsilon=0}) that there are actually two critical coupling strengths marking transitions to states of definite phase:
\begin{equation}
\lambda_\eta^\pm=\frac1{1\pm\eta}\sqrt{\omega\omega_0}.
\label{eqn:critical_points_kappa=0}
\end{equation}
Moreover, photon numbers for solutions bifurcating from both critical points, $\lambda_\eta^+$  and $\lambda_\eta^-$, follow the same form, that of Eq.~(\ref{eqn:photon_number_eta=0}):
\begin{equation}
\frac{\langle a^\dagger a\rangle_\eta^\pm}N=\frac{\omega_0}{4\omega}\frac{\lambda^4-(\lambda_\eta^\pm)^4}{\lambda^2(\lambda_\eta^\pm)^2}.
\label{eqn:photon_number_kappa=0}
\end{equation}
The transition at $\lambda_\eta^+$ corresponds to the extension of the Dicke phase transition of Ref.~\cite{hepp&lieb_1973a} discussed in Refs.~\cite{hepp&lieb_1973b} and \cite{carmichael_etal_1973}: the zero photon state becomes unstable and is replaced by a stable state of nonzero photon number. The transition at $\lambda_\eta^-$, not identified before to our knowledge, marks a restabilization of the zero photon state and the birth of an unstable state of nonzero photon number. It provides the fulcrum upon which the unification of the coherently driven extension of the Dicke phase transition and the breakdown of photon blockade turns.
\subsection{Dissipative realization}
\label{sec:dissipative_realization}
While Dicke's paper \cite{dicke_1954} generated enormous interest in superradiance as a transient, away-from-equilibrium process \cite{gross&haroche_1982}, the Dicke quantum phase transition of Hepp and Lieb was, for many years, largely seen as academic---beyond the reach of experiments due to a needed coupling strength on the order of the transition frequency, and, on the theory side, suspect because of approximations used in the Dicke model \cite{rzazewski_etal_1975,bialynicki-birula&rzazewski_1979,gawedzki&rzazewski_1981,rzazewski&wodkiewicz_1991}.  Dissipative realizations of the Dicke Hamiltonian as an effective Hamiltonian overcome these obstacles by replacing a transition from a ground to an excited state by one between a pair of ground states. Specifically, we have the scheme introduced by Dimer \emph{et al.} \cite{dimer_etal_2007,baden_etal_2014} in mind; although there are essentially parallel setups, where internal states are replaced by momentum states of a Bose-Einstein condensate \cite{baumann_etal_2010,baumann_etal_2011}.

\begin{figure}
\begin{center}
\includegraphics[width=2.75in]{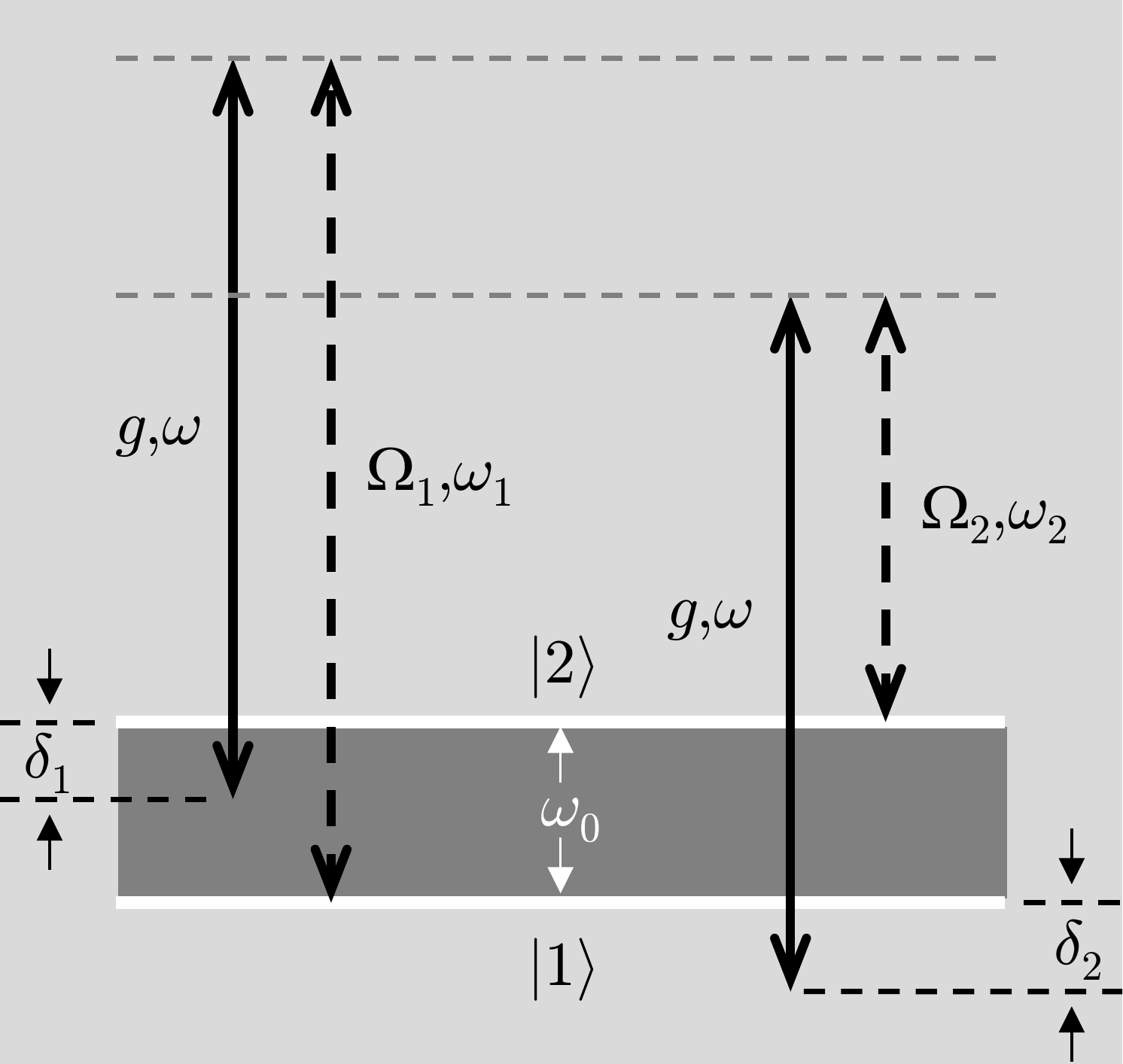}
\end{center}
\caption{Schematic of the open system realization of the model Hamiltonian, Eq.~(\ref{eqn:hamiltonian_raman_model}). A pair of ground states, denoted $|1\rangle$ and $|2\rangle$, are coupled to an optical cavity mode, frequency $\omega$, via far-from-resonance Raman transitions, where bold dashed arrows represent external laser drives while the transfer of photons to and from the cavity mode, coupling strength $g$, is represented by bold solid arrows; $\Omega_{1,2}$ and $\omega_{1,2}$ are drive amplitudes and frequencies, and $\delta_{1,2}$ are detunings; excited states are assumed far from resonance and not shown.}
\label{fig:fig1}
\end{figure}

We consider a pair of Raman transitions between states $|1\rangle$ and $|2\rangle$---the two-state system---as sketched in Fig.~\ref{fig:fig1}, where one leg of each transition is driven by a laser field, with amplitudes and frequencies $\Omega_{1,2}$ and $\omega_{1,2}$, and the other creates and annihilates cavity photons of frequency $\omega$, with coupling strength to the cavity mode $g$. Adopting this configuration, with the excited states (not shown) adiabatically eliminated, and in an interaction picture---free Hamiltonian $\omega_+a^\dagger a+\omega_-J_z$, $\omega_{\pm}=(\omega_1\pm\omega_2)/2$---an effective Hamiltonian is realized in the form of Eq.~(\ref{eqn:hamiltonian_counter_rotating}):
\begin{eqnarray}
H_\eta^\prime&=&\Delta a^\dagger a+\Delta_0J_z\notag\\
&&+\frac{\lambda}{\sqrt N}(aJ_++a^\dagger J_-)+\eta\frac{\lambda}{\sqrt N}(a^\dagger J_++aJ_-),
\label{eqn:hamiltonian_raman_model}
\end{eqnarray}
with effective frequencies
\begin{eqnarray}
\Delta&=&\omega-\omega_+=\frac{\delta_1+\delta_2}2,\\
\label{eqn:delta}
\Delta_0&=&\omega_0-\omega_-=\frac{\delta_1-\delta_2}2,
\label{eqn:delta_zero}
\end{eqnarray}
where $\delta_1$ and $\delta_2$ are Raman detunings (Fig.~\ref{fig:fig1}), and the coupling constants $\lambda$ and $\eta\lambda$ follow from the strength of the Raman coupling (see Ref.~\cite{dimer_etal_2007}). We consider an initial state $|0\rangle|1\rangle$, with $|0\rangle$ the cavity mode vacuum, in which case the Raman driving is a source of photons through the counter-rotating interaction, an external drive that is off-set by the cavity loss; thus, the dissipative realization of the generalized Dicke Hamiltonian, Eq.~(\ref{eqn:hamiltonian_counter_rotating}), is modeled by the master equation
\begin{equation}
\frac{d\rho}{dt}=-i[H_\eta^\prime,\rho]+\kappa{\mathcal L}[a]\rho,
\label{eqn:master_equation}
\end{equation}
where $\kappa$ is the loss rate and ${\mathcal L}[\xi]\,\cdot=2\xi \cdot \xi^\dagger-\xi^\dagger \xi\cdot-\cdot \xi^\dagger\xi$. We show (Sec.~\ref{sec:epsilon=0}) that in the presence of dissipation, for $\eta<\eta_\kappa$,
\begin{equation}
\eta_\kappa\equiv\frac{\kappa}{|\Delta|}\left[1+\sqrt{1+\frac{\kappa^2}{\Delta^2}}\mkern3mu\right]^{-1},
\label{eqn:eta_critical}
\end{equation}
there is no critical coupling strength, while for $\eta\geq\eta_\kappa$, there are two that for $\kappa\to0$ reduce to Eq.~(\ref{eqn:critical_points_kappa=0}):
\begin{equation}
\lambda_\eta^\pm\equiv\frac{\sqrt{|\Delta\Delta_0|}}{1-\eta^2}\left[1+\eta^2\mp2\eta\sqrt{1-\frac{(1-\eta^2)^2}{4\eta^2}\frac{\kappa^2}{\Delta^2}}\mkern3mu\right]^{1/2}.
\label{eqn:lambda_critical}
\end{equation}
Photon numbers generalizing Eq.~(\ref{eqn:photon_number_kappa=0}) are recovered from the mean-field steady state in Sec.~\ref{sec:epsilon=0} [Eq.~(\ref{eqn:photon_number_epsilon=0})].

\subsection{Extended model with coherent drive}
Equations~(\ref{eqn:hamiltonian_raman_model}) and (\ref{eqn:master_equation}) set out a driven and dissipative model where the driving of the field mode is mediated by externally driven Raman transitions; the dissipative realization of the effective rotating and counter-rotating interactions amounts to a \emph{nonlinear} driving of the field mode. In studies of the so-called breakdown of photon blockade  \cite{alsing&carmichael_1991,carmichael_2015,armen_etal_2009,fink_etal_2017}, the mode is subject to a coherent drive, i.e., \emph{linear} driving by an external field. We now extend our model by adding a coherent drive of amplitude $\sqrt N\epsilon$ and frequency $\omega_d$---a detuning $\omega_d-\omega_+$ in the interaction picture of Eq.~(\ref{eqn:hamiltonian_raman_model}). Choosing $\omega_1$ and $\omega_2$ so that $\omega_+=\omega_d$, the master equation then becomes
\begin{equation}
\frac{d\rho}{dt}=-i[H_\eta^\prime,\rho]-i\sqrt N\epsilon[a^\dagger+a,\rho]+\kappa{\mathcal L}[a]\rho,
\label{eqn:master_equation_drive}
\end{equation}
where, from Eq.~(\ref{eqn:delta}), $\Delta=\omega-\omega_d$ is now the detuning of the field mode from the drive.

The next section explores the parameter dependence of the mean-field steady states of Eq.~(\ref{eqn:master_equation_drive}). In particular, we  connect the breakdown of photon blockade, realized for $\eta=0$, to the coherently driven extension of the Dicke quantum phase transition. We show that an $\eta$-dependent critical point organizes behavior as a function of drive strength; we then establish a link through the previously unreported phase of the generalized model presented in Ref.~\cite{hepp&lieb_1973b}, i.e., the second critical coupling strength $\lambda_\eta^-$.

\section{Mean-Field Steady States}
\label{sec:mean-field}
The mean-field Maxwell-Bloch equations derived from the master equation, Eq.~(\ref{eqn:master_equation_drive}), are:
\begin{eqnarray}
\frac{d\alpha}{dt}&=&-(\kappa+i\Delta)\alpha-i\frac{\lambda}{\sqrt N}\frac12(\beta+\eta\beta^*)-i\sqrt N\epsilon,
\label{eqn:mean-field_alpha}\\
\frac{d\beta}{dt}&=&-i\Delta_0\beta+2i\frac{\lambda}{\sqrt N}(\alpha+\eta\alpha^*)\zeta,
\label{eqn:mean-field_beta}\\
\frac{d\zeta}{dt}&=&-i\frac{\lambda}{\sqrt N}\left[(\alpha\beta^*-\alpha^*\beta)-\eta(\alpha\beta-\alpha^*\beta^*)\right],
\label{eqn:mean-field_zeta}
\end{eqnarray}
with $\alpha\equiv\langle a\rangle$, $\beta\equiv2\langle J_-\rangle$, and $\zeta\equiv2\langle J_z\rangle$. We first outline a general approach to their steady state solution, where, introducing intensive variables
\begin{equation}
\bar\alpha\equiv\alpha/\sqrt N,\qquad\bar\beta\equiv\beta/N,\qquad\bar\zeta\equiv\zeta/N,
\label{eqn:scaling}
\end{equation}
Eqs.~(\ref{eqn:mean-field_alpha}) and (\ref{eqn:mean-field_beta}) require
\begin{eqnarray}
\bar\beta_x&=&2\lambda\frac{1+\eta}{\Delta_0}\bar\alpha_x\bar\zeta,
\label{eqn:steady-state_beta1}\\
\bar\beta_y&=&2\lambda\frac{1-\eta}{\Delta_0}\bar\alpha_y\bar\zeta,
\label{eqn:steady-state_beta2}
\end{eqnarray}
with $\bar\alpha_x$ and $\bar\alpha_y$ satisfying the simultaneous equations:
\begin{eqnarray}
\kappa\bar\alpha_x-\left[\Delta+\lambda^2\frac{(1-\eta)^2}{\Delta_0}\bar\zeta\right]\bar\alpha_y&=&0,
\label{eqn:steady-state_alpha1}\\
\kappa\bar\alpha_y+\left[\Delta+\lambda^2\frac{(1+\eta)^2}{\Delta_0}\bar\zeta\right]\bar\alpha_x&=&-\epsilon.
\label{eqn:steady-state_alpha2}
\end{eqnarray}
We may then solve Eqs.~(\ref{eqn:steady-state_beta1})--(\ref{eqn:steady-state_alpha2}) for $|\bar\beta|^2$ in terms of $\bar\zeta$ and impose the conservation law $\bar\zeta^2+|\bar\beta|^2=1$; hence we find an autonomous equation satisfied by $\bar\zeta$,
\begin{equation}
(1-\bar\zeta^2)[P(\bar\zeta)]^2=\frac{4\epsilon^2}{\lambda^2(1+\eta)^2}\bar\zeta^2Q(\bar\zeta),
\label{eqn:6th-order_polynomial}
\end{equation}
with $P(\bar\zeta)$ and $Q(\bar\zeta)$ both quadratic:
\begin{equation}
P(\bar\zeta)=\bar\zeta^2+2\frac{\Delta\Delta_0(1+\eta^2)}{\lambda^2(1-\eta^2)^2}\bar\zeta+\frac{\Delta_0^2(\kappa^2+\Delta^2)}{\lambda^4(1-\eta^2)^2},
\label{eqn:p_quadratic}
\end{equation}
and
\begin{equation}
Q(\bar\zeta)=\bar\zeta^2+2\frac{\Delta\Delta_0}{\lambda^2(1-\eta)^2}\bar\zeta+\frac{\Delta_0^2\kappa^2}{\lambda^4(1-\eta^2)^2}+\frac{\Delta^2\Delta_0^2}{\lambda^4(1-\eta)^4}.
\label{eqn:q_quadratic}
\end{equation}
Steady-state solutions for $\bar\zeta$ are seen to be roots of a 6th-order polynomial, with a possible six distinct solutions for any setting of the parameters: $\eta$, $\Delta$, $\Delta_0$, $\lambda$, $\epsilon$, and $\kappa$. In the following, for the most part, we set $\Delta_0=\Delta$ and keep $\kappa/\lambda$ fixed; we then explore the parameter dependence in the $(\Delta/\lambda,\epsilon/\lambda)$-plane for different choices of $\eta$. To start, we recover the results summarized in Secs.~\ref{sec:Dicke_rotating_wave} and \ref{sec:Dicke_counter_rotating} from our general solution scheme.
\subsection{Zero drive: $\epsilon=0$}
\label{sec:epsilon=0}
In the absence of a coherent drive, the right-hand side of Eq.~(\ref{eqn:6th-order_polynomial}) is zero, and the 6th-order polynomial satisfied by $\bar\zeta$ reduces to
\begin{equation}
(1-\bar\zeta^2)[P(\bar\zeta)]^2=0.
\label{eqn:6th-order_polynomial_epsilon=0}
\end{equation}
Equations~(\ref{eqn:steady-state_alpha1}) and (\ref{eqn:steady-state_alpha2}) are replaced by the homogeneous system
\begin{equation}
\left(
\begin{matrix}
\Delta_0\kappa&-\Delta\Delta_0-\lambda^2(1-\eta)^2\bar\zeta\\
\noalign{\vskip4pt}
\Delta\Delta_0+\lambda^2(1+\eta)^2\bar\zeta&\Delta_0\kappa
\end{matrix}
\mkern3mu\right)\mkern-4mu\left(
\begin{matrix}
\bar\alpha_x\\
\noalign{\vskip4pt}
\bar\alpha_y
\end{matrix}
\right)=0.
\label{eqn:steady-state_alpha_epsilon=0}
\end{equation}
Noting then that the determinant of this homogeneous system is $\lambda^4(1-\lambda^2)^2P(\bar\zeta)$, the condition for nontrivial solutions for $\bar\alpha$ is $P(\bar\zeta)=0$. Thus, the roots $\bar\zeta=\pm1$ of Eq.~(\ref{eqn:6th-order_polynomial_epsilon=0}) correspond to the trivial solution, $\bar\alpha=0$, while the roots of $P(\bar\zeta)=0$,
\begin{equation}
\bar\zeta_\pm=-\frac{\Delta\Delta_0}{\lambda^2(1-\eta^2)^2}\mkern-2mu\left[1+\eta^2\mp2\eta\sqrt{1-\frac{(1-\eta^2)^2}{4\eta^2}
\frac{\kappa^2}{\Delta^2}}\mkern3mu\right],
\label{eqn:nontrivial_zeta_epsilon=0}
\end{equation}
yield nontrivial solutions for $\bar\alpha$. The latter are physically acceptable if $\bar\zeta_\pm$ are real and $|\bar\zeta_\pm|\leq1$; the first condition is satisfied if $\eta\geq\eta_\kappa$, $\eta_\kappa$ defined in Eq.~(\ref{eqn:eta_critical}), and the second gives the critical coupling strengths, $\lambda_\eta^\pm$, defined in Eq.~(\ref{eqn:lambda_critical}); for $\eta\geq\eta_\kappa$ and $\lambda_\eta^+\leq\lambda\leq\lambda_\eta^-$, $\bar\zeta_+$ is the only acceptable root, while $\bar\zeta_+$ and $\bar\zeta_-$ are both acceptable if $\lambda\geq\lambda_\eta^-$.

Note that $\Delta$ and $\Delta_0$ are detunings and therefore two cases arise, one with $\Delta\Delta_0$ positive and $\bar\zeta_\pm<0$, and the other with $\Delta\Delta_0$ negative and $\bar\zeta_\pm>0$. Considering steady states only, there is no physical difference between the cases as a quick inspection of Eqs.~(\ref{eqn:mean-field_alpha})-(\ref{eqn:mean-field_zeta}) shows---simply reverse the signs of $\Delta_0$ and $\bar\zeta$ in Eq.~(\ref{eqn:mean-field_beta}); steady state stability can change, though. We always illustrate results with $\Delta_0=\Delta$, whence $\Delta\Delta_0$ is positive.

By eliminating $\Delta_0\kappa$ from the homogeneous system, Eq.~(\ref{eqn:steady-state_alpha_epsilon=0}), we may solve for
\begin{eqnarray}
(\bar\alpha_x^\pm)^2&=&-|\bar\alpha_\pm|^2\frac{\Delta\Delta_0+\lambda^2(1-\eta)^2\bar\zeta_\pm}{4\lambda^2\eta\bar\zeta_\pm},
\label{eqn:nontrivial_alphax_epsilon=0}\\
(\bar\alpha_y^\pm)^2&=&+|\bar\alpha_\pm|^2\frac{\Delta\Delta_0+\lambda^2(1+\eta)^2\bar\zeta_\pm}{4\lambda^2\eta\bar\zeta_\pm},
\label{eqn:nontrivial_alphay_epsilon=0}
\end{eqnarray}
and hence, using Eqs.~(\ref{eqn:steady-state_beta1}) and (\ref{eqn:steady-state_beta2}), and the conservation law $\bar\zeta^2+|\bar\beta|^2=1$, find
\begin{equation}
|\bar\alpha_\pm|^2=-\frac{\Delta_0}{4\Delta}\frac{1-\bar\zeta_\pm^2}{\bar\zeta_\pm}.
\label{eqn:photon_number_epsilon=0}
\end{equation}
This result gives back Eq.~(\ref{eqn:photon_number_kappa=0}), with $\omega\to\Delta$ and $\omega_0\to\Delta_0$, when $\kappa=0$.

\begin{figure}[t]
\begin{center}
\includegraphics[width=1.7in]{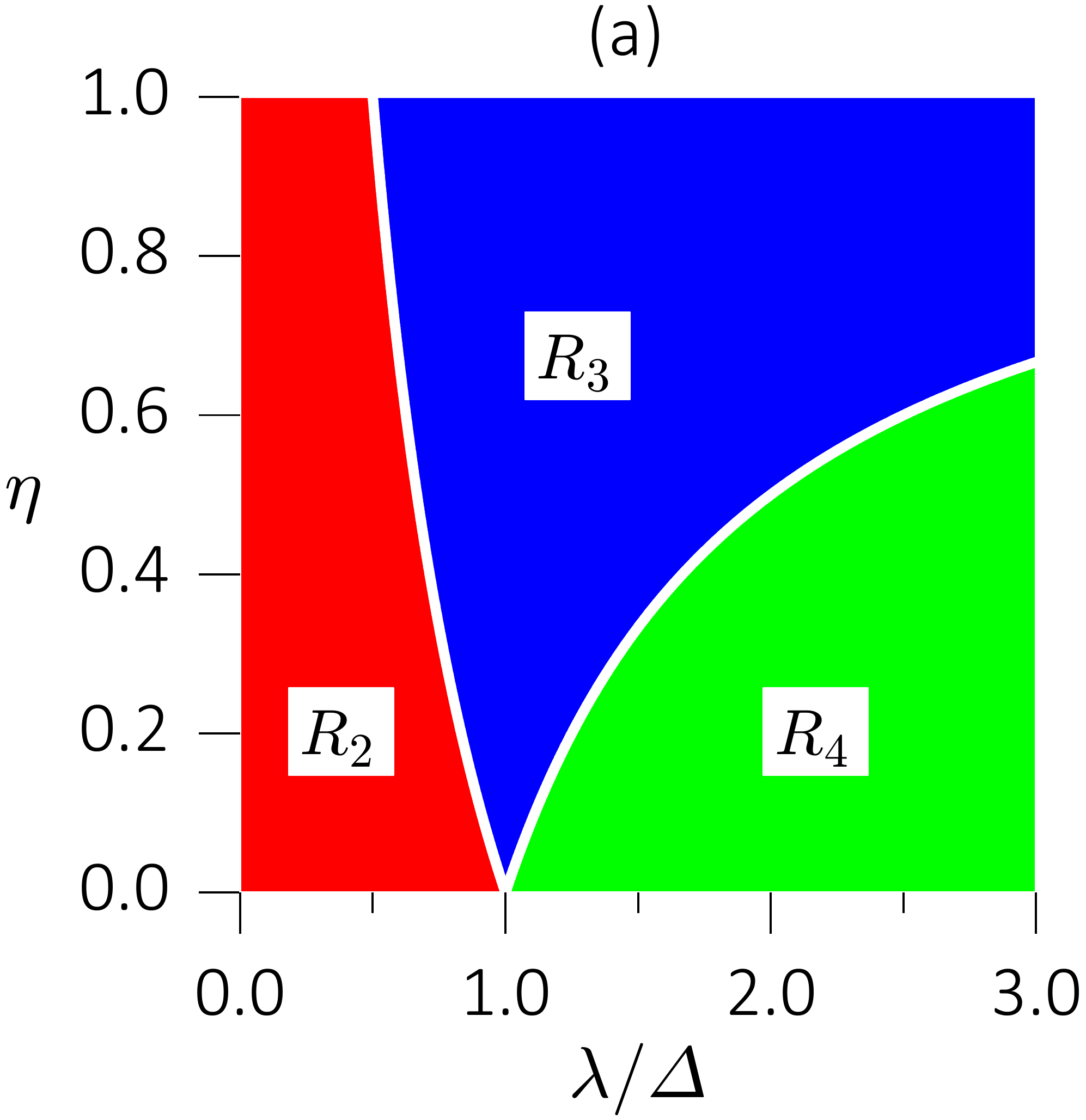}\includegraphics[width=1.7in]{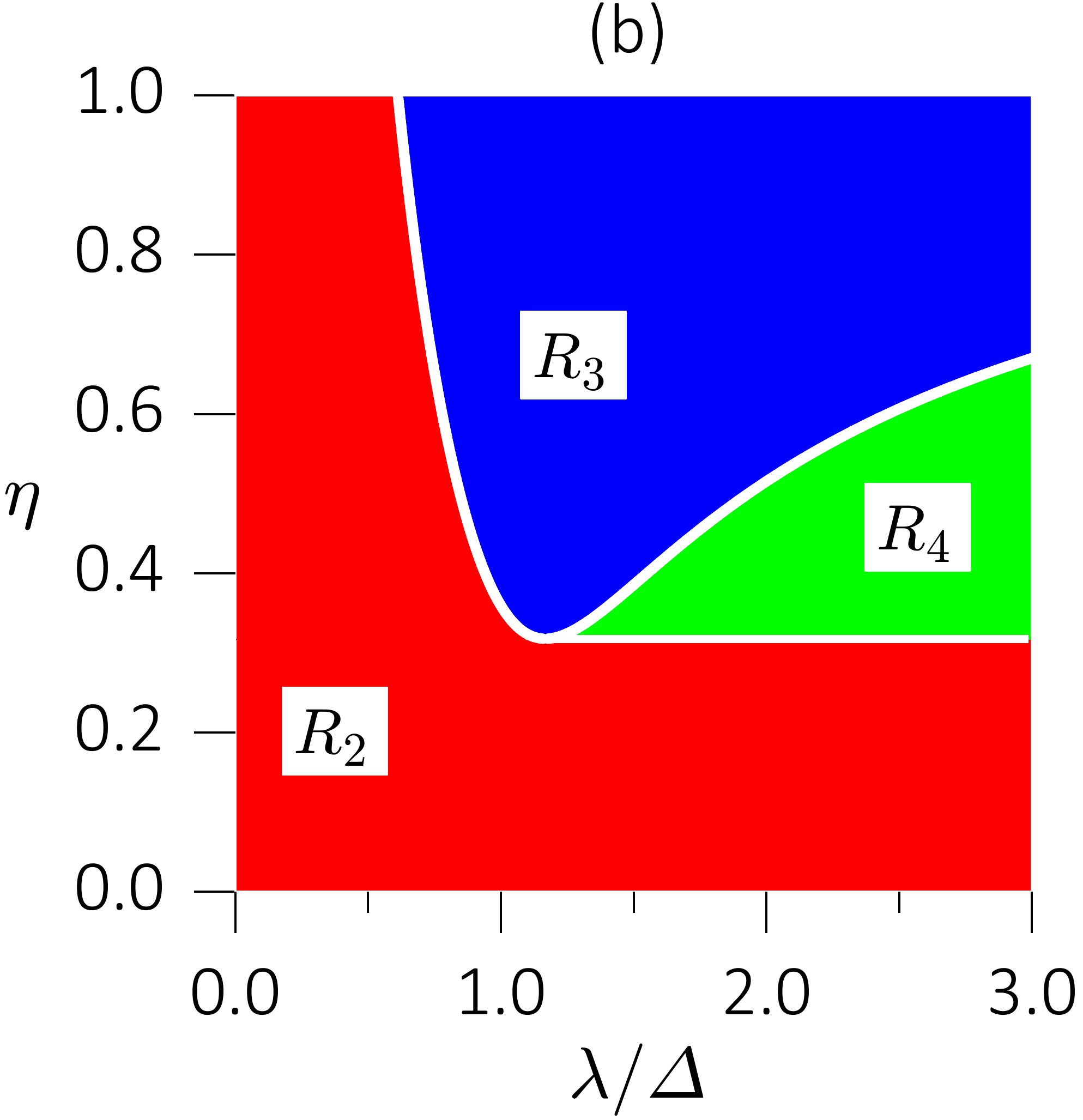}
\vskip0.1in
\includegraphics[width=1.7in]{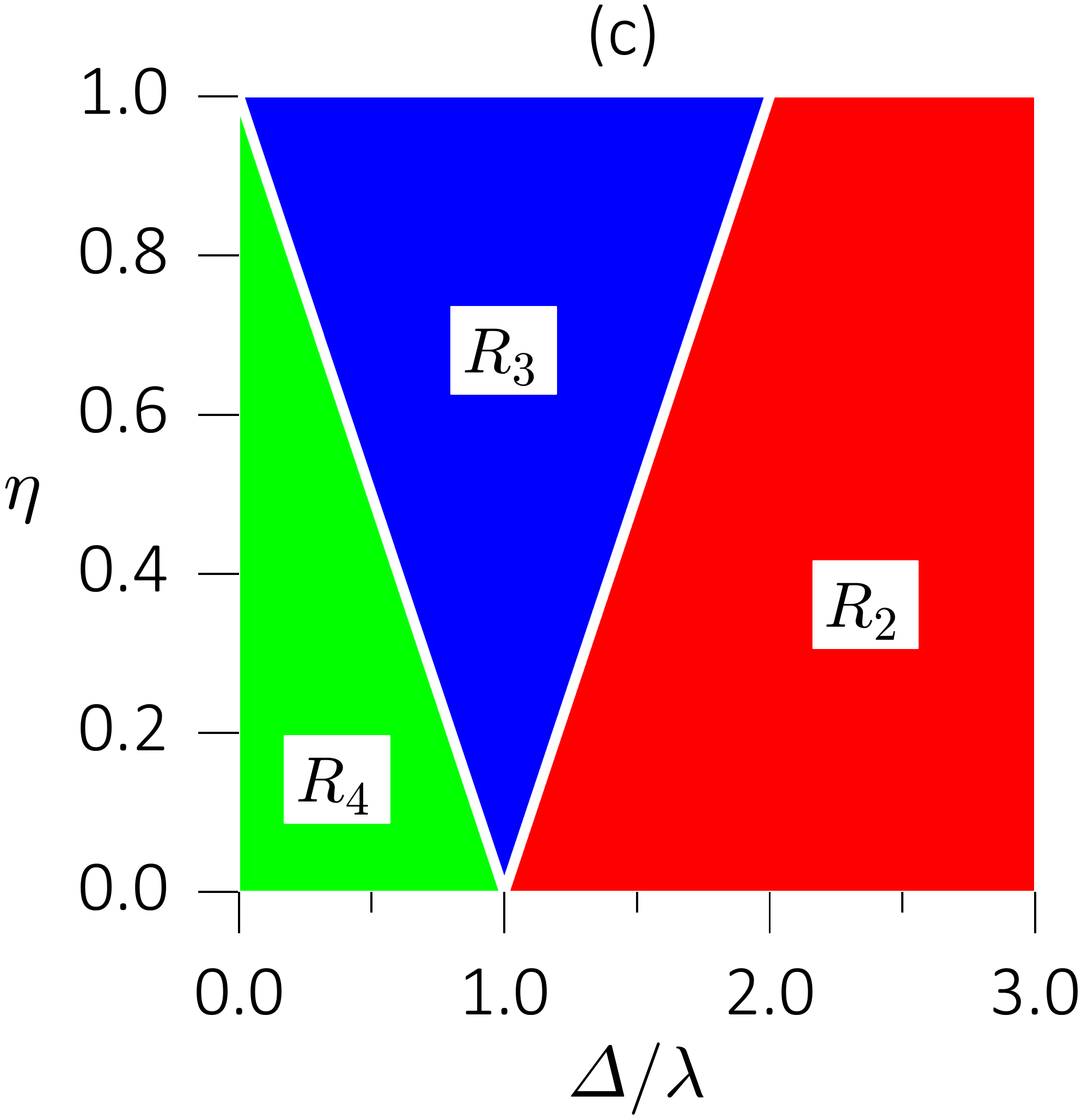}\includegraphics[width=1.7in]{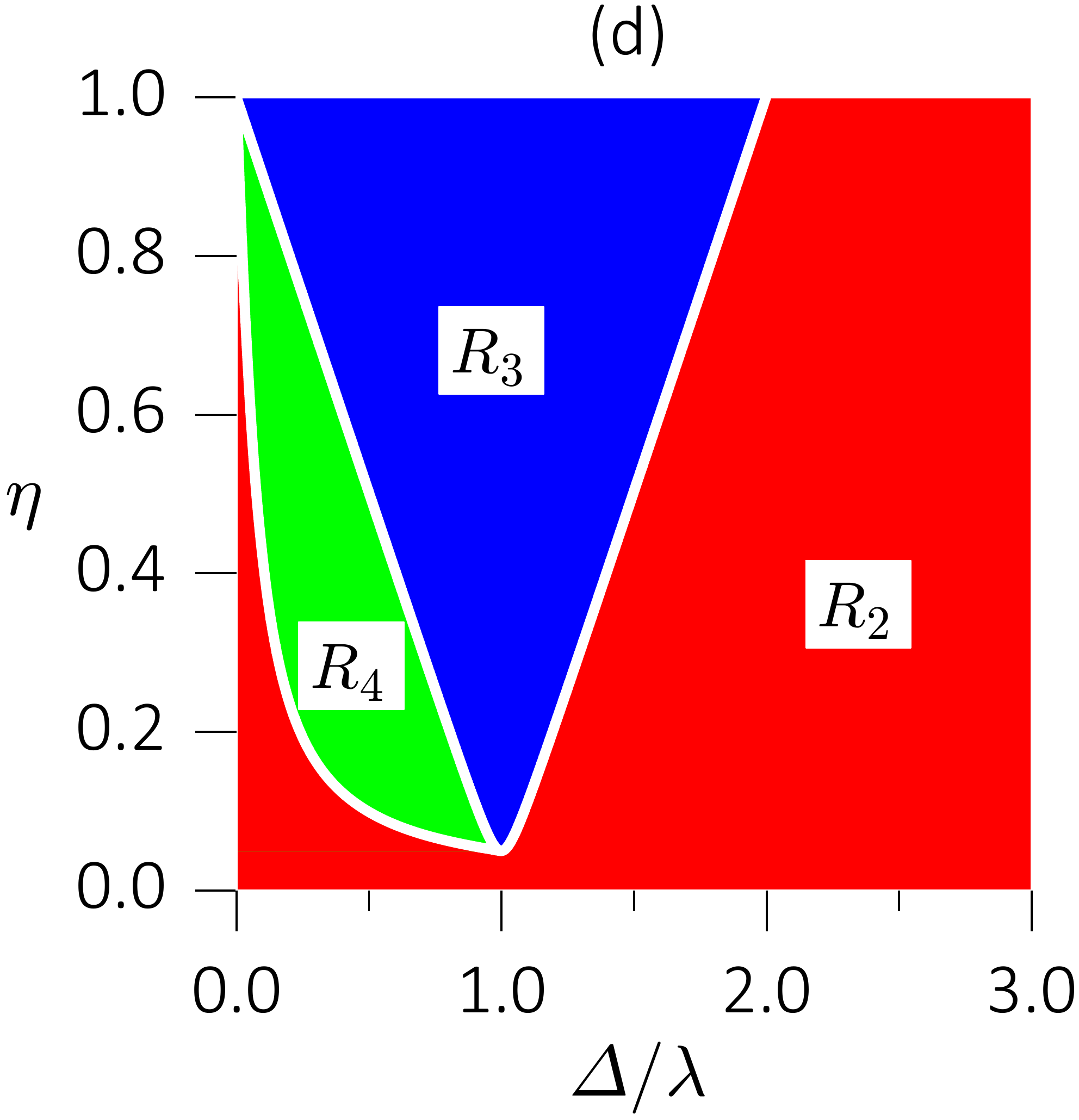}
\end{center}
\caption{Mean-field phase diagram for zero drive and $\Delta_0=\Delta$: (a) $\kappa/\Delta=0$, (b) $\kappa/\Delta=0.7$, (c) $\kappa/\lambda=0$, and (d) $\kappa/\lambda=0.1$. The cut through parameter space is the $(\eta,\lambda/\Delta)$-plane in (a) and (b), and the $(\eta,\Delta/\lambda)$-plane in (c) and (d).}
\label{fig:fig2}
\end{figure}

Figure \ref{fig:fig2} displays four cross-sections of the parameter space for $\epsilon=0$ and $\Delta_0=\Delta$, each subdivided according to the number of distinct steady-state solutions. Frames (a) and (c) apply to the non-dissipative model ($\kappa=0$), while frames (b) and (d) include cavity mode loss. Two complementary perspectives are provided: first, in frames (a) and (b), where the cut is the ($\lambda/\Delta$,$\eta$)-plane, and then, in frames (c) and (d), where the ($\Delta/\lambda$,$\eta$)-plane is shown. The first view envisages the coupling strength $\lambda$, at fixed detuning $\Delta$, as the control parameter, the historical view suggested by Refs.~\cite{hepp&lieb_1973a,wang&hioe_1973,hepp&lieb_1973b,carmichael_etal_1973}; the second envisages $\Delta$ as the control parameter, with $\lambda$ fixed, which is more natural for experiments in optics and the perspective carried through the remainder of the paper. To connect with Secs.~\ref{sec:Dicke_rotating_wave} and \ref{sec:Dicke_counter_rotating}, we note the following points:
\begin{enumerate}[(i)]
\item
The Dicke quantum phase transition in the rotating-wave approximation, originally proposed by Hepp and Lieb \cite{hepp&lieb_1973a}, maps to the line $\eta=0$ in frames (a) and (c). The critical point $\lambda/\Delta=\Delta/\lambda=1$ marks a transition from the trivial solution to one with photon number $|\alpha_\pm|^2=(\Delta^4-\lambda^4)/4\lambda^2\Delta^2$ [Eqs.~(\ref{eqn:photon_number_eta=0}) and (\ref{eqn:photon_number_epsilon=0})], where $\bar\zeta_\pm=-\Delta^2/\lambda^2$ is a double root of $P(\bar\zeta)=0$;  $\bar\beta/\bar\alpha=-2\Delta/\lambda$, but there is no preferred phase for $\bar\beta$, since Eqs.~(\ref{eqn:nontrivial_alphax_epsilon=0}) and (\ref{eqn:nontrivial_alphay_epsilon=0}) reduce to the tautology $0=0$.
\item
The $\eta=0$ transition does not occur in the presence of dissipation, as in frames (b) and (d) the $\eta=0$ axis bounds only the $R_2$ region.
\item
The critical point on the line $\eta=0$ [frames (a) and (c)] splits into a pair of critical points when $\eta>0$, subdividing the plane into regions of two, three, and four distinct solutions (two, four, and six solutions when double roots of $[P(\bar\zeta)]^2=0$ are considered). The transition at $\lambda_{\eta=1}^+=\Delta/2$ from region $R_2$ to $R_3$ recovers the renormalized critical point \cite{carmichael_etal_1973} when the rotating-wave approximation is lifted---the $R_2/R_3$ boundary carries that renormalization through as a function of $\eta$. To our knowledge, the critical point defining the $R_3/R_4$ boundary has not been reported before, although Hepp and Lieb do discuss a model that embraces our inclusion of the parameter $\eta$ \cite{hepp&lieb_1973b}. The transition between regions $R_3$ and $R_4$ is central to the unification we present with a coherent drive included (Sec.~\ref{sec:coherent_drive_intermediate_eta}).
\item
Contrasting the situation in (i), nontrivial solutions in regions $R_3$ and $R_4$ assign $\bar\beta$ and $\bar\alpha$ a definite phase, through Eqs.~(\ref{eqn:steady-state_beta1}), (\ref{eqn:steady-state_beta2}), (\ref{eqn:nontrivial_alphax_epsilon=0}), and (\ref{eqn:nontrivial_alphay_epsilon=0}).
\item
While the map from frame (b) to frame (d) appears straightforward, the map from frame (c) to frame (d) is not: a diagram with two boundaries at fixed $\eta$ now acquires three, as the $R_2/R_4$ boundary bends up to meet $\eta=1$. This follows from the term $\kappa^2/\Delta^2$ under the square root in Eq.~(\ref{eqn:nontrivial_zeta_epsilon=0}): when $\kappa\neq0$, $\bar\zeta_\pm$ are complex for $\eta>\eta_\kappa$, a $\Delta$-dependent condition at fixed $\kappa$ [Eq.~(\ref{eqn:eta_critical})].
\end{enumerate}

\begin{figure}[t]
\begin{center}
\includegraphics[width=3.4in]{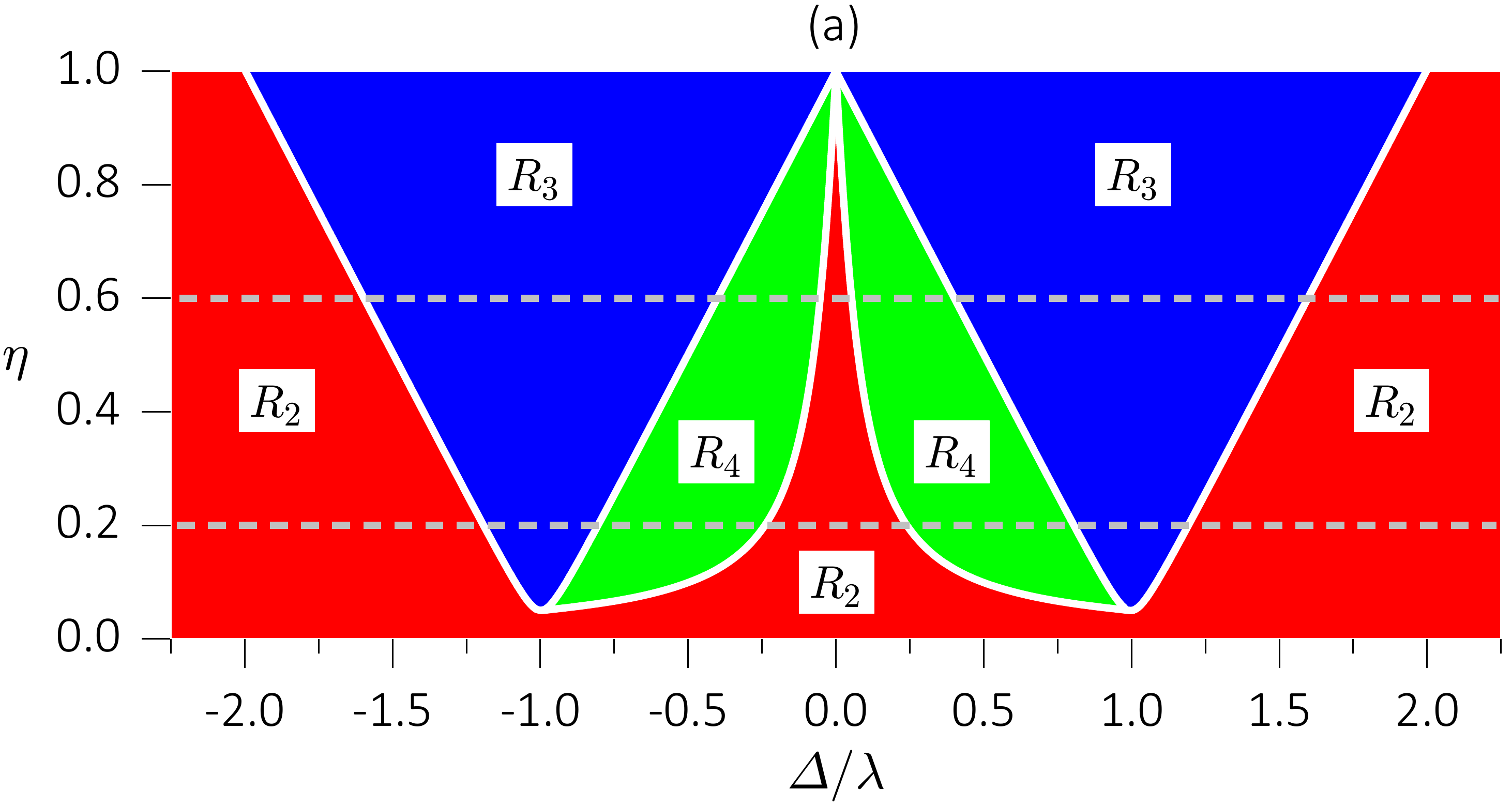}
\vskip0.1in
\includegraphics[width=1.65in]{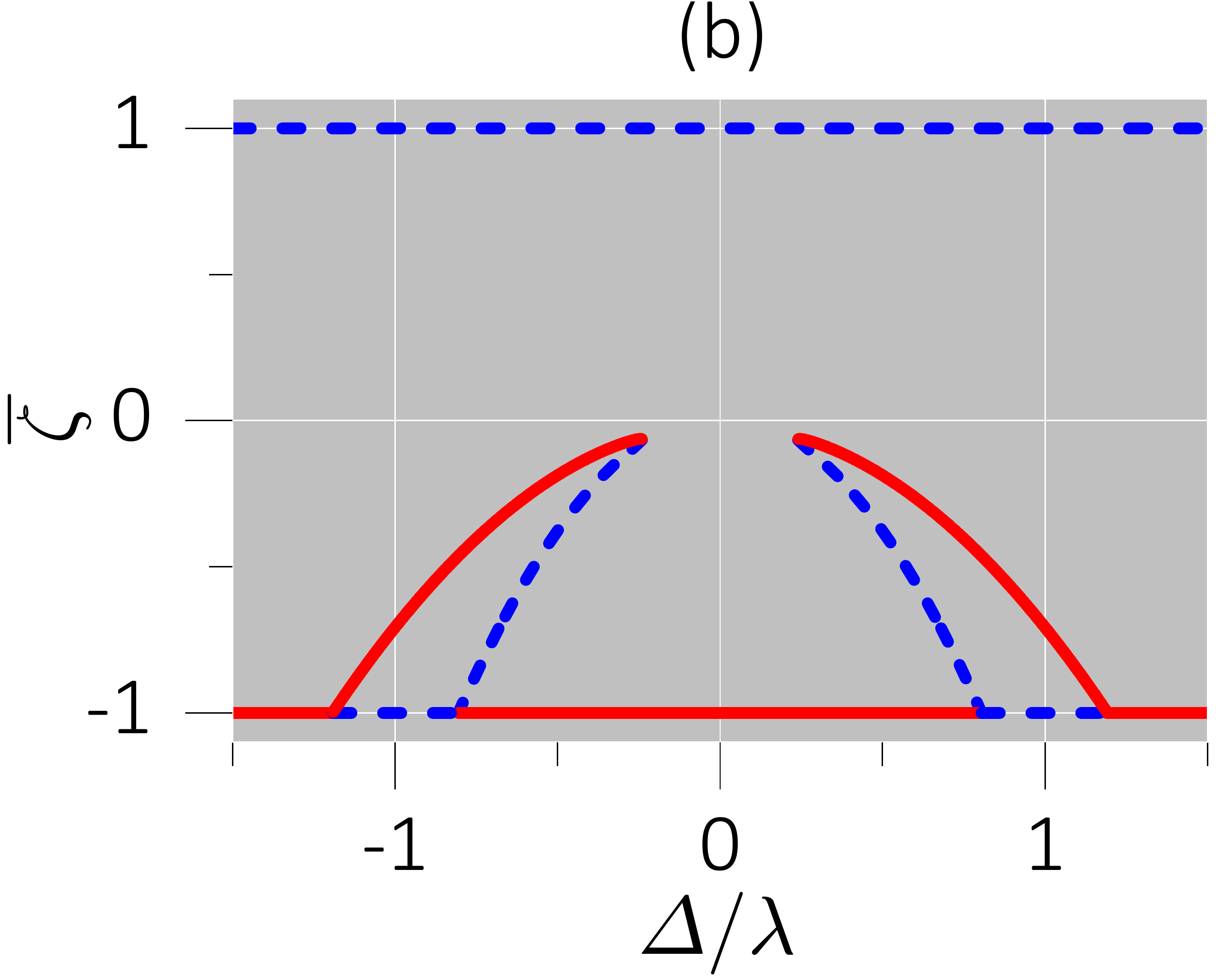}\hskip0.075in\includegraphics[width=1.65in]{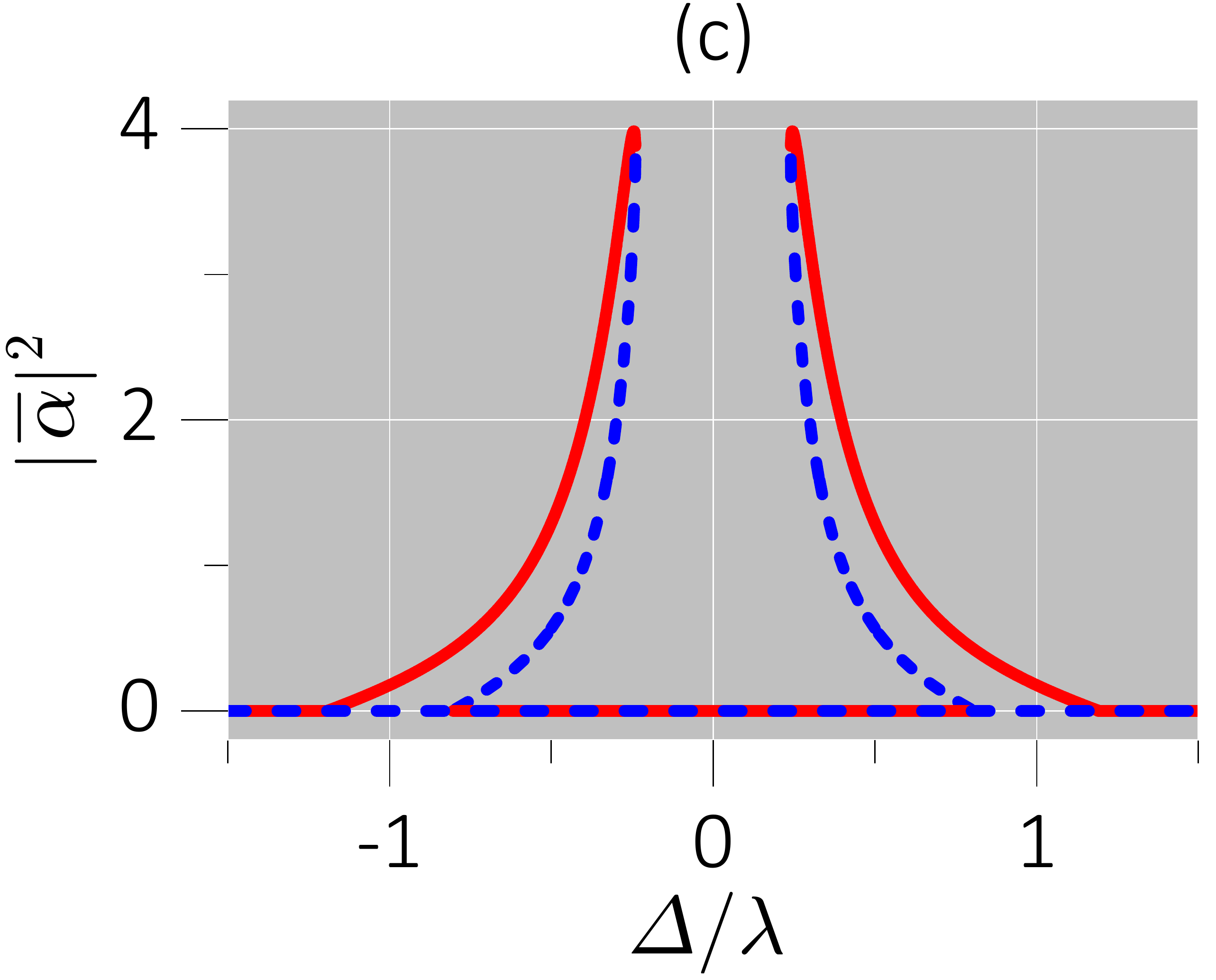}
\vskip0.1in
\includegraphics[width=1.65in]{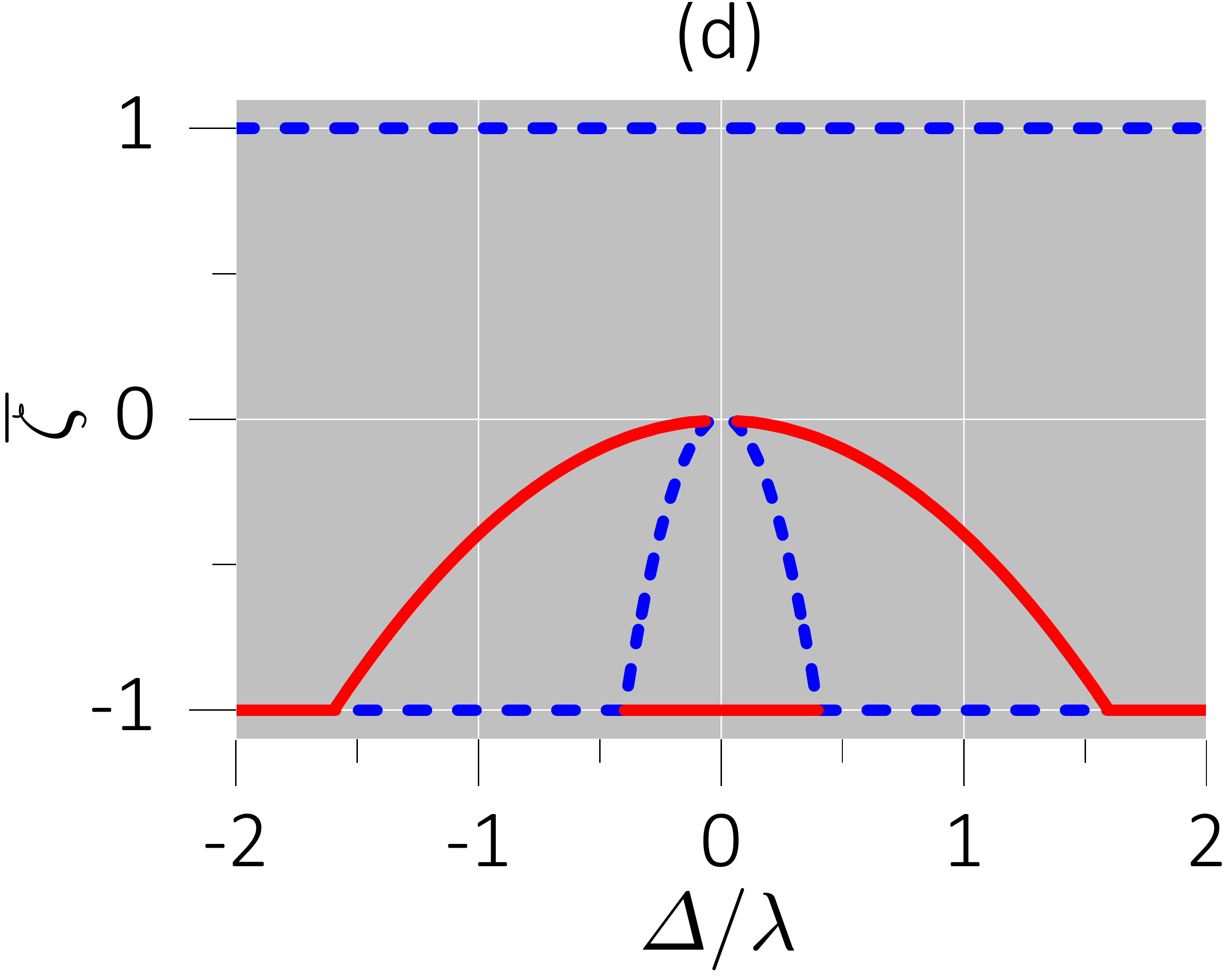}\hskip0.075in\includegraphics[width=1.65in]{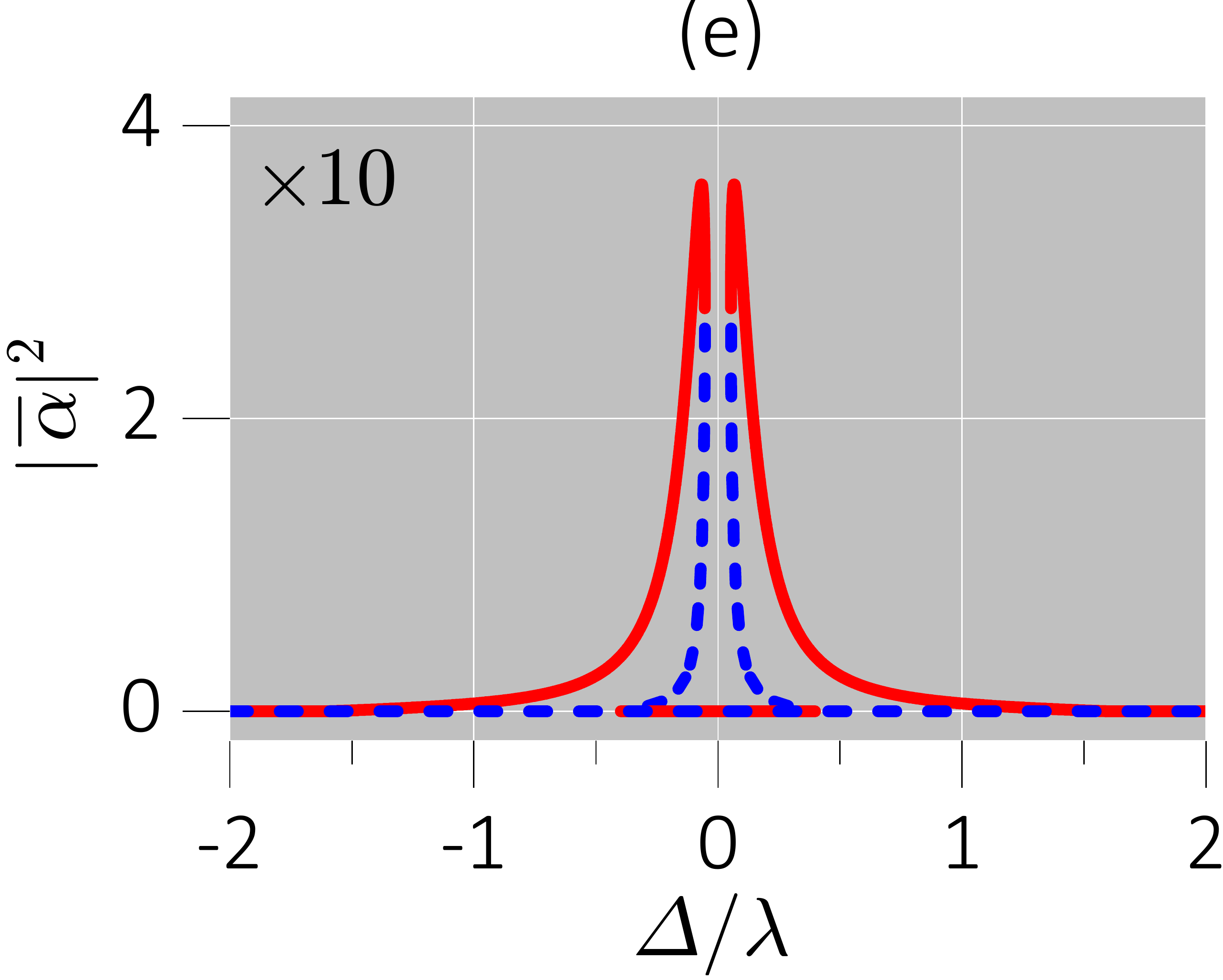}
\end{center}
\caption{Mean-field steady states for zero drive and $\Delta_0=\Delta$: $\kappa/\lambda=0.1$ and $\eta=0.2$ [(b),(c)] and $\eta=0.6$ [(d),(e)]. The two sweeps through the phase diagram are indicated by dashed lines in (a); solid red (dashed blue) lines indicate locally stable (unstable) steady states in (b)-(e).}
\label{fig:fig3}
\end{figure}

Figure \ref{fig:fig3} further illustrates the parameter dependence of the mean-field steady states in the absence of a drive. The symmetrical presentation of the phase diagram in frame (a) is modelled after Ref.~\cite{carmichael_2015} (Figs.~1 and 2) and carried through in  Figs.~\ref{fig:fig4}, \ref{fig:fig5}, and \ref{fig:fig7}. Frames (b)-(e) show steady states and their stability as a function of detuning for $\eta=0.2$ and $\eta=0.6$; they illustrate how the regions in frame (a) interconnect as solutions track smoothly with the changing detuning and bifurcate at boundaries:
\begin{description}
\item[Region $R_2$]
Solutions $\bar\zeta=\pm1$ only; the solution $\bar\zeta=-1$ ($+1$) is stable (unstable). Two solutions in total.
\item[Region $R_3$]
Solutions $\bar\zeta=\pm1$ and the root $\bar\zeta_+$ of $P(\bar\zeta)=0$; the solutions $\bar\zeta=\pm1$ are both unstable and $\bar\zeta^+$ is stable. Three solutions in total.
\item[Region $R_4$]
Solutions $\bar\zeta=\pm1$ and the  roots $\bar\zeta_+$ and $\bar\zeta_-$ of $P(\bar\zeta)=0$; the solutions $\bar\zeta=-1$ (+1) and $\bar\zeta_+$ ($\bar\zeta_-$) are stable (unstable). Four solutions in total.
\end{description}
\noindent

\subsection{Critical drive strength: $\Delta_0=0$}
\label{sec:critical_drive}
We turn now to the dependence on the coherent drive strength, where we begin by identifying the critical point that organizes behavior as function of $\epsilon$. To this end, we must first give special consideration to $\Delta_0=0$, a limit not readily recovered from our general solution scheme, due to the $\Delta_0$ in the denominator of Eqs.~(\ref{eqn:steady-state_beta1}) and (\ref{eqn:steady-state_beta2}); we essentially review an analysis presented by Alsing and Carmichael \cite{alsing&carmichael_1991}, but extended here to arbitrary $\eta$.

From Eqs.~(\ref{eqn:p_quadratic}) and (\ref{eqn:q_quadratic}), when $\Delta_0=0$, $P(\bar\zeta)=Q(\bar\zeta)=\bar\zeta^2$, and the 6th-order polynomial satisfied by $\bar\zeta$ becomes
\begin{equation}
(1-\bar\zeta^2)\bar\zeta^4=\left(\epsilon/\epsilon_{\rm crit}\right)^2\bar\zeta^4,
\label{eqn:6th-order_polynomial_Delta_0=0}
\end{equation}
with
\begin{equation}
\epsilon_{\rm crit}\equiv\frac12\lambda(1+\eta),
\label{eqn:critical_drive}
\end{equation}
where the significance of $\epsilon_{\rm crit}$ as a critical drive strength is elaborated below.
Equations (\ref{eqn:steady-state_beta1}) and (\ref{eqn:steady-state_beta2}) carry over in the form
\begin{equation}
\bar\alpha_x\bar\zeta=\bar\alpha_y\bar\zeta=0,
\label{eqn:alpha_zeta_Delta_0=0}
\end{equation}
and Eqs.~(\ref{eqn:steady-state_alpha1}) and (\ref{eqn:steady-state_alpha2}) as
\begin{eqnarray}
\kappa\bar\alpha_x-\Delta\bar\alpha_y-\lambda\frac12(1-\eta)\bar\beta_y&=&0,
\label{eqn:steady-state_alpha1_Delta_0=0}\\
\kappa\bar\alpha_y+\Delta\bar\alpha_x+\lambda\frac12(1+\eta)\bar\beta_x&=&-\epsilon.
\label{eqn:steady-state_alpha2_Delta_0=0}
\end{eqnarray}
Working then from Eq.~(\ref{eqn:alpha_zeta_Delta_0=0}), we can identify two distinct classes of solutions, one holding below $\epsilon_{\rm crit}$ and the other above.

\subsubsection{Solutions with $\bar\alpha_x=\bar\alpha_y=0$ ($\epsilon\leq\epsilon_{\rm crit}$)}
Equation (\ref{eqn:alpha_zeta_Delta_0=0}) may be satisfied with $\bar\alpha_x=\bar\alpha_y=0$, which, from Eqs.~(\ref{eqn:steady-state_alpha1_Delta_0=0}) and (\ref{eqn:steady-state_alpha2_Delta_0=0}), requires
\begin{equation}
\bar\beta_x=-\epsilon/\epsilon_{\rm crit},\qquad\bar\beta_y=0,
\end{equation}
and hence, from the conservation law $\bar\zeta^2+|\bar\beta|^2=1$,
\begin{equation}
\bar\zeta=\pm\sqrt{1-\left(\epsilon/\epsilon_{\rm crit}\right)^2}.
\label{eqn:zeta_below_Delta_0=0}
\end{equation}
The same result follows directly from Eq.~(\ref{eqn:6th-order_polynomial_Delta_0=0}) under the assumption $\bar\zeta\neq0$. This solution is physically acceptable for $\epsilon\leq\epsilon_{\rm crit}$, though larger drives require Eq.~(\ref{eqn:alpha_zeta_Delta_0=0}) to be satisfied in another way.

\subsubsection{Solutions with $\bar\zeta=0$ ($\epsilon\geq\epsilon_{\rm crit}$)}
Equation (\ref{eqn:alpha_zeta_Delta_0=0}) may also be satisfied with $\bar\zeta=0$, which leaves only the phase of $\bar\beta$ to be determined:
\begin{equation}
\bar\beta=e^{i\phi}.
\end{equation}
From Eq.~(\ref{eqn:mean-field_zeta}), the phase of $\bar\alpha$ must satisfy
\begin{equation}
{\rm Im}\big[\bar\alpha(e^{-i\phi}-\eta e^{i\phi})\big]=0,
\end{equation}
and also, from Eq.~(\ref{eqn:mean-field_alpha}),
\begin{equation}
\bar\alpha=-i\frac{\epsilon+\epsilon_{\rm crit}(e^{i\phi}+\eta e^{-i\phi})/(1+\eta)}{\kappa+i\Delta}.
\end{equation}
The phase $\phi$ is therefore a solution of the transcendental equation
\begin{equation}
\epsilon\cos\phi+\epsilon_{\rm crit}=\frac{\Delta\sin\phi}{\kappa(1-\eta^2)}[\epsilon(1+\eta)^2+\epsilon_{\rm crit}4\eta\cos\phi].
\end{equation}
If we then take $\Delta=0$ as well as $\Delta_0=0$ (and $\eta\neq1$), we arrive at the much simpler equation
\begin{equation}
\phi=\cos^{-1}(-\epsilon_{\rm crit}/\epsilon),
\end{equation}
with solution $\phi=\pi$ for $\epsilon=\epsilon_{\rm crit}$ and two solutions for the phase of $\bar\beta$ above $\epsilon_{\rm crit}$. This prediction of a bistability in phase above $\epsilon_{\rm crit}$ recovers the so-called Spontaneous Dressed-State Polarization of Alsing and Carmichael \cite{alsing&carmichael_1991} (see also \cite{kilin&krinitskaya_1991}) but generalized to $\eta\neq0$.

\subsection{Rotating-wave approximation with coherent drive: $\eta=0$}
\label{sec:rotating_wave_coherent_drive_eta=0}
We now begin to lay out the connection between the breakdown of photon blockade and the coherently driven extension of the Dicke quantum phase transition. In this section, we introduce the breakdown of photon blockade  as the coherently driven extension of Sec.~\ref{sec:epsilon=0} in the limit $\eta=0$. In so doing, we introduce a completely new region of nontrivial steady states, one disconnected and distinct from regions $R_3$ and $R_4$ of Figs.~\ref{fig:fig2} and \ref{fig:fig3}. What follows recovers results from Ref.~\cite{carmichael_2015}.

Returning to the 6th-order polynomial satisfied by $\bar\zeta$, Eq.~(\ref{eqn:6th-order_polynomial}), with $\eta$ zero, $Q(\bar\zeta)=P(\bar\zeta)$, and the polynomial takes the simpler form
\begin{equation}
(1-\bar\zeta^2)[P(\bar\zeta)]^2=\bar\epsilon^2\bar\zeta^2P(\bar\zeta),
\label{eqn:6th-order_polynomial_eta=0}
\end{equation}
with
\begin{equation}
P(\bar\zeta)=(\bar\Delta_0\bar\kappa)^2+(\bar\Delta_0\bar\Delta+\bar\zeta)^2,
\label{eqn:p_quadratic_eta=0}
\end{equation}
where we have introduced parameters scaled by $\epsilon_{\rm crit}$:
\begin{equation}
\bar\epsilon\equiv\epsilon/\epsilon_{\rm crit},\qquad (\bar\kappa,\bar\Delta,\bar\Delta_0)\equiv(\kappa,\Delta,\Delta_0)/2\epsilon_{\rm crit}.
\label{eqn:scaled_parameters}
\end{equation}
The roots of $P(\bar\zeta)=0$ are nonphysical (complex) when $\eta=0$ [Eq.~(\ref{eqn:nontrivial_zeta_epsilon=0})] and therefore $P(\bar\zeta)$ may be cancelled on both sides of Eq.~(\ref{eqn:6th-order_polynomial_eta=0}), which means there are at most four distinct solutions.

Turning then to the field, the homogeneous system, Eq.~(\ref{eqn:steady-state_alpha_epsilon=0}), is replaced by
\begin{equation}
\left(
\begin{matrix}
\bar\kappa&-\bar\Delta-\bar\Delta_0^{-1}\bar\zeta\\
\noalign{\vskip2pt}
\bar\Delta+\bar\Delta_0^{-1}\bar\zeta&\bar\kappa
\end{matrix}
\mkern3mu\right)\mkern-4mu\left(
\begin{matrix}
\bar\alpha_x\\
\noalign{\vskip2pt}
\bar\alpha_y
\end{matrix}
\right)=\left(
\begin{matrix}
0\\
\noalign{\vskip2pt}
-\bar\epsilon/2
\end{matrix}
\right)
\label{eqn:inhomogeneous_system_eta=0}
\end{equation}
with solution for the field amplitude ($\bar\Delta_0\neq0$)
\begin{equation}
\bar\alpha=-i\frac{\bar\epsilon/2}{\bar\kappa+i\left(\bar\Delta+\bar\Delta_0^{-1}\bar\zeta\right)}.
\label{eqn:steady-state_alpha3_eta=0}
\end{equation}
Thus, the field mode responds to coherent driving as a resonator in the presence of a nonlinear dispersion, where the dispersion is defined by solutions to Eq.~(\ref{eqn:6th-order_polynomial_eta=0}). If we then note that $P(\bar\zeta)=\bar\Delta_0^2\bar\epsilon^2/4|\bar\alpha|^2$ [Eqs.~(\ref{eqn:p_quadratic_eta=0}) and (\ref{eqn:steady-state_alpha3_eta=0})], whence, from Eq.~(\ref{eqn:6th-order_polynomial_eta=0}),
\begin{equation}
\bar\zeta=\pm\frac{|\bar\Delta_0|}{\left(\bar\Delta_0^2+4|\bar\alpha|^2\right)^{1/2}},
\end{equation}
we recover the autonomous equation of state for the field mode \cite{carmichael_2015}:
\begin{equation}
\bar\alpha=-i\frac{\bar\epsilon/2}{\bar\kappa+i\left[\bar\Delta\pm\hbox{sgn}\left(\bar\Delta_0)(\bar\Delta_0^2+4|\bar\alpha|^2\right)^{-1/2}\right]}.
\label{eqn:state_equation_eta=0}
\end{equation}

\begin{figure}[t]
\begin{center}
\includegraphics[width=3.4in]{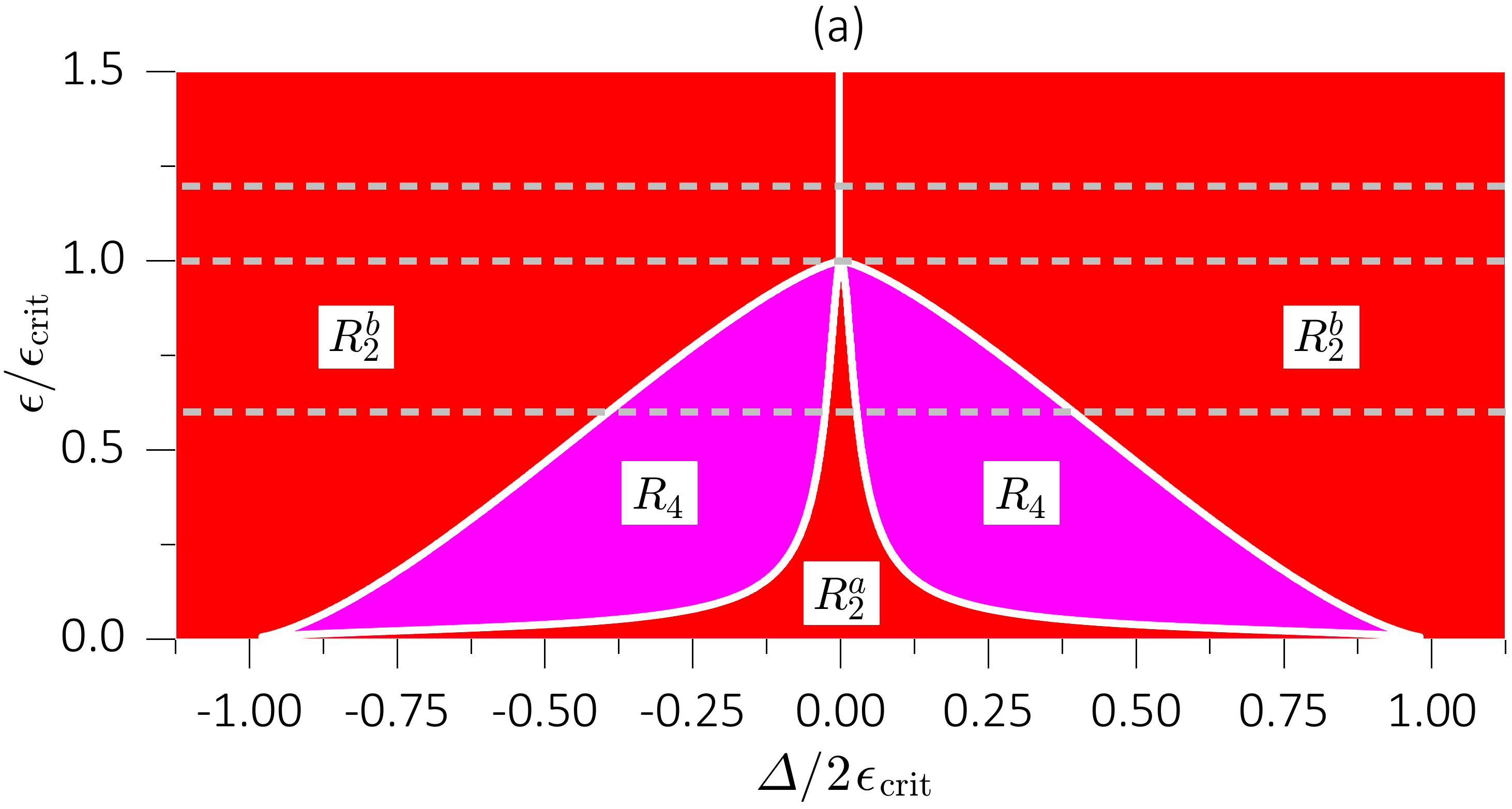}
\vskip0.1in
\includegraphics[width=1.65in]{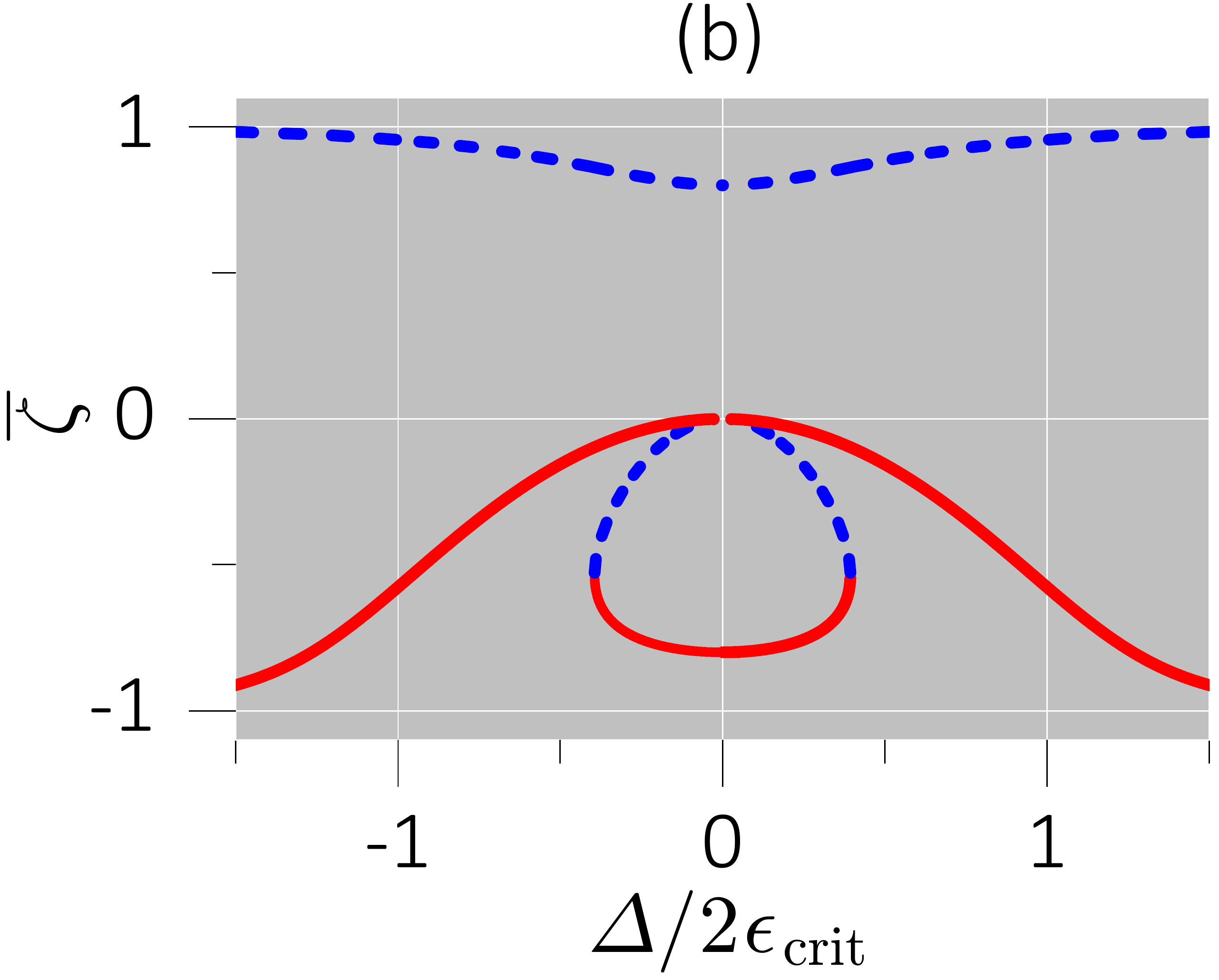}\hskip0.075in\includegraphics[width=1.65in]{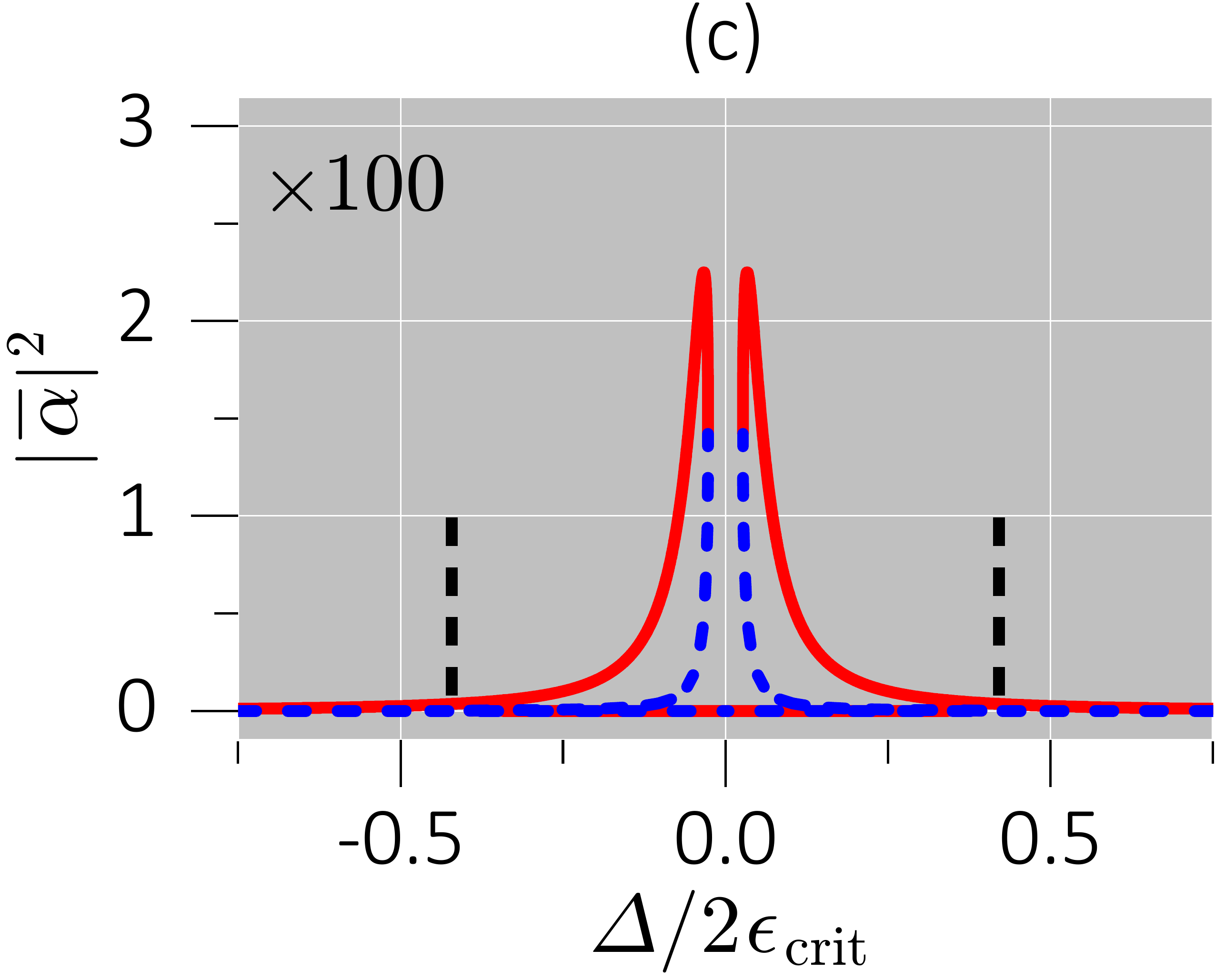}
\vskip0.1in
\includegraphics[width=1.65in]{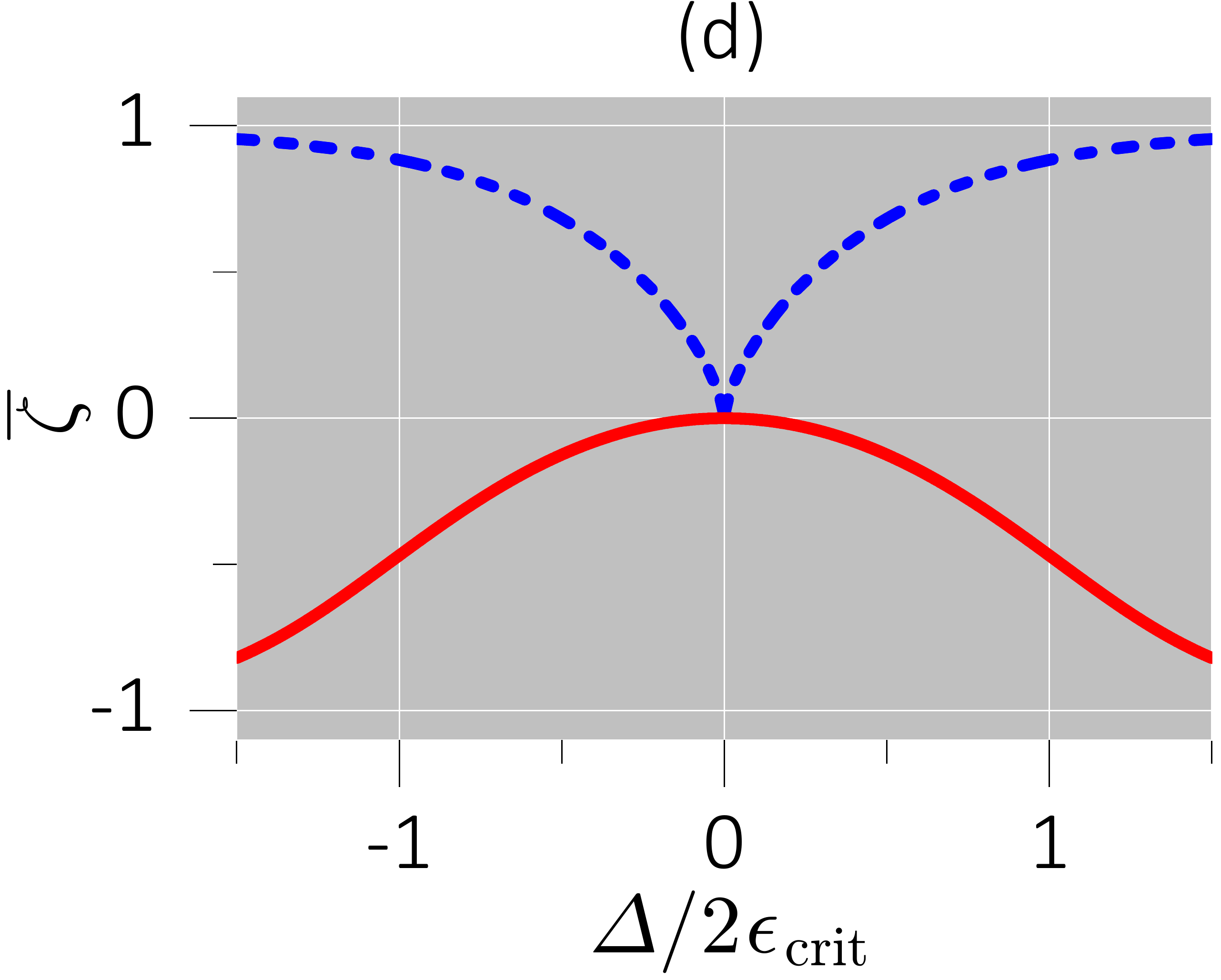}\hskip0.075in\includegraphics[width=1.65in]{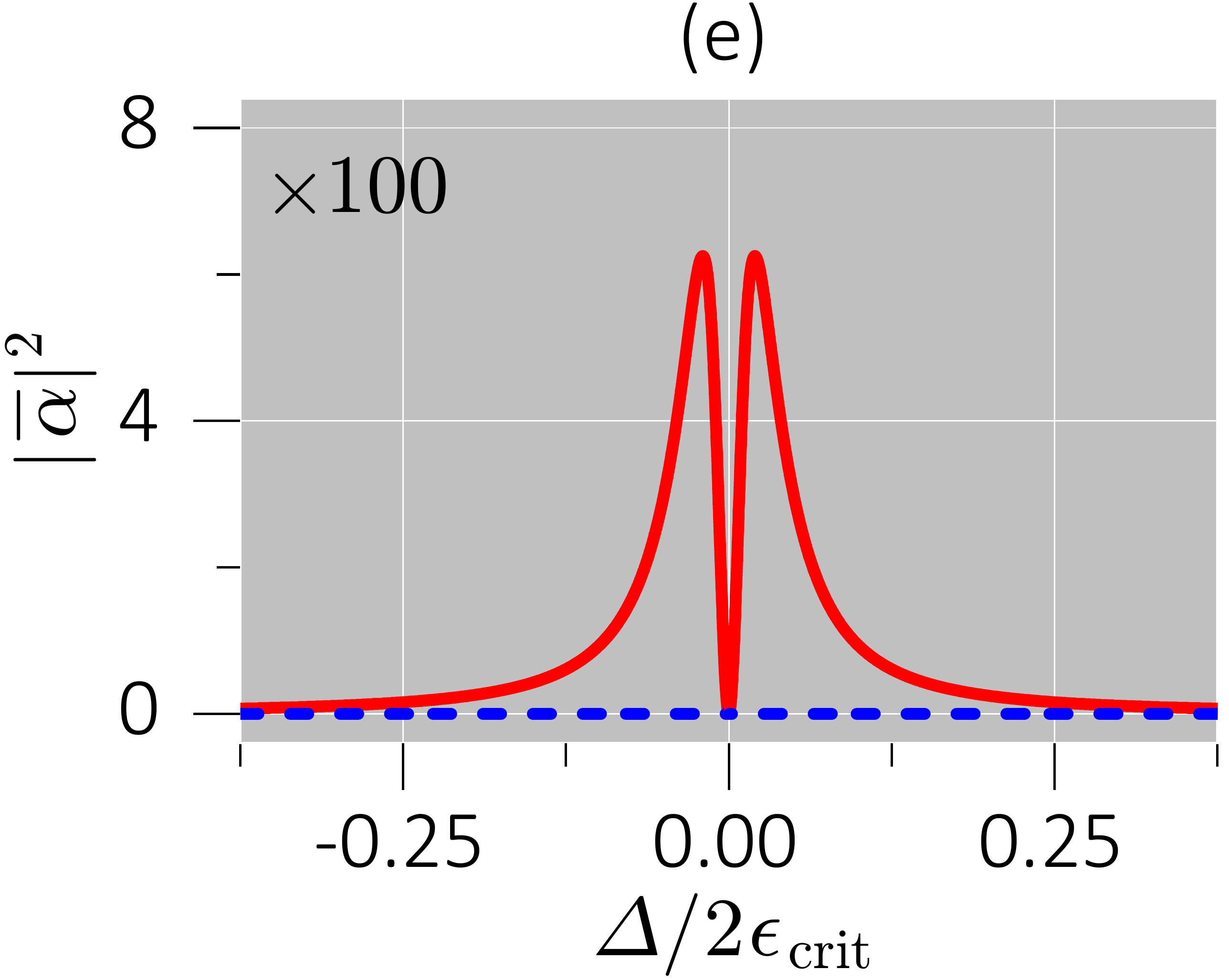}
\vskip0.1in
\includegraphics[width=1.65in]{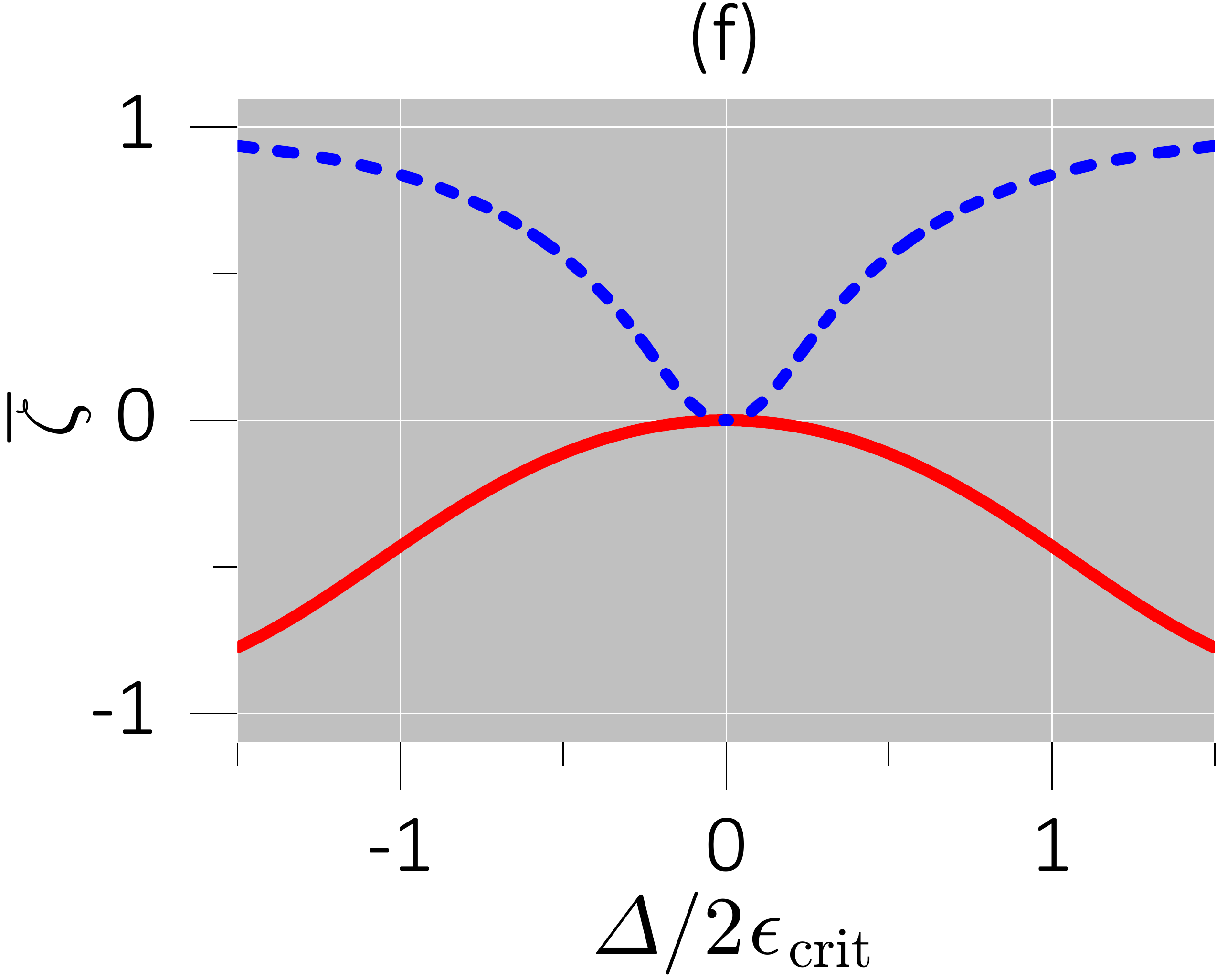}\hskip0.075in\includegraphics[width=1.65in]{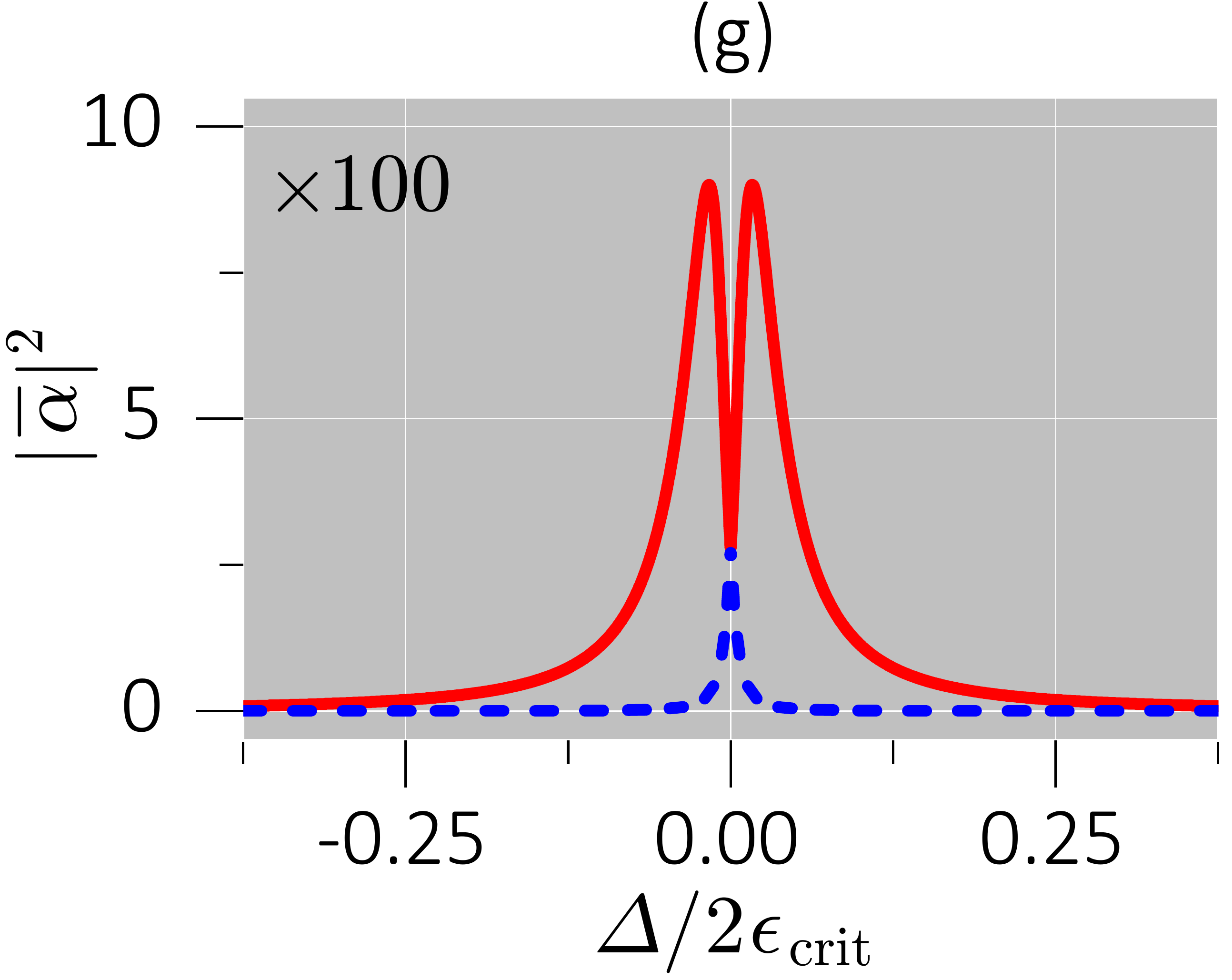}
\end{center}
\caption{Mean-field steady states for $\eta=0$ and $\Delta_0=\Delta$: $\kappa/\lambda=0.02$ and $\epsilon/\epsilon_{\rm crit}=0.6$ [(b),(c)], $\epsilon/\epsilon_{\rm crit}=1.0$ [(d),(e)], and $\epsilon/\epsilon_{\rm crit}=1.2$ [(f),(g)]. The three sweeps through the phase diagram are indicated by dashed lines in (a); solid red (dashed blue) lines indicate stable (unstable) steady states in (b)-(g); dashed black lines demark the range of bistability in (c).}
\label{fig:fig4}
\end{figure}

Figure \ref{fig:fig4} illustrates the results for mean-field steady states obtained from Eqs.~(\ref{eqn:6th-order_polynomial_eta=0}) and (\ref{eqn:state_equation_eta=0}) when $\Delta_0=\Delta$. The phenomenology follows that mapped out in Fig.~4 of Ref.~\cite{carmichael_2015}, where regions of two and four distinct solutions [frame (a)] interconnect through the frequency pulling of vacuuum Rabi resonances located  at $\Delta/2\epsilon_{\rm crit}=\pm1$ for $\epsilon/\epsilon_{\rm crit}\to0$:
\begin{description}
\item[Region $R_2^a$]
Two solutions that approach $\bar\zeta=\pm1$ in the limit of zero drive; the solution approaching $\bar\zeta=-1$ ($+1$) is stable (unstable). Two solutions in total.
\item[Region $R_4$]
Two solutions that approach $\bar\zeta=\pm1$ in the limit of zero drive and two additional solutions that arise from the bistable folding of the solution that approaches $\bar\zeta=-1$; the solution approaching $\bar\zeta=-1$ ($+1$) is stable (unstable), and the two additional solutions are stable and unstable. Four solutions in total.
\item[Region $R_2^b$]
Two solutions that approach $\bar\zeta=\pm1$ in the limit of large detuning; the solution approaching $\bar\zeta=-1$ ($+1$) is stable (unstable). Two solutions in total.
\end{description}
We emphasize that regions $R_2^a$ and $R_2^b$ comprise a single connected region of two distinct solutions in frame (a) of Fig.~\ref{fig:fig4}; region $R_4$ does not touch the $\Delta/2\epsilon_{\rm crit}$ axis, although it comes close when $\kappa/\lambda$ is small. We note also that regions $R_4$ of Fig.~\ref{fig:fig3} and $R_4$ of Fig.~\ref{fig:fig4} are distinct and do not share a common boundary; their interface occurs away from $\eta=0$ and is discussed in Sec.~\ref{sec:coherent_drive_intermediate_eta}.

\subsection{The quantum Rabi Hamiltonian with coherent drive: $\eta=1$}
\label{sec:Dicke_coherent_drive_eta=1}
Taking now the opposite limit, $\eta=1$, we meet with a region of nontrivial steady states that is contiguous with $R_3$ of Figs.~\ref{fig:fig2} and \ref{fig:fig3}. The new region supports four distinct solutions, while $R_3$ supports only three. Nonetheless, the boundary forms a continuous interface since one solution in $R_3$ corresponds to a double root of Eq.~(\ref{eqn:6th-order_polynomial_epsilon=0})---a root of $[P(\bar\zeta)]^2=0$; the coherent drive lifts this degeneracy and splits one distinct solution into two.

In order to avoid the divergence of $P(\bar\zeta)$ and $Q(\bar\zeta)$ as $\eta\to1$, we take Eqs.~(\ref{eqn:p_quadratic}) and (\ref{eqn:q_quadratic}) over in the form
\begin{equation}
(1-\eta)^2P(\bar\zeta)=4\bar\Delta\bar\Delta_0\bar\zeta+4\bar\Delta_0^2(\bar\kappa^2+\bar\Delta^2),
\end{equation}
and
\begin{equation}
(1-\eta)^4Q(\bar\zeta)=16\bar\Delta^2\Delta_0^2,
\end{equation}
in which case the 6th-order polynomial in $\bar\zeta$, Eq.~(\ref{eqn:6th-order_polynomial}), simplifies as
\begin{equation}
(1-\bar\zeta^2)\mkern-2mu\left[\bar\zeta+\frac{\bar\Delta_0}{\bar\Delta}(\bar\kappa^2+\bar\Delta^2)\right]^2=\bar\epsilon^2\bar\zeta^2,
\label{eqn:6th-order_polynomial_eta=1}
\end{equation}
again a 4th-order polynomial with two or four physically acceptable solutions. In the $\bar\epsilon\to0$ limit, the range of four solutions is confined by the inequality
\begin{equation}
\frac{|\bar\Delta_0|}{|\bar\Delta|}(\bar\kappa^2+\bar\Delta^2)\leq1,
\end{equation}
which recovers the $\lambda_{\eta\to1}^+$ threshold of Eq.~(\ref{eqn:lambda_critical}). Note also that, as advertised, the root  $\bar\zeta=-(\bar\Delta_0/\bar\Delta)(\bar\kappa^2+\bar\Delta^2)$ on the $\bar\epsilon=0$ boundary is a double root; thus the region $R_4$ of Fig.~\ref{fig:fig5}---four distinct roots in the interior---interfaces continuously with the three distinct roots of region $R_3$ in Figs.~\ref{fig:fig2} and \ref{fig:fig3}.

Turning to the field, from Eqs.~(\ref{eqn:steady-state_alpha1}) and (\ref{eqn:steady-state_alpha2}), Eq.~(\ref{eqn:inhomogeneous_system_eta=0}) ($\eta=0$) is replaced by
\begin{equation}
\left(
\begin{matrix}
\bar\kappa&-\bar\Delta\\
\noalign{\vskip2pt}
\bar\Delta+\bar\Delta_0^{-1}\bar\zeta&\bar\kappa
\end{matrix}
\mkern3mu\right)\mkern-4mu\left(
\begin{matrix}
\bar\alpha_x\\
\noalign{\vskip2pt}
\bar\alpha_y
\end{matrix}
\right)=\left(
\begin{matrix}
0\\
\noalign{\vskip2pt}
-\bar\epsilon/2
\end{matrix}
\right),
\label{eqn:inhomogeneous_system_eta=1}
\end{equation}
where the coupling through $\bar\zeta$ is no longer symmetrical in the off-diagonals of the matrix on the left-hand side, and is therefore not serving the function of a nonlinear dispersion. Indeed, the physical interpretation for $\eta=1$ says the coupling through $\bar\zeta$ belongs on the right-hand side of Eq.~(\ref{eqn:inhomogeneous_system_eta=1}) where it acts as a nonlinear drive. The interpretation is made particularly clear if we write
\begin{equation}
\bar\beta=\bar\Delta_0^{-1}2\bar\alpha_x\bar\zeta,
\end{equation}
Eqs.~(\ref{eqn:steady-state_beta1}) and (\ref{eqn:steady-state_beta2}), and then, from $\bar\zeta^2+|\bar\beta|^2=1$,
\begin{equation}
\bar\zeta=\pm|\bar\Delta_0|(\bar\Delta_0^2+4\bar\alpha_x^2)^{-1/2}.
\end{equation}
Now, moving the term  $\Delta_0^{-1}\bar\alpha_x\bar\zeta$ to the right-hand side of Eq.~(\ref{eqn:inhomogeneous_system_eta=1}), the equation is rewritten as
\begin{equation}
\left(
\begin{matrix}
\bar\kappa&-\bar\Delta\\
\noalign{\vskip2pt}
\bar\Delta&\bar\kappa
\end{matrix}
\mkern3mu\right)\mkern-4mu\left(
\begin{matrix}
\bar\alpha_x\\
\noalign{\vskip2pt}
\bar\alpha_y
\end{matrix}
\right)=\left(
\begin{matrix}
0\\
\noalign{\vskip2pt}
-\bar\epsilon/2\mp\bar\alpha_x(\bar\Delta_0^2+4\bar\alpha_x^2)^{-1/2}
\end{matrix}
\right),
\label{eqn:alpha_eta=1}
\end{equation}
where, if we can assume $4\bar\alpha_x^2\gg\bar\Delta_0^2$, we find two solutions with the amplitude of the coherent drive simply changed from $\bar\epsilon$ to $\bar\epsilon\pm1$:
\begin{equation}
\bar\alpha=-i\frac{(\bar\epsilon\pm1)/2}{\bar\kappa+i\bar\Delta},
\label{eqn:alpha_eta=1_approx}
\end{equation}
and $\bar\zeta=\pm|\bar\Delta_0|/|\bar\alpha_x|$, $\bar\beta=\pm{\rm sgn}(\bar\Delta_0){\rm sgn}(\bar\alpha_x)$.

More generally, Fig.~\ref{fig:fig5} shows the dependence of mean-field steady states on drive amplitude and detuning for $\eta=1$ and $\bar\Delta_0=\bar\Delta$; frames (b)-(g) illustrate results for three sweeps through a parameter space that divides into just two separate regions [frame (a)]:

\begin{figure}[t]
\begin{center}
\includegraphics[width=3.4in]{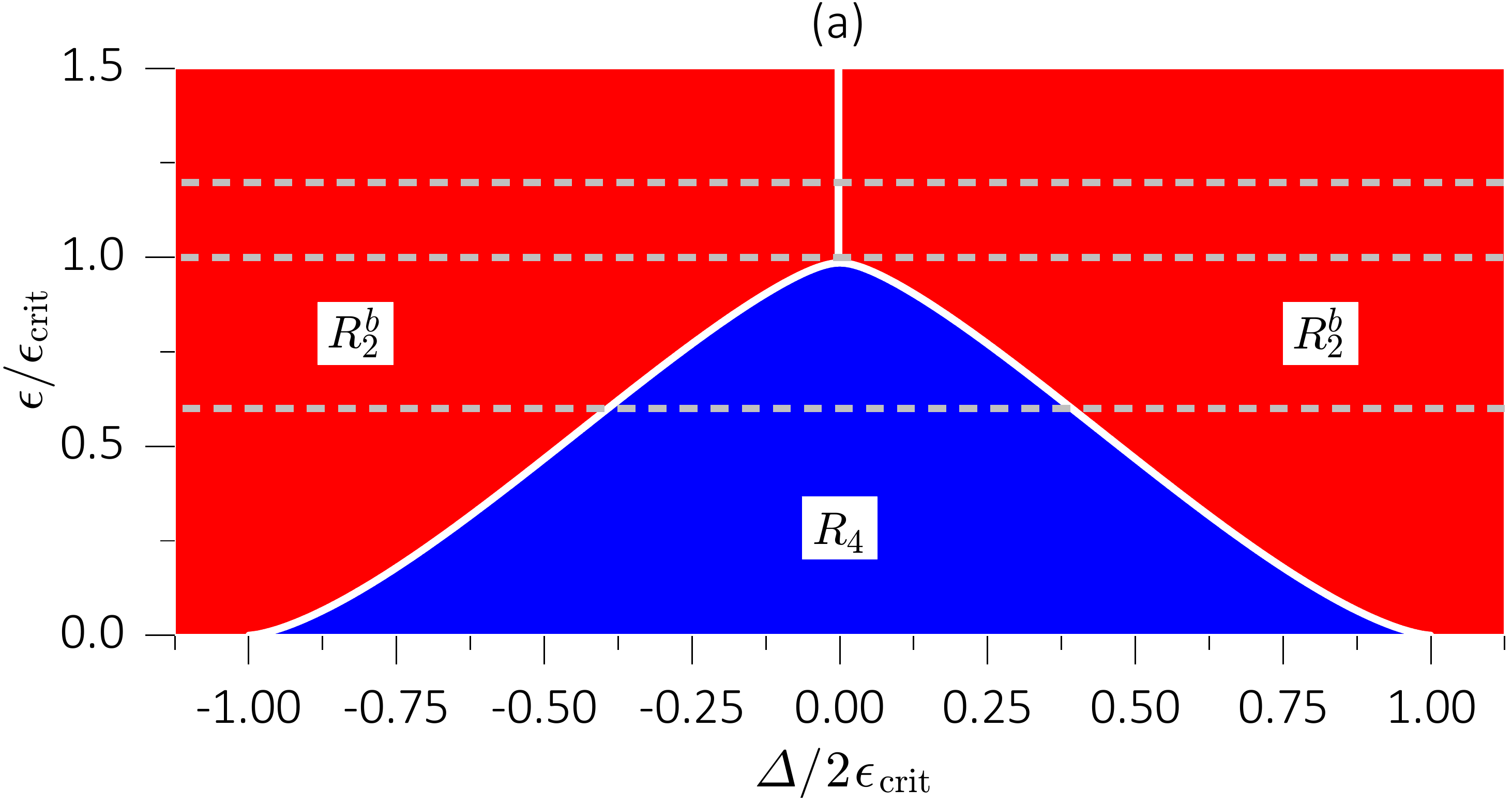}
\vskip0.1in
\includegraphics[width=1.65in]{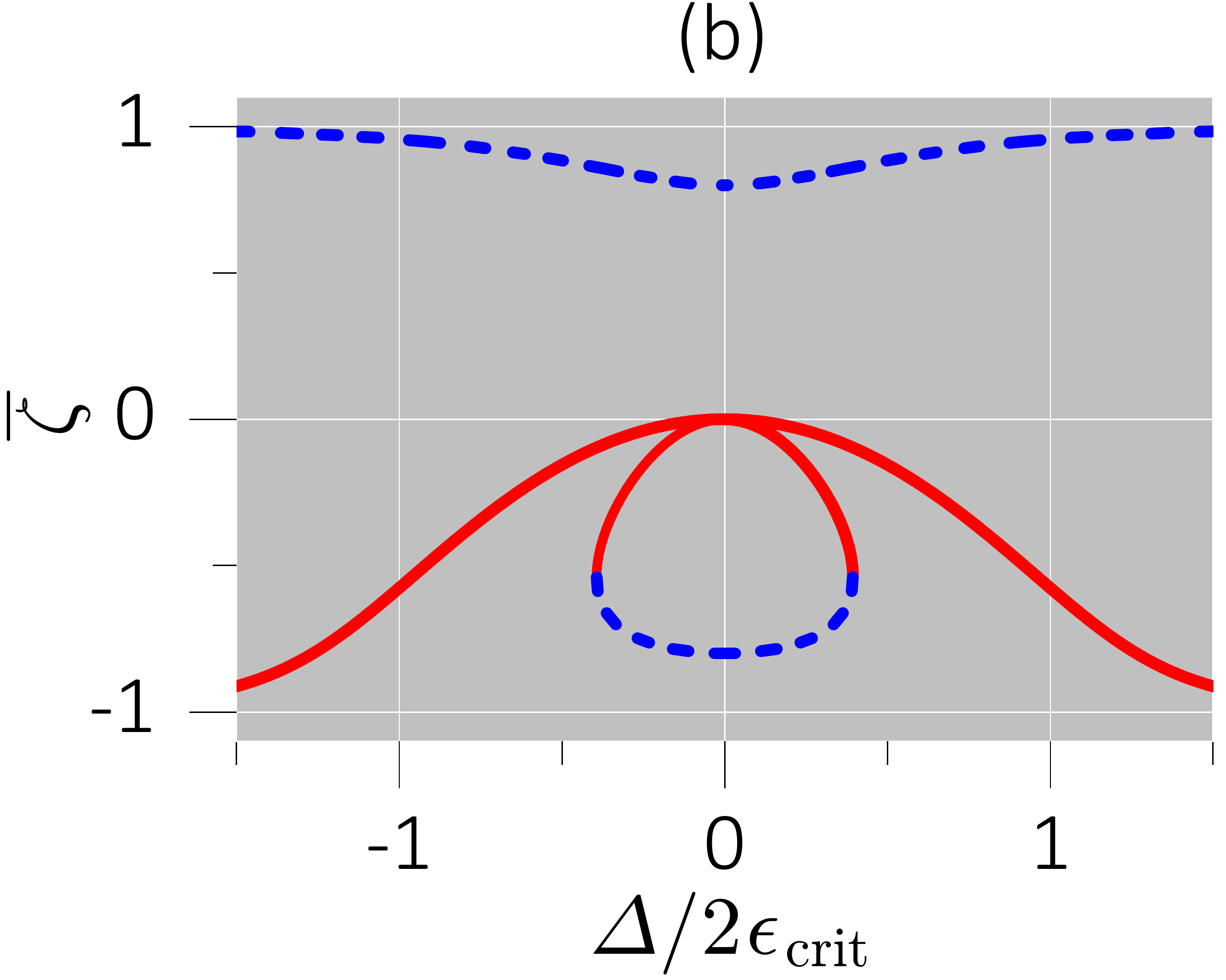}\hskip0.075in\includegraphics[width=1.65in]{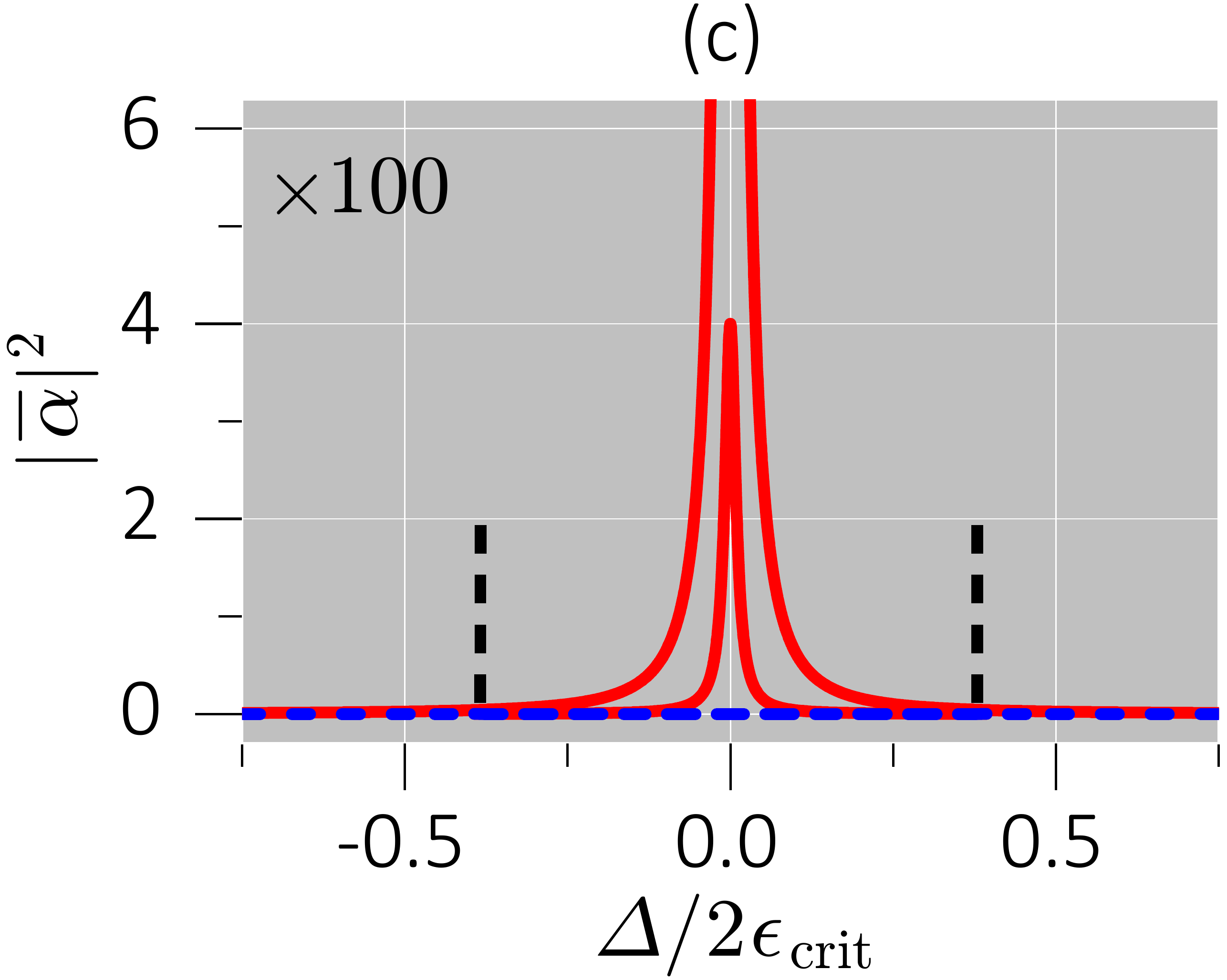}
\vskip0.1in
\includegraphics[width=1.65in]{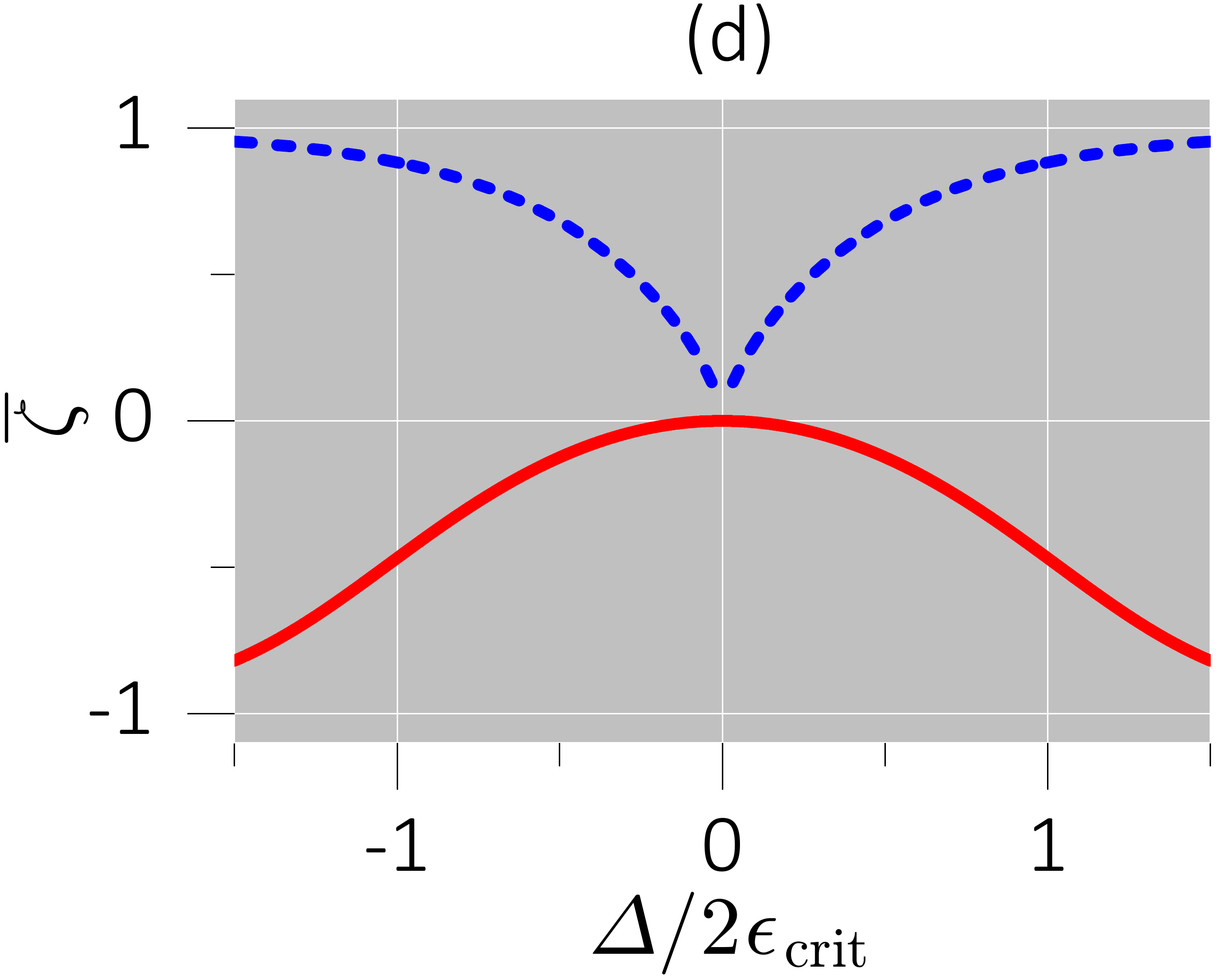}\hskip0.075in\includegraphics[width=1.65in]{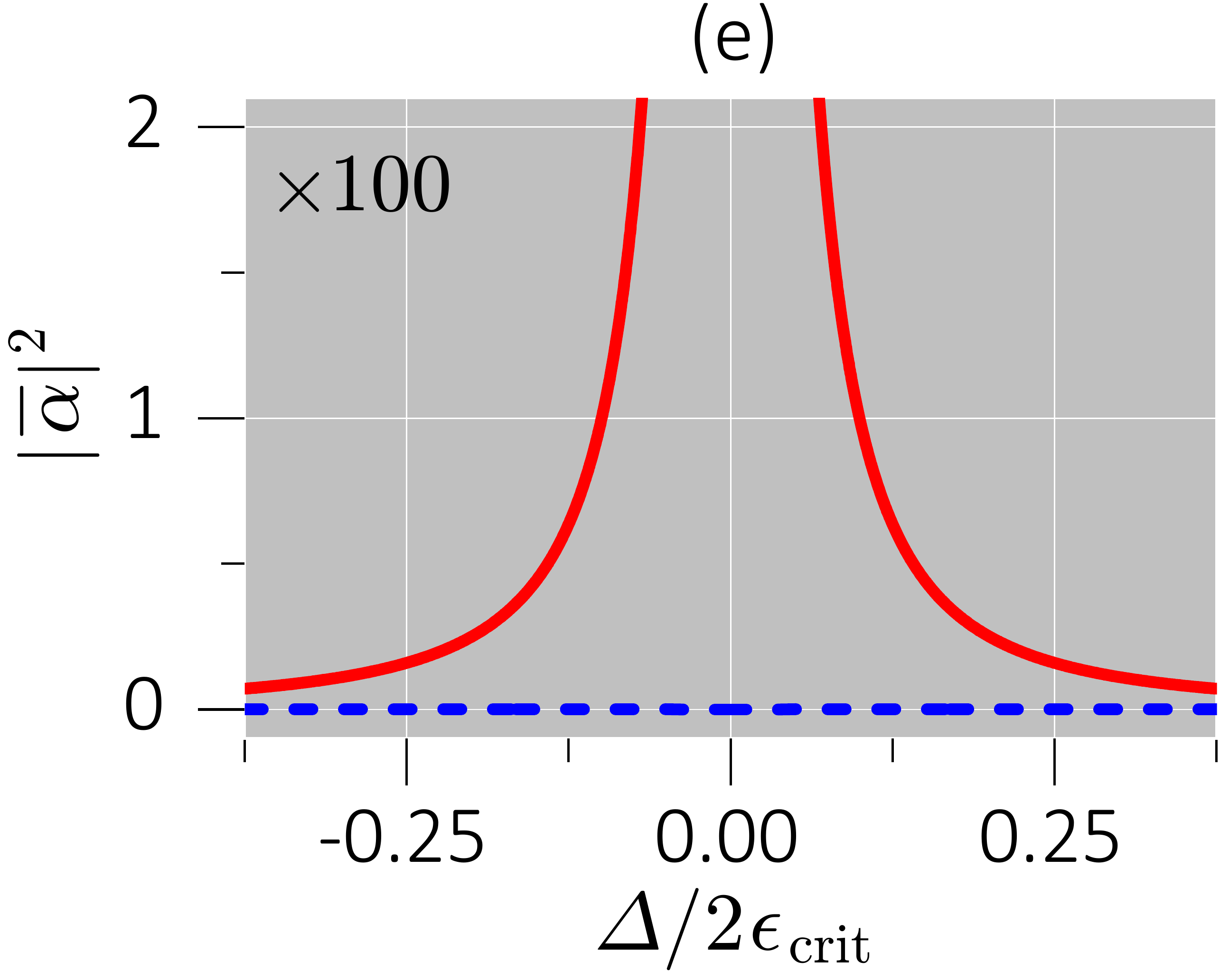}
\vskip0.1in
\includegraphics[width=1.65in]{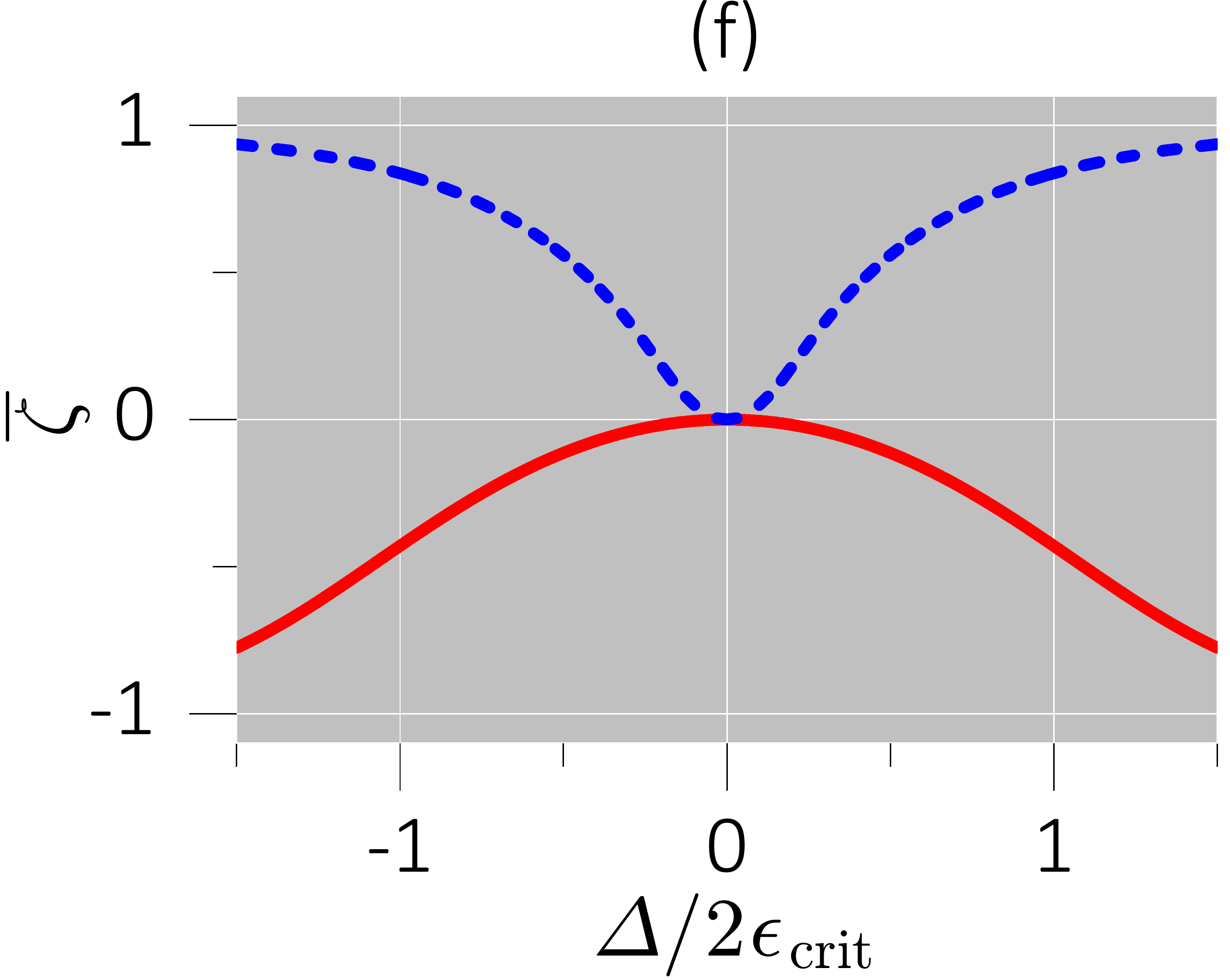}\hskip0.075in\includegraphics[width=1.65in]{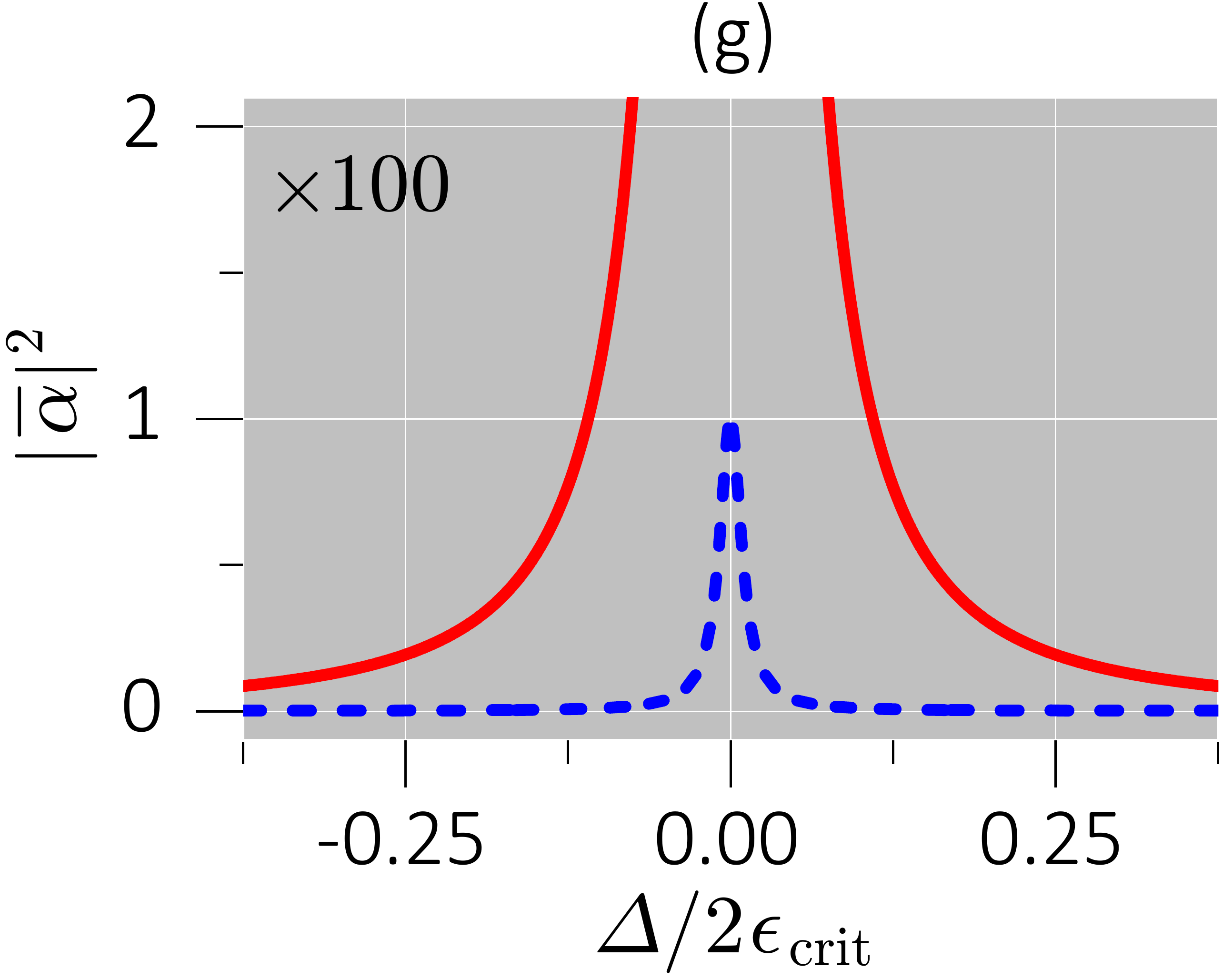}
\end{center}
\caption{Mean-field steady states for $\eta=1$ and $\Delta_0=\Delta$: $\kappa/\lambda=0.02$  and $\epsilon/\epsilon_{\rm crit}=0.6$ [(b),(c)], $\epsilon/\epsilon_{\rm crit}=1.0$ [(d),(e)], and $\epsilon/\epsilon_{\rm crit}=1.2$ [(f),(g)]. The three sweeps through the phase diagram are indicated by dashed lines in (a); solid red (dashed blue) lines indicate stable (unstable) steady states in (b)-(g); dashed black lines demark the range of bistability in (c).}
\label{fig:fig5}
\end{figure}
\begin{description}
\item[Region $R_4$]
Two solutions that approach $\bar\zeta=\pm1$ in the limit of zero drive and two that approach the root $\bar\zeta_+=-\bar\kappa^2-\bar\Delta^2$ of $[P(\bar\zeta)]^2=0$; the solutions that approach $\bar\zeta=\bar\zeta_+$ ($\pm1$) are stable (unstable); the solution that approaches $\bar\zeta=-1$ links in a closed loop to one of the solutions approaching $\bar\zeta_+$. Four solutions in total.
\item[Region $R_2^b$]
Two solutions that approach $\bar\zeta=\pm1$ in the limit of large detuning; the solution approaching $\bar\zeta=-1$ ($+1$) is stable (unstable). Two solutions in total.
\end{description}
We note the following additional points:
\begin{enumerate}[(i)]
\item
Two of the four solutions in region $R_4$ are consistent with the assumption adopted above Eq.~(\ref{eqn:alpha_eta=1_approx}) [frame (c) of Fig.~\ref{fig:fig5}] so long as $\bar\kappa\ll1$; the remaining two solutions satisfy Eq.~(\ref{eqn:alpha_eta=1}) but do not admit the approximation leading to Eq.~(\ref{eqn:alpha_eta=1_approx}).
\item
The boundary between regions $R_4$ and $R_2^b$ in frame (a) of Fig.~\ref{fig:fig5} follows the curve
\begin{equation}
\bar\epsilon=\left\{1-\left[\frac{|\bar\Delta_0|}{|\bar\Delta|}(\bar\kappa^2+\bar\Delta^2)\right]^{2/3}\right\}^{3/2}.
\end{equation}
The boundary is a line of double roots of Eq.~(\ref{eqn:6th-order_polynomial_eta=1}), and the curve may be found by equating derivatives on the left- and right-hand sides of this equation.
\item
The critical point $\epsilon_{\rm crit}$ [Eq.~(\ref{eqn:critical_drive})] organizes behavior as a function of drive strength and detuning in much the same way as it does for  $\eta=0$.
\item
The closed loop in frame (b) of Fig.~\ref{fig:fig5} is similar to the loop in frame (b) of Fig.~\ref{fig:fig4}; both shrink with increasing drive strength to eventually vanish at the critical point---frames (d) of Figs.~\ref{fig:fig4} and \ref{fig:fig5}. Note, though, that the stabilities are interchanged; this change is clearly reflected in the accompanying plots of $|\bar\alpha|^2$ [frames (c) of Figs.~\ref{fig:fig4} and \ref{fig:fig5}].
\item
The stable solutions displayed in frames (c), (e), and (g) of Fig.~\ref{fig:fig5} are all single nearly Lorentzian peaks; the splitting in the corresponding frames of Fig.~\ref{fig:fig4} does not occur.
\end{enumerate}

\subsection{Intermediate regime: $0<\eta<1$}
\label{sec:coherent_drive_intermediate_eta}
Summarizing what we have learned: with no counter-rotating interaction, the dissipative Dicke system shows no phase transition as a function of coupling strength [$\eta=0$ in frames (b) and (d) of Fig.~\ref{fig:fig1}], although the breakdown of photon blockade takes place in the presence of a coherent drive (Fig.~\ref{fig:fig4}); the dissipative system does, however, show the standard phase transition when $\eta=1$, where it is deformed by a coherent drive and vanishes with increasing drive strength at a renormalized photon-blockade-breakdown critical point (Fig.~\ref{fig:fig5}).

In this section we unify these limiting cases by letting $\eta$ vary continuously between 0 and 1. We show how the previously unreported phase of the Dicke system, i.e., region $R_4$ of Figs.~\ref{fig:fig2} and \ref{fig:fig3}, underlies this unification.

We begin with the interface between frame (a) of Fig.~\ref{fig:fig3} and frame (a) of Fig.~\ref{fig:fig5}, where regions of three and four distinct solutions connect on the boundary $\bar\epsilon=0$, $\eta=1$: moving off the boundary with a perturbation $\bar\epsilon\to\delta\bar\epsilon$ lifts the degeneracy of a double root of Eq.~(\ref{eqn:6th-order_polynomial_epsilon=0}), and thus provides the link between regions. Something similar is encountered on the $\bar\epsilon=0$ boundary with $\eta_\kappa<\eta<1$ (e.g., along the lines $\eta=0.6$ and $\eta=0.2$ in Fig.~\ref{fig:fig3}); however, now two regions, $R_3$ and $R_4$, link to contiguous regions under the perturbation $\bar\epsilon\to\delta\bar\epsilon$. Since $R_4$ accommodates \emph{two} double roots of Eq.~(\ref{eqn:6th-order_polynomial_epsilon=0}), we predict its linkage to a contiguous region of six distinct solutions in the presence of a coherent drive.

We illustrate this situation in Fig.~\ref{fig:fig6} where we plot the function $\sqrt{1-\bar\zeta^2}P(\bar\zeta)$---the square root of the left-hand side of Eq.~(\ref{eqn:6th-order_polynomial_epsilon=0})---for four detunings along the $\eta=0.2$ sweep of Fig.~\ref{fig:fig3}: frames (a), (b), (c), (d) refer, in sequence, to points in regions $R_2$, $R_3$, $R_4$, $R_2$ along the sweep---moving inwards from either end; they show examples of two, three, four, and again two distinct roots. The trivial roots, $\bar\zeta=\pm1$, appear in every plot, while the additional roots [frames (b) and (c)] are double roots of $[P(\bar\zeta)]^2=0$. The perturbation $\bar\epsilon\to\delta\bar\epsilon$ replaces each dashed line in the figure by a pair of curves $\pm\delta\bar\epsilon|\bar\zeta|\sqrt{Q(\bar\zeta)}$, and thus lifts the degeneracy of each double root. [It is readily shown that $Q(\bar\zeta)>0$.]

Figure~\ref{fig:fig7} shows how the results displayed in Figs.~\ref{fig:fig3}-\ref{fig:fig5} are generalized for $\eta=0.2$ and $\bar\Delta_0=\bar\Delta$. The parameter space is now divided into a larger number of regions, integrating those already met in the three limiting cases:
\begin{figure}[t]
\begin{center}
\includegraphics[width=1.7in]{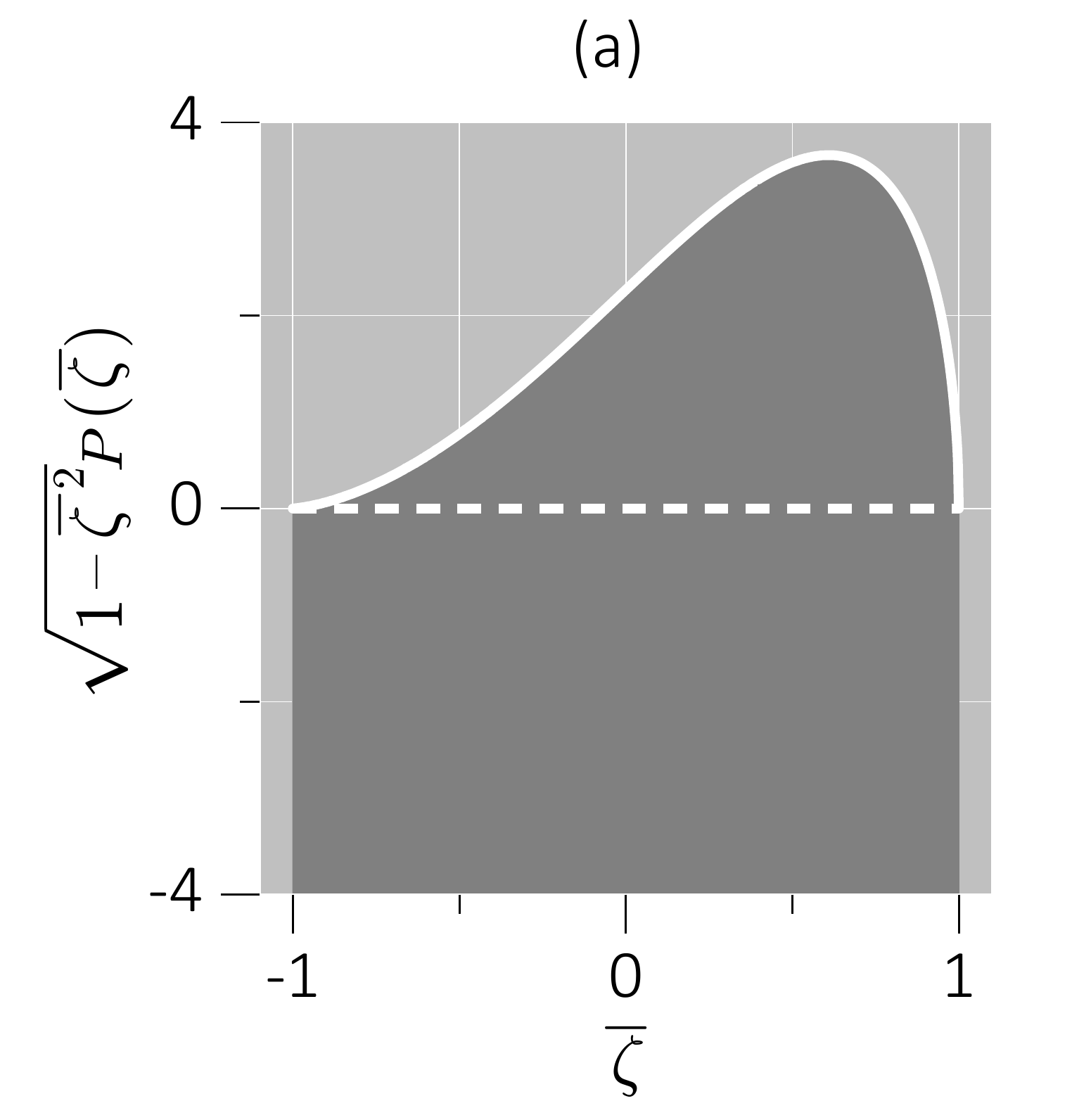}\includegraphics[width=1.7in]{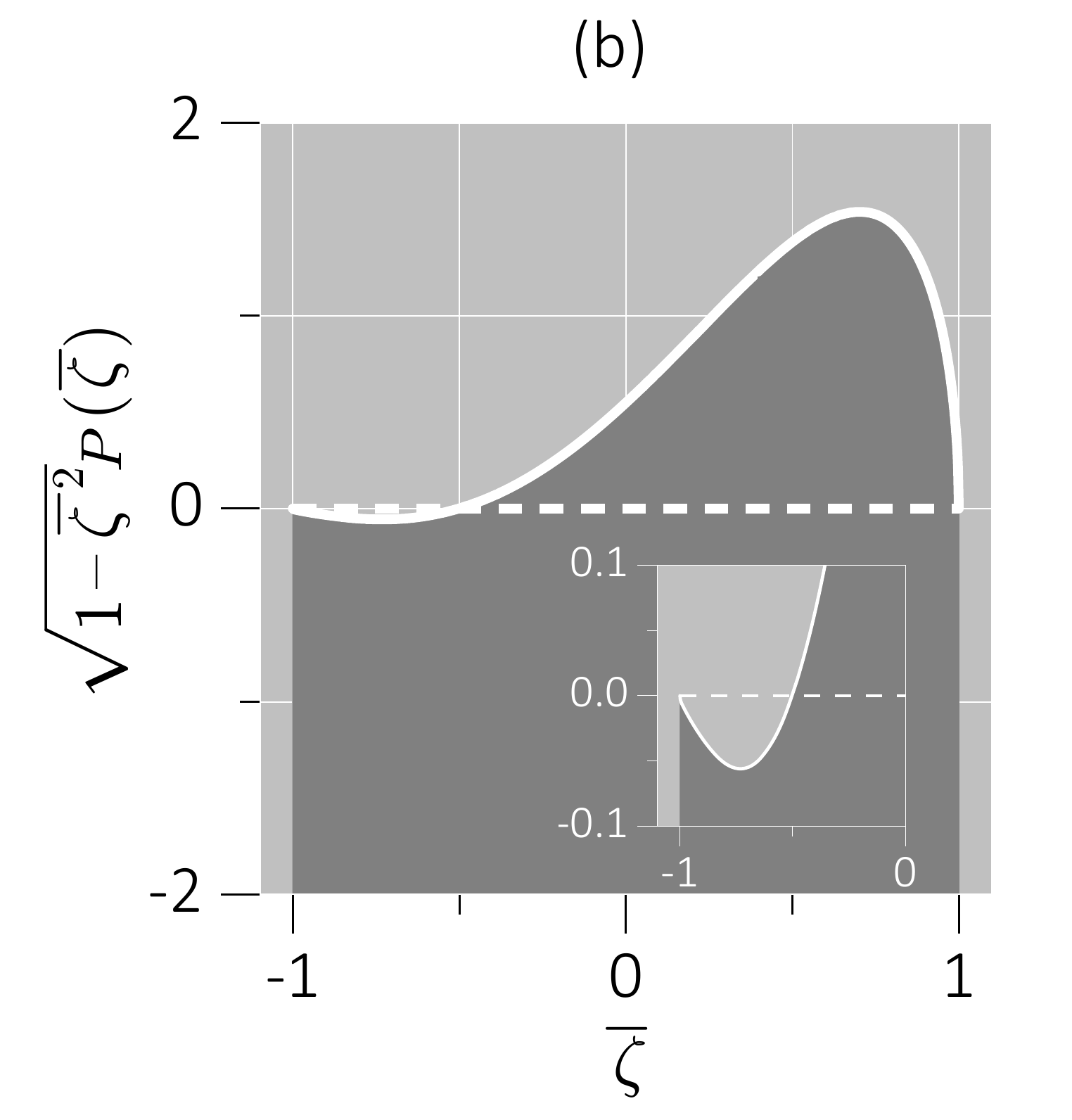}
\vskip0.1in
\includegraphics[width=1.7in]{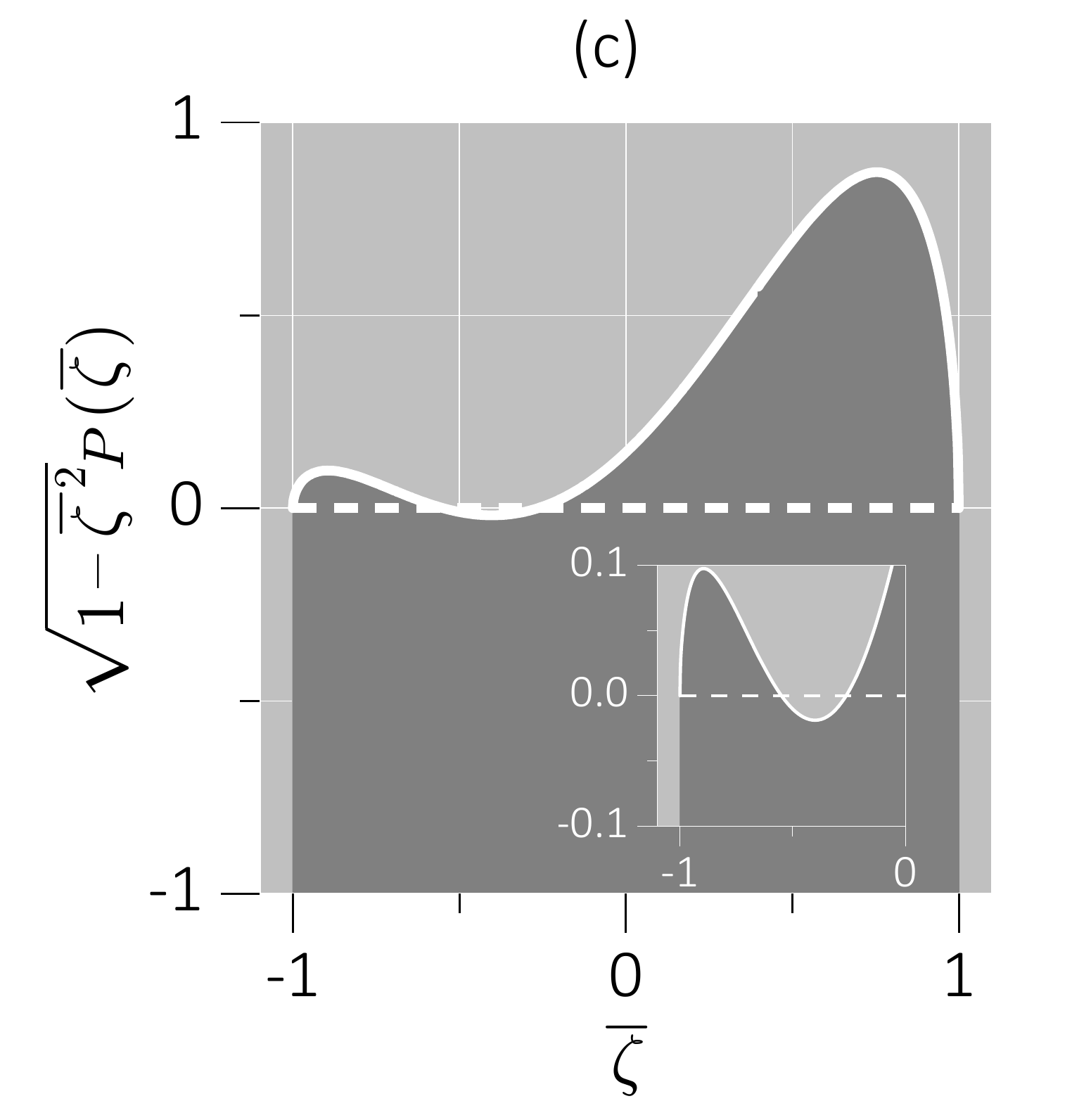}\includegraphics[width=1.7in]{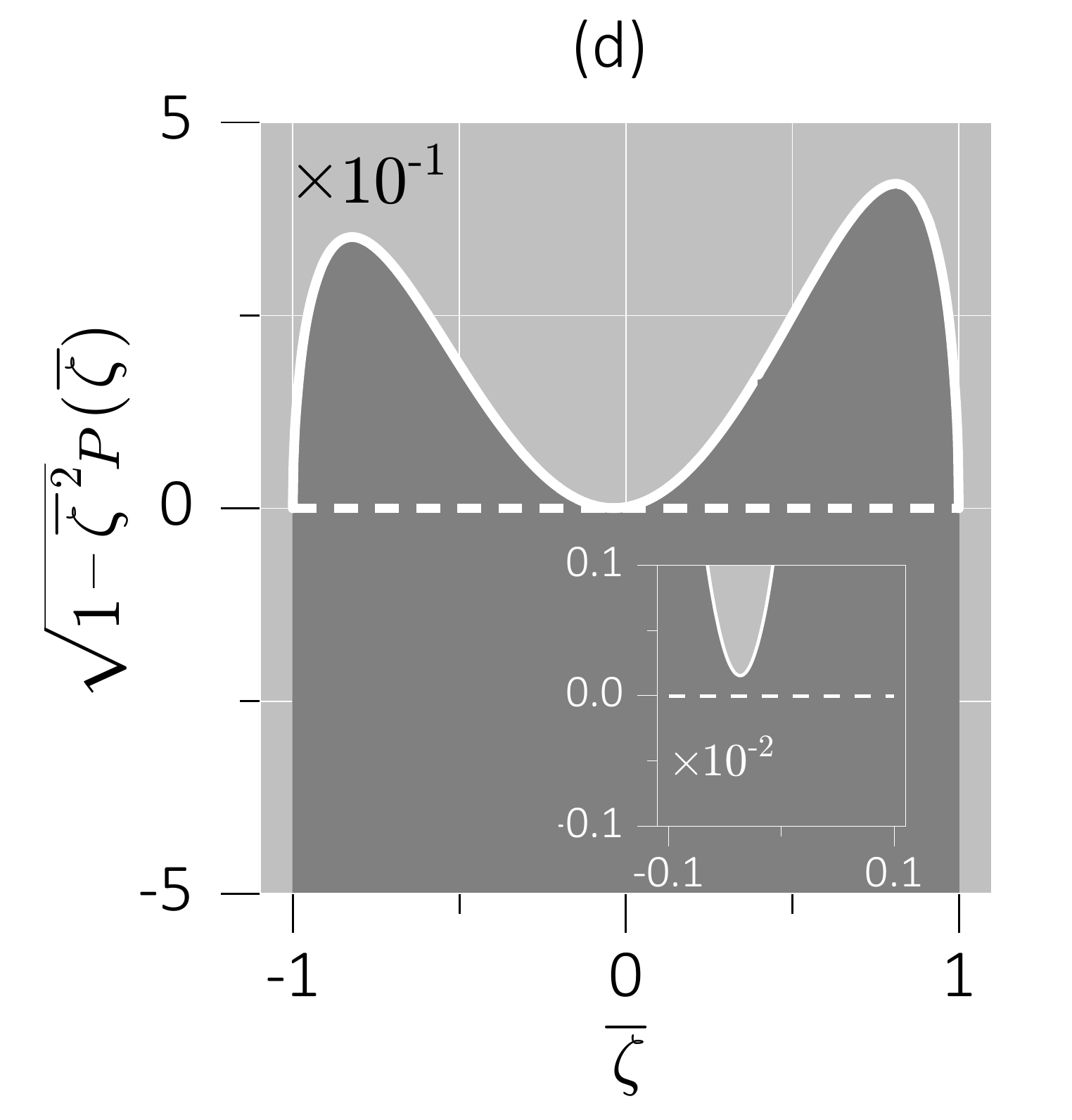}
\end{center}
\caption{The square root of the left-hand side of Eq.~(\ref{eqn:6th-order_polynomial_epsilon=0}) as a function of $\bar\zeta$ for $\eta=0.2$ and $\Delta_0=\Delta$: $\kappa/2\epsilon_{\rm crit}=1/12$ and $|\Delta|/2\epsilon_{\rm crit}=1.0$, $0.7$, $0.5$, $0.15$ [(a)-(d)]. Zeros of this function (crossings of the black dashed lines) are roots of Eq.~(\ref{eqn:6th-order_polynomial_epsilon=0}).}
\label{fig:fig6}
\end{figure}

\begin{description}
\item[Region $R_2^a$]
Two solutions that approach $\bar\zeta=\pm1$ in the limit of zero drive; the solution approaching $\bar\zeta=-1$ ($+1$) is stable (unstable). Two solutions in total.
\item[Region $R_6$]
Two solutions that approach $\bar\zeta=\pm1$ in the limit of zero drive and four additional solutions---two that approach each of the double roots, $\bar\zeta_\pm$, of $[P(\bar\zeta)]^2=0$. The solutions approaching $\bar\zeta=-1$ and $\bar\zeta_+$ ($+1$ and $\bar\zeta_-$) are stable (unstable). Six solutions in total.
\item[Region $R_4^a$]
Two solutions that approach $\bar\zeta=\pm1$ in the limit of zero drive and two that approach the double root $\bar\zeta_+$ of $[P(\bar\zeta)]^2=0$; the solutions approaching $\bar\zeta_+$ ($\pm1$) are stable (unstable). Four solutions in total.
\item[Region $R_4^b$]
Two solutions that approach $\bar\zeta=\pm1$ in the limit of zero drive and two additional solutions that arise from the bistable folding of the solution that approaches $\bar\zeta=-1$; the solution approaching $\bar\zeta=-1$ ($+1$) is stable (unstable), and the two additional solutions are one stable/unstable. Four solutions in total.
\item[Region $R_2^b$]
Two solutions that approach $\bar\zeta=\pm1$ in the limit of large detuning; the solution approaching $\bar\zeta=-1$ ($+1$) is stable (unstable). Two solutions in total.
\end{description}

Frames (b)-(e) of Fig.~\ref{fig:fig7} show how the corresponding plots in Fig.~\ref{fig:fig3} change when the degeneracy of the double roots ($\bar\epsilon=0$) is lifted ($\bar\epsilon\neq0$) to link regions $R_3$ and $R_4$ of Fig.~\ref{fig:fig3} to regions $R_4^a$ and $R_6$, respectively, of Fig.~\ref{fig:fig7}. (Note, however, that $\kappa/\lambda$ takes different values in the figures, so region boundaries do not line up.) The change is clearly seen, for example, comparing frame (b) of Fig.~\ref{fig:fig3} with frames (b) and (d) of Fig.~\ref{fig:fig7}: a single stable upper branch---Fig.~\ref{fig:fig3}---is split into two stable upper branches---Fig.~\ref{fig:fig7}; and a single unstable branch connecting upper and lower stable branches in Fig.~\ref{fig:fig3}  splits into two unstable branches in Fig.~\ref{fig:fig7} [near overlapping dashed lines in frame (d)]. In this way features met separately in the limiting cases of Figs.~\ref{fig:fig4} and \ref{fig:fig5} are linked together.

Finally, for larger amplitudes of the drive---e.g., adding sweeps at $\bar\epsilon=0.6$, 1.0, and 1.2 in frame (a) of Fig.~\ref{fig:fig7}---mean-field steady states follow the breakdown of photon blockade, as in frames (b)-(g) of Fig.~\ref{fig:fig4}.

\begin{figure}[t]
\begin{center}
\includegraphics[width=3.4in]{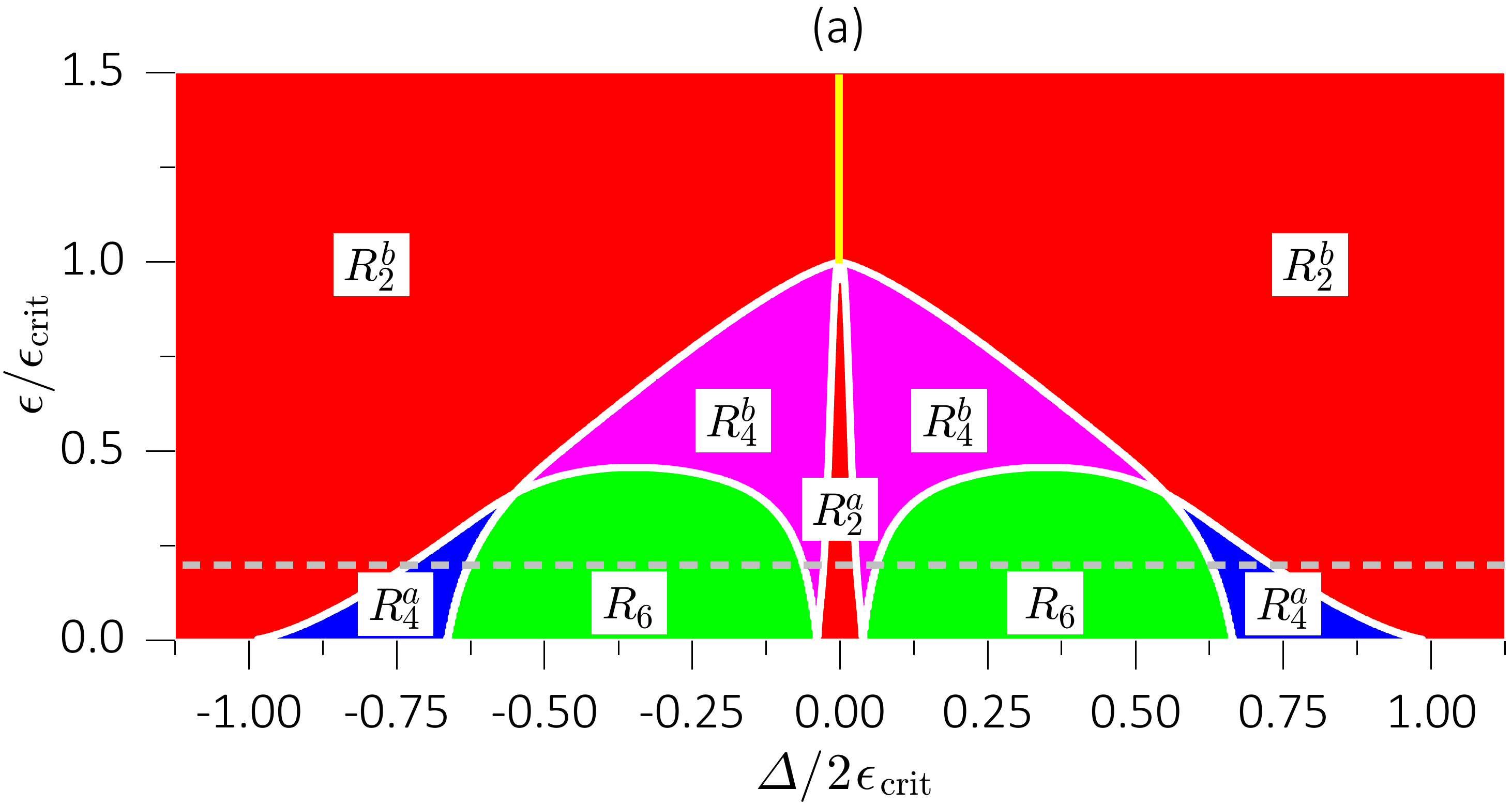}
\vskip0.1in
\includegraphics[width=1.65in]{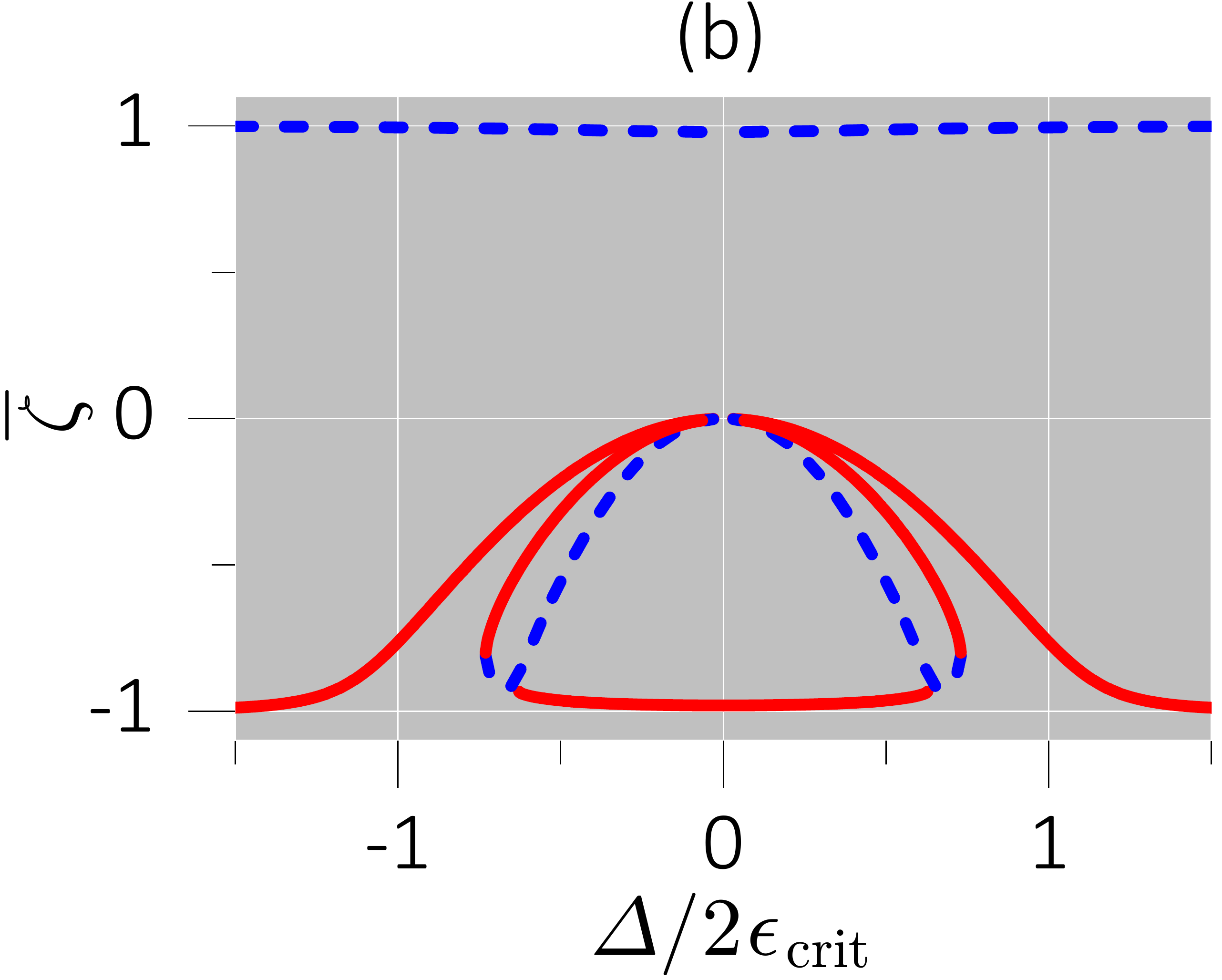}\hskip0.075in\includegraphics[width=1.65in]{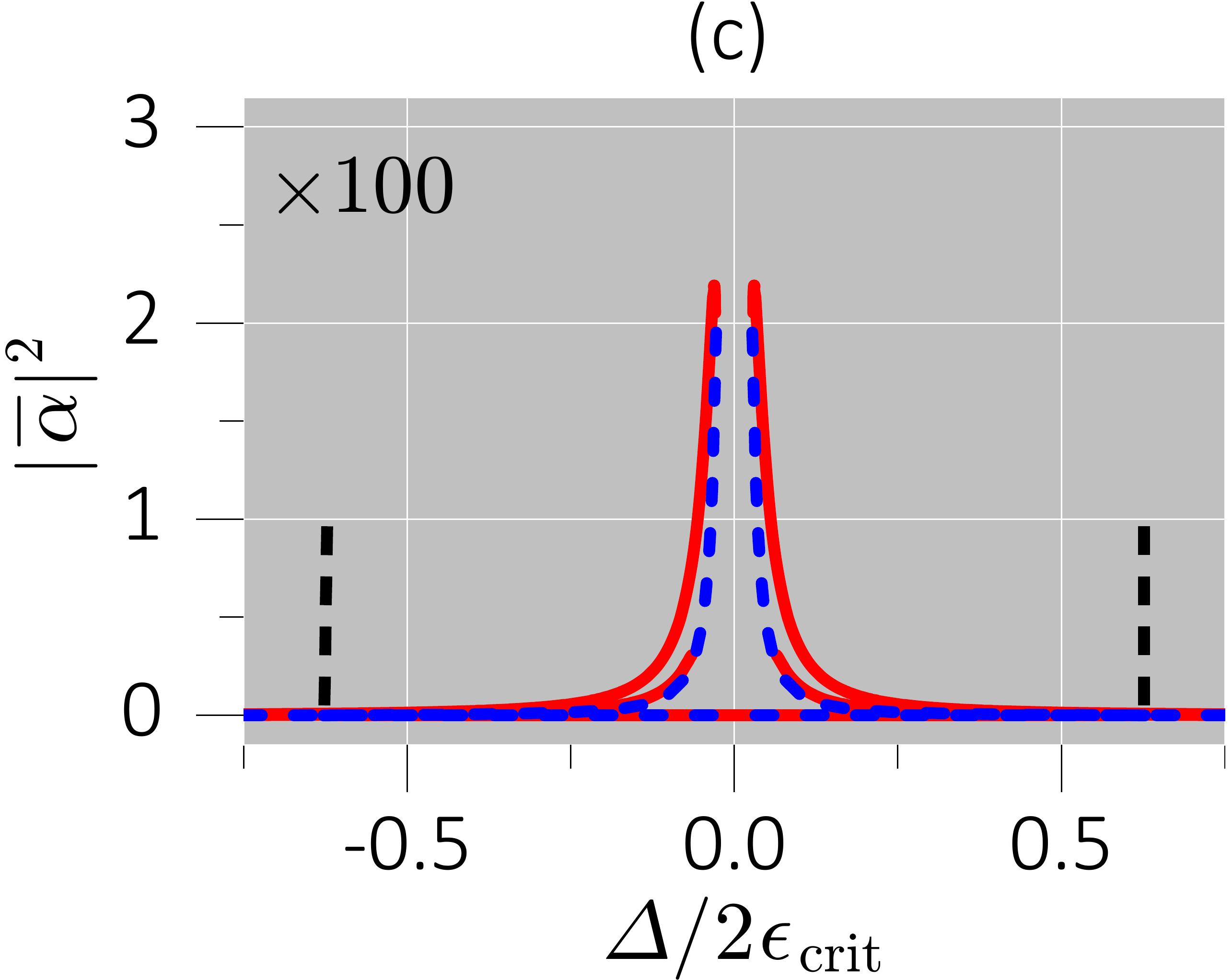}
\vskip0.1in
\includegraphics[width=1.65in]{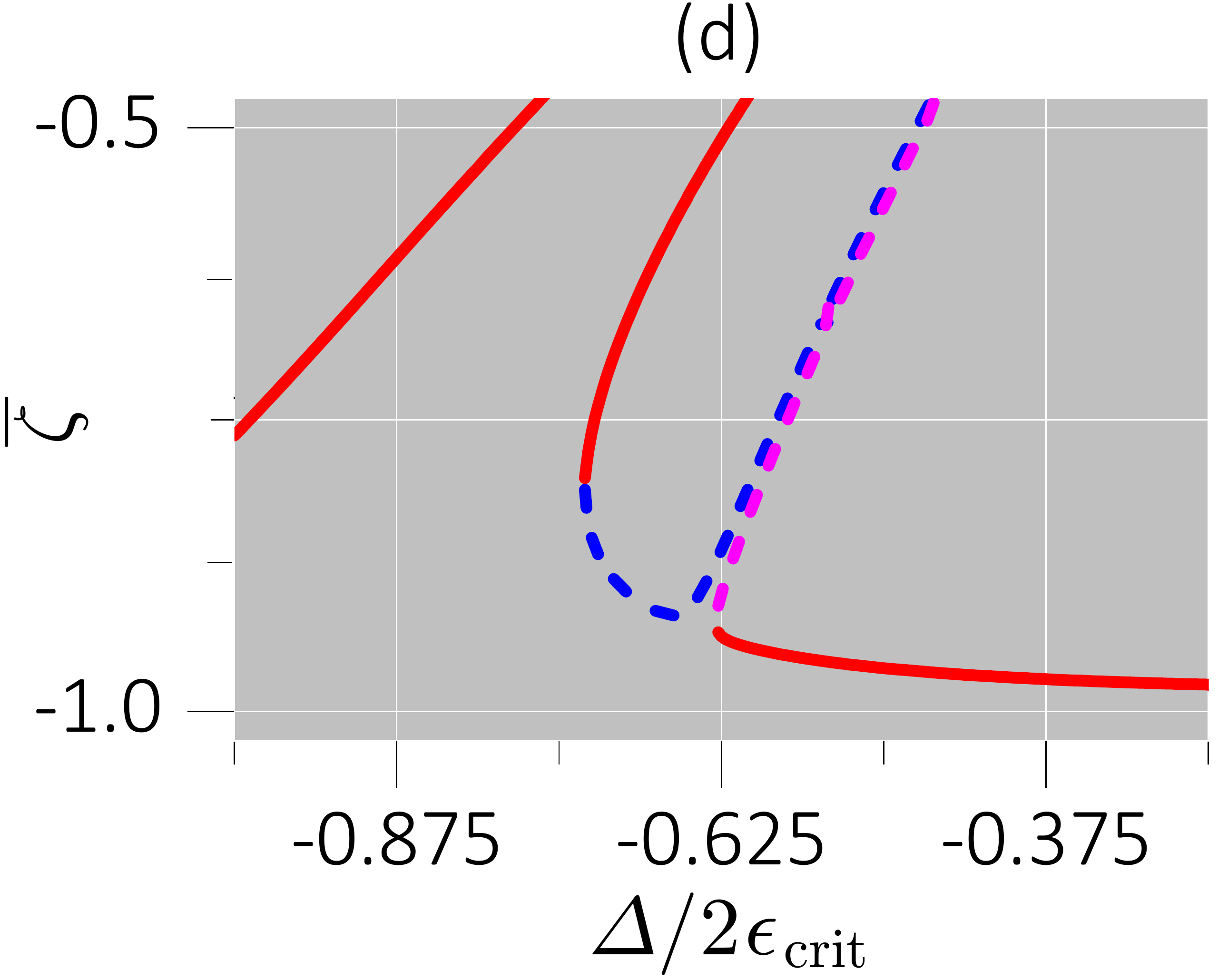}\hskip0.075in\includegraphics[width=1.65in]{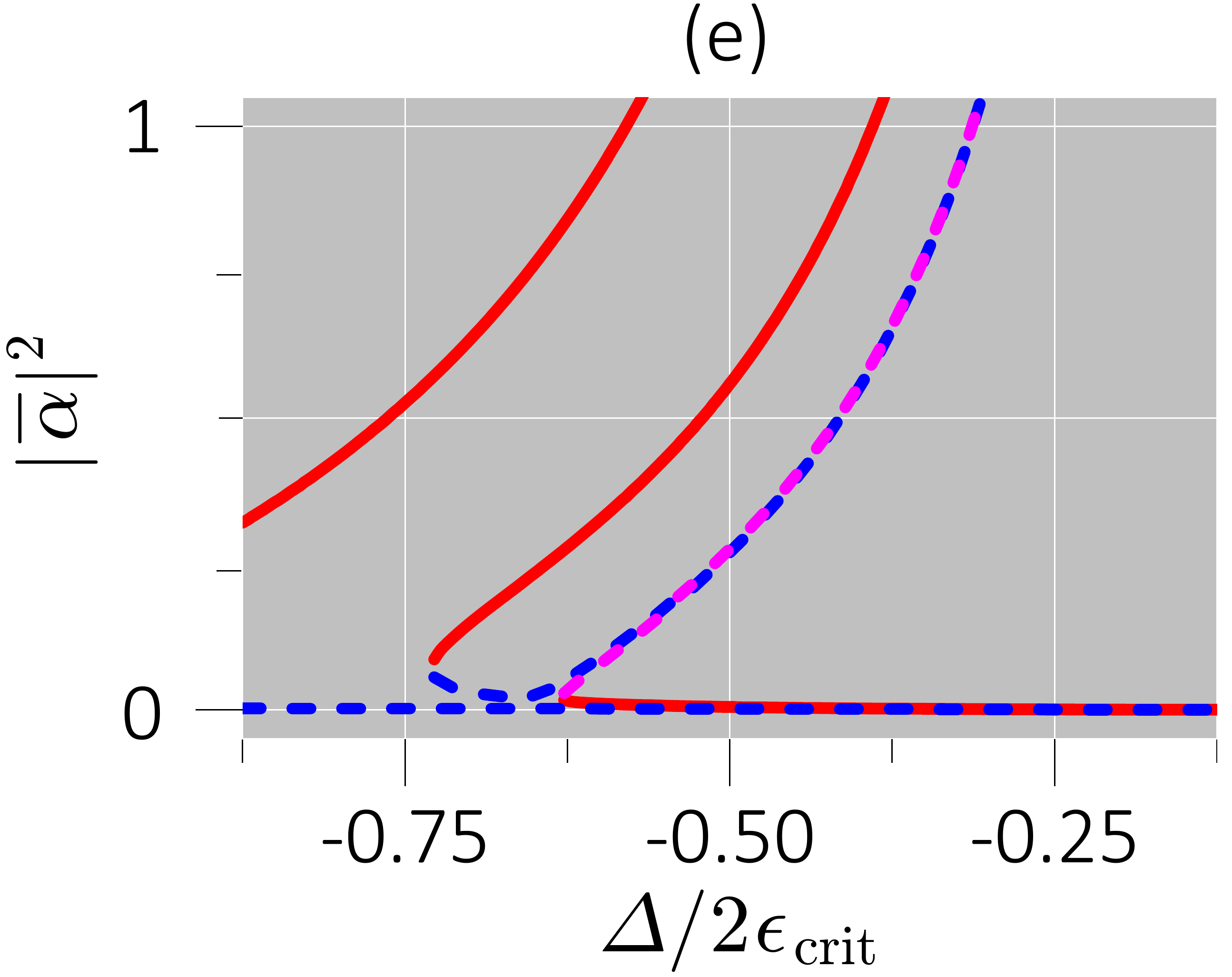}
\end{center}
\caption{Mean-field steady states for $\eta=0.2$ and $\Delta_0=\Delta$: $\kappa/\lambda=0.02$ and $\epsilon/\epsilon_{\rm crit}=0.2$; [(d),(e)] expands the view in [(b),(c)]. The sweep through the phase diagram is indicated by the dashed line in (a); solid red (dashed blue and magenta) lines indicate stable (unstable) steady states in (b)-(e); dashed black lines demark the range of bistability or tristability in (c).}
\label{fig:fig7}
\end{figure}

\section{Quantum Fluctuations: One Two-State System}
\label{sec:quantum_fluctuations}
While the mean field analysis may be highly suggestive of what to expect from an experimental realization of our generalized Jaynes-Cummings-Rabi model, an account in these terms is incomplete---fluctuations are neglected. We encounter coexisting steady states, for example, and although both are stable under small perturbations when Maxwell-Bloch equations are solved, what of the stability once quantum fluctuations are introduced?

It is beyond the scope of this work to address questions like this in any detail. We limit ourselves here to a few observations about the full quantum treatment for the case $N=1$, where a number of calculations are feasible, some analytical and some numerical, to parallel results for the breakdown of photon blockade \cite{carmichael_2015}. While it may seem that $N=1$ takes us very far from a many-particle limit where contact with mean-field results may be made, this is not generally the case: it is shown in Ref.~\cite{carmichael_2015} that the many-photon limit is a strong-coupling limit, and many of the figures from Sec.~\ref{sec:mean-field} have photon numbers ranging in the hundreds for $N=1$---after the scaling of Eq.~(\ref{eqn:scaling}) is undone.

In this section, we show that the $\eta$-dependence of the critical drive strength (Sec.~\ref{sec:critical_drive}) follows from the quasi-energy spectrum, extending the previous calculation of the spectrum for $\eta=0$~\cite{alsing_etal_1992} to the general case. We then address the role of multi-photon resonances in the limit of small $\eta$, where we uncover behavior similar to multi-photon blockade \cite{shamailov_etal_2010} under weak coherent driving, but only for even numbers of photons absorbed. Finally, we use quantum trajectories to explore the accessibility of co-existing mean-field steady states in the presence of fluctuations.

\subsection{Quasienergies for $\Delta_0=\Delta=0$}

Ever since the seminal work of Jaynes and Cummings \cite{jaynes&cummings_1963} (see also \cite{paul_1963})), the energy spectrum of a single two-state system interacting with one mode of the radiation field has been a fundamental element of quantum optics models and physical understanding. The level scheme is remarkably simple when compared with extensions to the quantum Rabi model \cite{braak_2011} and generalizations to include a counter-rotating interaction after the manner of Sec.~\ref{sec:Dicke_counter_rotating} \cite{tomka_etal_2014}. Alsing \emph{et al}. \cite{alsing_etal_1992} showed that the simplicity carries over to the driven Jaynes-Cummings Hamiltonian when the two-state system and radiation mode are resonant with the drive. In this case, a Bogoliubov transformation diagonalizes the interaction picture Hamiltonian, so that quasienergies are recovered. The critical drive $\epsilon_{\rm crit}$ is then the point at which all quasienergies collapse to zero. In this section we show that the method employed by Alsing \emph{et al}. carries through for arbitrary $\eta$, and the collapse to zero reproduces Eq.~(\ref{eqn:critical_drive}).

We consider the Hamiltonian $H_\eta^{\prime\prime}=H_\eta^\prime+\sqrt N\epsilon(a^\dagger+a)$, where $H_\eta^\prime$ is given by Eq.~(\ref{eqn:hamiltonian_raman_model}). Taking the coherent drive on resonance and considering just one two-state system, the Hamiltonian is
\begin{equation}
H_\eta^{\prime\prime}=\lambda(a\sigma_++a^\dagger\sigma_-)+\eta\lambda(a^\dagger\sigma_++a\sigma_-)+\epsilon(a^\dagger+a).
\label{eqn:hamiltonian_zero_detuning_one_atom}
\end{equation}
We seek solutions to the eigenvalue problem $H_\eta^{\prime\prime}|\psi_E\rangle=E|\psi_E\rangle$, where $E$ is a quasienergy and
\begin{equation}
|\psi_E\rangle=|\psi^{(1)}_E\rangle|1\rangle+|\psi^{(2)}_E\rangle|2\rangle,
\label{eqn:eigenket}
\end{equation}
with the kets $|\psi^{(1,2)}_E\rangle$ expanded over the Fock states, $|n\rangle$, $n=1,2,\ldots$, of the field mode; we must find allowed values of $E$ and the corresponding field kets.

It is straightforward to show that the field kets satisfy the homogeneous system of equations
\begin{equation}
\left(
\begin{matrix}
\epsilon(a^\dagger+a)-E&\lambda(a^\dagger+\eta a)\\
\noalign{\vskip4pt}
\lambda(\eta a^\dagger+a)&\epsilon(a^\dagger+a)-E
\end{matrix}
\mkern2mu\right)\mkern-4mu\left(
\begin{matrix}
|\psi^{(1)}_E\rangle\\
\noalign{\vskip4pt}
|\psi^{(2)}_E\rangle
\end{matrix}
\right)=0,
\label{eqn:quasienergy_homogeneous_system}
\end{equation}
whence multiplication on the left by ${\rm diag}(\eta a^\dagger+a,a^\dagger+\eta a)$ takes us to the coupled equations:
\begin{eqnarray}
-\epsilon(1-\eta)|\psi^{(1)}_E\rangle&=&[\epsilon(a^\dagger+a)-E](\eta a^\dagger+a)|\psi^{(1)}_E\rangle\notag\\
\noalign{\vskip2pt}
&&+\lambda[aa^\dagger+\eta(a^{\dagger 2}+a^2)+\eta^2a^\dagger a]|\psi^{(2)}_E\rangle,\notag\\
&&\label{eqn:eigenket_again1}\\
\epsilon(1-\eta)|\psi^{(2)}_E\rangle&=&[\epsilon(a^\dagger+a)-E](a^\dagger+\eta a)|\psi^{(2)}_E\rangle\notag\\
\noalign{\vskip2pt}
&&+\lambda[a^\dagger a+\eta(a^{\dagger 2}+a^2)+\eta^2aa^\dagger]|\psi^{(1)}_E\rangle.\notag\\
&&\label{eqn:eigenket_again2}
\end{eqnarray}
We then use Eq.~(\ref{eqn:quasienergy_homogeneous_system}) to substitute for $(\eta a^\dagger+a)|\psi^{(1)}_E\rangle$ and $(a^\dagger+\eta a)|\psi^{(2)}_E\rangle$, respectively, on the right-hand sides of Eqs.~(\ref{eqn:eigenket_again1}) and (\ref{eqn:eigenket_again2}), and thus obtain the more compact form:
\begin{eqnarray}
\left(O(E)-\lambda^2\frac{1-\eta^2}2\right)|\psi^{(1)}_E\rangle-\epsilon\lambda(1-\eta)|\psi^{(2)}_E\rangle&=&0,\mkern20mu
\label{eqn:eigenket_yet_again1}\\
\noalign{\vskip2pt}
\left(O(E)+\lambda^2\frac{1-\eta^2}2\right)|\psi^{(2)}_E\rangle+\epsilon\lambda(1-\eta)|\psi^{(1)}_E\rangle&=&0,
\label{eqn:eigenket_yet_again2}
\end{eqnarray}
with
\begin{eqnarray}
O(E)&=&\lambda^2(1+\eta^2)\frac{a^\dagger a+aa^\dagger}2+\lambda^2\eta\left(a^{\dagger 2}+a^2\right)\notag\\
&&-\left[\epsilon\left(a^\dagger+a\right)-E\right]^2.
\label{eqn:o_operator}
\end{eqnarray}
Since the coefficients of the second terms on the left-hand side are constants, Eqs.~(\ref{eqn:eigenket_yet_again1}) and (\ref{eqn:eigenket_yet_again2}) can now be readily uncoupled, and yield the autonomous equation
\begin{equation}
O_+(E)O_-(E)|\psi^{(1,2)}_E\rangle=0,
\label{eqn:oplus_by_ominus_equation}
\end{equation}
where
\begin{equation}
O_{\pm}(E)=O(E)\pm\lambda^2\frac{1-\eta^2}2\Lambda^{1/2},
\label{eqn:oplusminus_operator}
\end{equation}
with
\begin{equation}
\Lambda=1-\frac1{(1+\eta)^2}\frac{4\epsilon^2}{\lambda^2}.
\label{eqn:Lambda_definition}
\end{equation}

Note now that $O_+(E)$ and $O_-(E)$ commute, and so the general solution to Eq.~(\ref{eqn:oplus_by_ominus_equation}) expands as
\begin{equation}
|\psi_{E}^{(1,2)}\rangle=c_+^{(1,2)}|\psi_E^{(+)}\rangle+c_-^{(1,2)}|\psi_E^{(-)}\rangle,
\label{eqn:oplus_by_ominus_solution}
\end{equation}
where $|\psi_E^{(+)}\rangle$ and $|\psi_E^{(-)}\rangle$ satisfy
\begin{equation}
O_{\pm}(E)|\psi_E^{(\pm)}\rangle=0.
\label{eqn:oplusminus_equation}
\end{equation}
Moreover, the operators $O_{\pm}(E)$ are quadratic in creation and annihilation operators, so the diagonalization may be completed by a Bogoliubov transformation: introduce parameters $\nu$, $\xi$, $\alpha(E)$, and $\mu_\pm(E)$, such that
\begin{equation}
O_\pm(E)=\nu U^\dagger[\xi,\alpha(E)]\frac{a^\dagger a+aa^\dagger}2U[\xi,\alpha(E)]+\mu_\pm(E),
\label{eqn:oplusminus_unitary}
\end{equation}
where the unitary $U[\xi,\alpha(E)]\equiv D[\alpha(E)]S(\xi)$ executes a displacement and then a squeeze,
\begin{equation}
a\buildrel U\over\to[a+\alpha(E)]\cosh\xi+[a^\dagger+\alpha(E)]\sinh\xi,
\label{eqn:a_transform}
\end{equation}
whence, from Eq.~(\ref{eqn:oplusminus_equation}),
\begin{equation}
\left(\frac{a^\dagger a+aa^\dagger}2+\frac{\mu_\pm(E)}\nu\right)\!\left\{U[\xi,\alpha(E)]|\psi_E^{(\pm)}\rangle\right\}=0.
\end{equation}
The number operator now acts on the left-hand side, and $|\psi_E^{(+)}\rangle$ and $|\psi_E^{(-)}\rangle$ are displaced and squeezed Fock states:
\begin{equation}
|\psi_{E_{n_\pm}}^{(\pm)}\rangle=U^\dagger[\xi,\alpha(E_{n_\pm})]|n_\pm\rangle,
\end{equation}
$n_\pm=0,1,2,\ldots$, where allowed quasienergies are indexed by the integers $n_\pm$ and must satisfy
\begin{equation}
n_\pm+\frac12+\frac{\mu_\pm(E_{n_\pm})}\nu=0.
\label{eqn:quasienergy_constraint}
\end{equation}
It remains to equate terms on both sides of Eq.~(\ref{eqn:oplusminus_unitary}) to fix the parameters of the Bogoliubov transformation, which yields
\begin{equation}
\nu=\lambda^2(1-\eta^2)\Lambda^{1/2},\qquad
\xi=\frac12\ln\left(\frac{1+\eta}{1-\eta}\Lambda^{1/2}\right),
\label{eqn:paramters1}
\end{equation}
and
\begin{eqnarray}
\alpha(E)&=&\frac{2\epsilon E}{\lambda^2(1+\eta)^2}\Lambda^{-1},\\
\noalign{\vskip4pt}
\mu_\pm(E)&=&\pm\frac\nu2-E^2\Lambda^{-1},
\label{eqn:parameters2}
\end{eqnarray}
and thus the allowed quasienergies follow from
\begin{equation}
n_\pm+\frac12\pm\frac12-E_{n_{\pm}}^2\frac1{\lambda^2(1-\eta^2)}\Lambda^{-3/2}=0.
\label{eqn:energies1}
\end{equation}

Equation (\ref{eqn:energies1}) is the targeted result, which reveals the generalized critical drive strength. It is helpful, however, for clarity, to recognize that $n_+$ and $n_-$ provide a double coverage of the nonnegative integers---traced to the \emph{two} components on the right-hand side of Eq.~(\ref{eqn:oplus_by_ominus_solution})---and to replace $n_\pm$ by a single index $n$: first, associate $n=0$ with $n_-=0$, from which Eq.~(\ref{eqn:energies1}) yields the quasienergy
\begin{equation}
E_0=0,
\label{eqn:energy_zero}
\end{equation}
with corresponding ket
\begin{equation}
|\psi_{E_0}^{(-)}\rangle=U^\dagger[\xi,\alpha(E_0)]|0\rangle;
\end{equation}
and second, associate $n=1,2,\dots$ with both $n_+=n-1$ and $n_-=n$, both of which, when substituted in Eq.~(\ref{eqn:energies1}), yield the quasienergy doublet
\begin{equation}
E_{n,\pm}=\pm\lambda\sqrt n\sqrt{1-\eta^2}\Lambda^{3/4},
\label{eqn:quasienergy_doublet}
\end{equation}
although with distinct corresponding kets:
\begin{eqnarray}
|\psi_{E_{n,\pm}}^{(+)}\rangle&=&U^\dagger[\xi,\alpha(E_{n,\pm})]|n-1\rangle,\\
\noalign{\vskip2pt}
|\psi_{E_{n,\pm}}^{(-)}\rangle&=&U^\dagger[\xi,\alpha(E_{n,\pm})]|n\rangle.
\end{eqnarray}

It is clear from Eq.~(\ref{eqn:quasienergy_doublet}) that all quasienergies collapse to zero for $n$ finite and $\Lambda=0$, a condition that returns, from Eq.~(\ref{eqn:Lambda_definition}), the critical drive strength $\epsilon_{\rm crit}$ [Eq.~(\ref{eqn:critical_drive})]. From this fully quantum mechanical point of view, $\epsilon_{\rm crit}$ marks a transition from a discrete quasienergy spectrum to a continuous one; the continuous side is recovered from the limit $\Lambda\to0$, $n\to\infty$, $\sqrt n\Lambda^{3/4}$ constant. Note that a continuous spectrum is also recovered in the limit $\eta\to1$, $n\to\infty$, $\sqrt n\sqrt{1-\eta^2}$ constant. A continuous spectrum is expected for $\eta=1$, since if we set $\eta=1$ in Eq.~(\ref{eqn:quasienergy_homogeneous_system}), $E$ is an eigenvalue of the quadrature operator $a^\dagger+a$.

The coefficients $c_\pm^{(1,2)}$ [Eq.~(\ref{eqn:oplus_by_ominus_solution})] follow from Eqs.~(\ref{eqn:quasienergy_homogeneous_system}) and (\ref{eqn:eigenket_yet_again1}), and normalization (see Ref.~\cite{alsing_etal_1992}).

\subsection{Multi-photon resonance}
With the focus on just one two-state system, Figs.~\ref{fig:fig4}, \ref{fig:fig5}, and \ref{fig:fig7} show photon numbers ranging from zero to a few thousand, and although numbers are smaller in Fig.~\ref{fig:fig3}, the range is similar when $\kappa/\lambda$ is set to $0.02$ instead of $0.1$. While we might expect mean-field theory to be broadly reliable for thousands, even hundreds of photons, it will surely miss important features when photon numbers are small. Indeed, photon blockade is a photon by photon effect, underpinned, not by a mean-field nonlinearity, but by a strongly anharmonic ladder of few-photon excited states; it breaks down through multi-photon absorption, where, in Fig.~4 of Ref.~\cite{carmichael_2015}, for example,  multi-photon resonances dominate the response to weak driving and the mean-field story of dispersive bistability is not picked up until $\epsilon/\epsilon_{\rm crit}\sim0.4$.

Recall now that in its dissipate realization (Sec.~\ref{sec:dissipative_realization}) our generalized model involves not one, but two external
drives---a linear drive of strength $\epsilon$, and a second, nonlinear drive of strength $\eta$. We show now that the multi-photon response to weak driving carries over, with minor modification, from linear to nonlinear driving.

Reinstating detuning and setting $\Delta_0=\Delta$, we consider the Hamiltonian $H^{\prime\prime\prime}_\eta=\Delta a^\dagger a+\Delta\sigma_z+H^{\prime\prime}_\eta$, where $H^{\prime\prime}_\eta$ is given by Eq.~(\ref{eqn:hamiltonian_zero_detuning_one_atom}). It is convenient for clarity, however, to adopt an interaction picture, where we define
\begin{eqnarray}
H_\eta^{\prime\prime}(t)&\equiv&U_0^\dagger(t)H^{\prime\prime}_\eta U_0(t)\notag\\
\noalign{\vskip2pt}
&=&H_{\rm JC}+H_{\epsilon}(t)+H_\eta(t),
\end{eqnarray}
$U_0(t)\equiv\exp[-i\Delta(a^\dagger a+\sigma_z)t]$, and thus isolate the Jaynes-Cummings interaction,
\begin{equation}
H_{\rm JC}=\lambda(a\sigma_++a^\dagger\sigma_-),
\end{equation}
which is perturbed by the linear drive
\begin{equation}
H_\epsilon(t)=\epsilon(a^\dagger e^{i\Delta t}+ae^{-i\Delta t}),
\end{equation}
and the nonlinear drive
\begin{equation}
H_\eta(t)=\eta\lambda(a^\dagger\sigma_+e^{2i\Delta t}+a\sigma_-e^{-2i\Delta t}).
\end{equation}
We also recall the eigenvalues and eigenkets of $H_{\rm JC}$:
\begin{equation}
E_0^{\rm JC}=0,\qquad E_{n,\pm}^{\rm JC}=\pm\lambda\sqrt n,
\end{equation}
$n=1,2,\ldots$, and
\begin{eqnarray}
|E_0^{\rm JC}\rangle&=&|0\rangle|1\rangle,\\
|E_{n,\pm}^{\rm JC}\rangle&=&\frac1{\sqrt2}\left(|n\rangle|1\rangle\pm|n-1\rangle|2\rangle\right),
\end{eqnarray}
where the first (second) ket refers to the field mode (two-state system) in each product on the right-hand side.

Note now that the perturbation $H_\epsilon(t)$ has non-zero matrix elements between neighboring pairs of kets in the $n$-step sequence
\begin{equation}
|E_0^{\rm JC}\rangle\rightarrow|E_{1,\pm}^{\rm JC}\rangle\rightarrow\cdots\rightarrow|E_{n-1,\pm}^{\rm JC}\rangle\rightarrow|E_{n,\pm}^{\rm JC}\rangle,
\end{equation}
$n=1,2,\ldots$, while $H_\eta(t)$ has non-zero matrix elements between pairs of kets in the $n/2$-step sequence
\begin{equation}
|E_0^{\rm JC}\rangle\rightarrow|E_{2,\pm}^{\rm JC}\rangle\rightarrow\cdots\rightarrow|E_{n-2,\pm}^{\rm JC}\rangle\rightarrow|E_{n,\pm}^{\rm JC}\rangle,
\end{equation}
$n=2,4,\ldots$. There are thus matrix elements to mediate multi-photon transitions from $|E_0^{\rm JC}\rangle$ to $|E_{n,\pm}^{\rm JC}\rangle$ driven by either perturbation, but with the qualification that $H_\eta(t)$ can only drive those with even $n$; resonance is achieved under the condition
\begin{equation}
\Delta=\pm\lambda/\sqrt n,
\end{equation}
which is met either by $n$ steps of $\Delta$ off-setting $\pm\lambda\sqrt n$, or $n/2$ steps of $2\Delta$.

\begin{figure}[t]
\begin{center}
\includegraphics[width=3.4in]{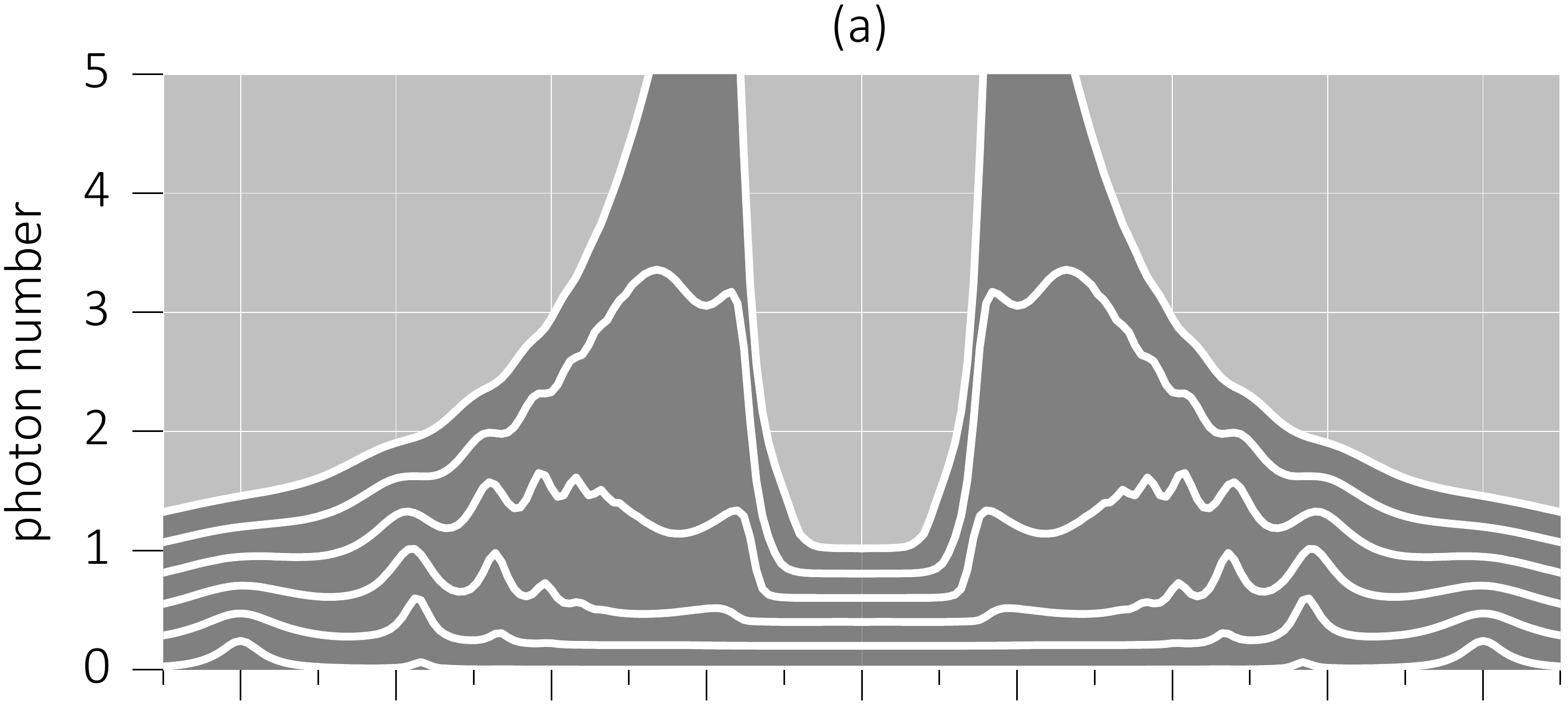}
\vskip0.1in
\includegraphics[width=3.4in]{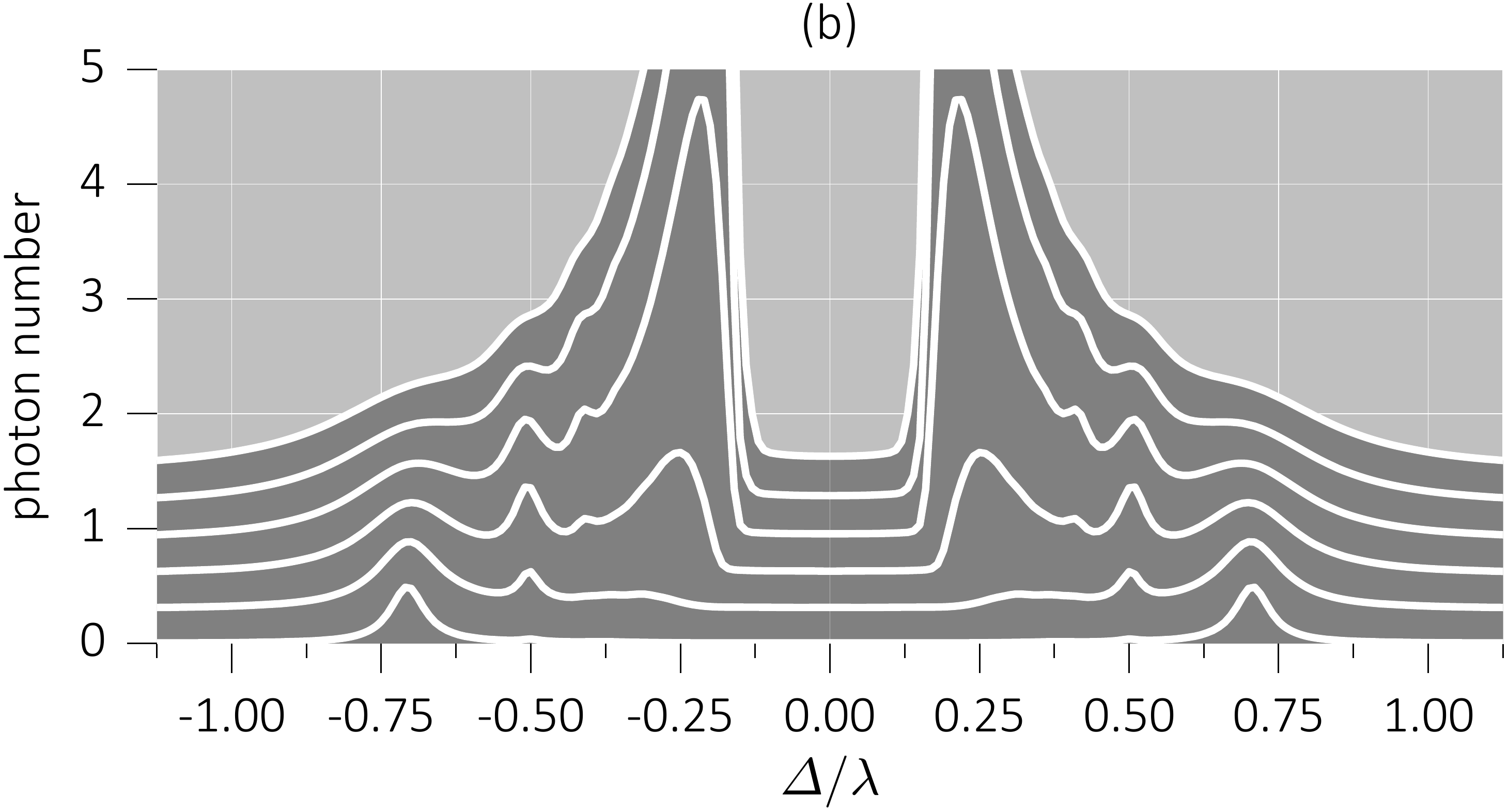}
\end{center}
\caption{Steady-state photon number expectation computed from the master equation, Eq.~(\ref{eqn:master_equation_drive}), for $N=1$, $\Delta_0=\Delta$, and $\kappa/\lambda=0.02$: (a) $\eta=0$ and $\epsilon/\epsilon_{\rm crit}=0.08$, $0.16$, $0.24$, $0.32$, $0.40$, $0.48$ (lower to upper) and (b) $\epsilon/\epsilon_{\rm crit}=0$ and $\eta=0.04$, $0.08$, $0.12$, $0.16$, $0.20$, $0.24$ (lower to upper); successive curves  are displace upwards by $0.2$ and $0.3$ in (a) and (b), respectively.}
\label{fig:fig8}
\end{figure}

Frame (a) of Fig.~\ref{fig:fig8} illustrates the breakdown of photon blockade from a fully quantum mechanical point of view; we identify up to six multi-photon resonances before they begin to merge and wash out due to power broadening at higher drives. This figure displays quantum corrections, for $N=1$, to the mean-field results of Fig.~\ref{fig:fig4}, where at high drives---$\epsilon/\epsilon_{\rm crit}=0.40$ and $0.48$---the layout of frame (a) of Fig.~\ref{fig:fig4} begins to appear with the photon number averaged over fluctuation-driven switching between the pair of coexisting mean-field steady states.

Frame (b) of Fig.~\ref{fig:fig8} shows the similar figure for driving through the nonlinear perturbation $H_\eta(t)$. Once again multi-photon resonances are seen, but only three of the previous six---those corresponding to the absorption of two, four, and six photons. The figure in this case adds quantum corrections to the mean-field results of Fig.~\ref{fig:fig3} (but note that $\kappa/\lambda$ is $0.02$ in Fig~\ref{fig:fig8} and $0.1$ in Fig.~\ref{fig:fig3}).

\subsection{Quantum induced switching between mean-field steady states}
While multi-photon resonances are completely beyond the scope of mean-field results, Fig.~\ref{fig:fig8} does provide a hint of  mean-field predictions once photon numbers rise above two or three, where, in the vicinity of zero detuning, we see clear evidence of regions $R_2^a$ in Fig.~\ref{fig:fig4} and $R_2$ in Fig.~\ref{fig:fig3}. In this section, we use quantum trajectory simulations to further trace connections between the mean-field theory and a full quantum treatment.

Note, first, that unlike the common situation for phase transitions of light, where the many-photon limit is a weak-coupling limit (Secs.~IVA and IVC of Ref.~\cite{carmichael_2015}), the photon number for our generalized Jaynes-Cummings Rabi model scales with $N(\lambda/\kappa)^2$---i.e., the many-photon limit is a strong-coupling limit; this is seen, for example, from Eq.~(\ref{eqn:alpha_eta=1_approx}), which, undoing the scaling of Eqs.~(\ref{eqn:scaling}) and (\ref{eqn:scaled_parameters}), reads
\begin{equation}
|\alpha|^2=N\left(\frac{\lambda}{\kappa}\right)^2\frac{(1+\eta)^2}4\frac{(\bar\epsilon\pm 1)^2}{1+(\Delta/\kappa)^2}.
\label{eqn:photon_number_eta=1_unscaled}
\end{equation}
The scaling is also apparent from a comparison between frames (c) and (e) of Fig.~3, and frames (c), (e), and (f) of Figs.~\ref{fig:fig4} and \ref{fig:fig5}, and frame (c) of Fig.~\ref{fig:fig7}: with $\lambda/\kappa=10$ in Fig.~\ref{fig:fig3}, photon numbers range from 4 to 40, while with five times larger coupling in Figs.~\ref{fig:fig4}, \ref{fig:fig5}, and \ref{fig:fig7} they range in the hundreds and thousands; indeed,  frames (c), (e), and (f) of Fig.~\ref{fig:fig5} rise to reach photon numbers of $6.4\times10^3$, $10^4$, and $1.21\times10^4$, respectively, at zero detuning [Eq.~({\ref{eqn:photon_number_eta=1_unscaled}}]. Such high numbers can be reached with just one two-state system, since, when the coupling is strong, there is no need for a large value of $N$ to offset a weak nonlinearity per photon.

\begin{figure}[htpb!]
\begin{center}
\includegraphics[width=3.4in]{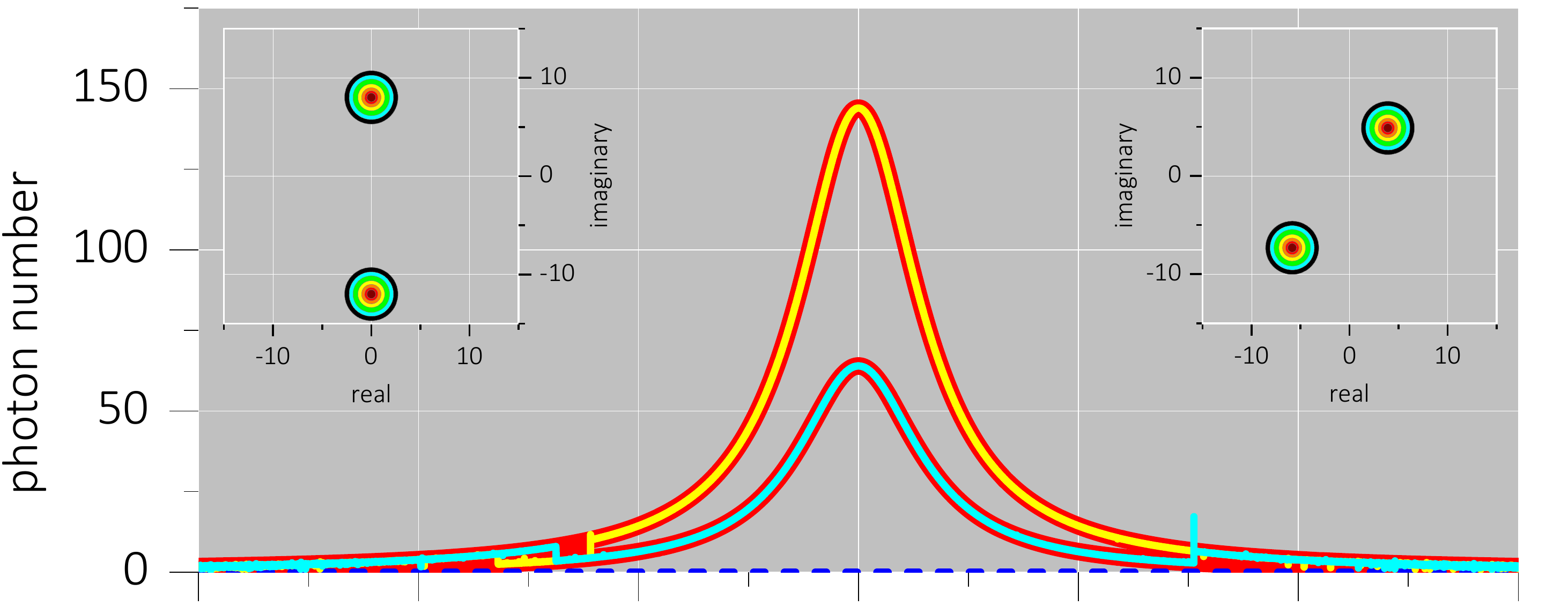}
\vskip0.2in
\includegraphics[width=3.4in]{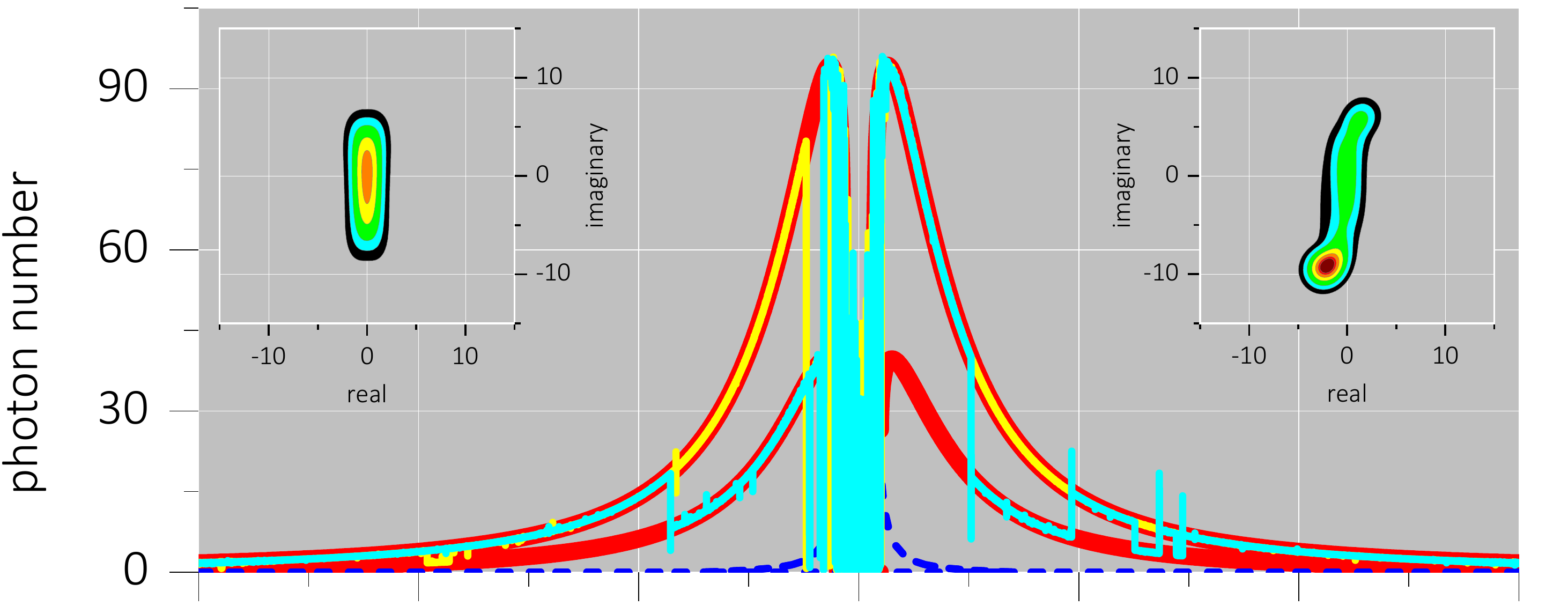}
\vskip0.2in
\includegraphics[width=3.4in]{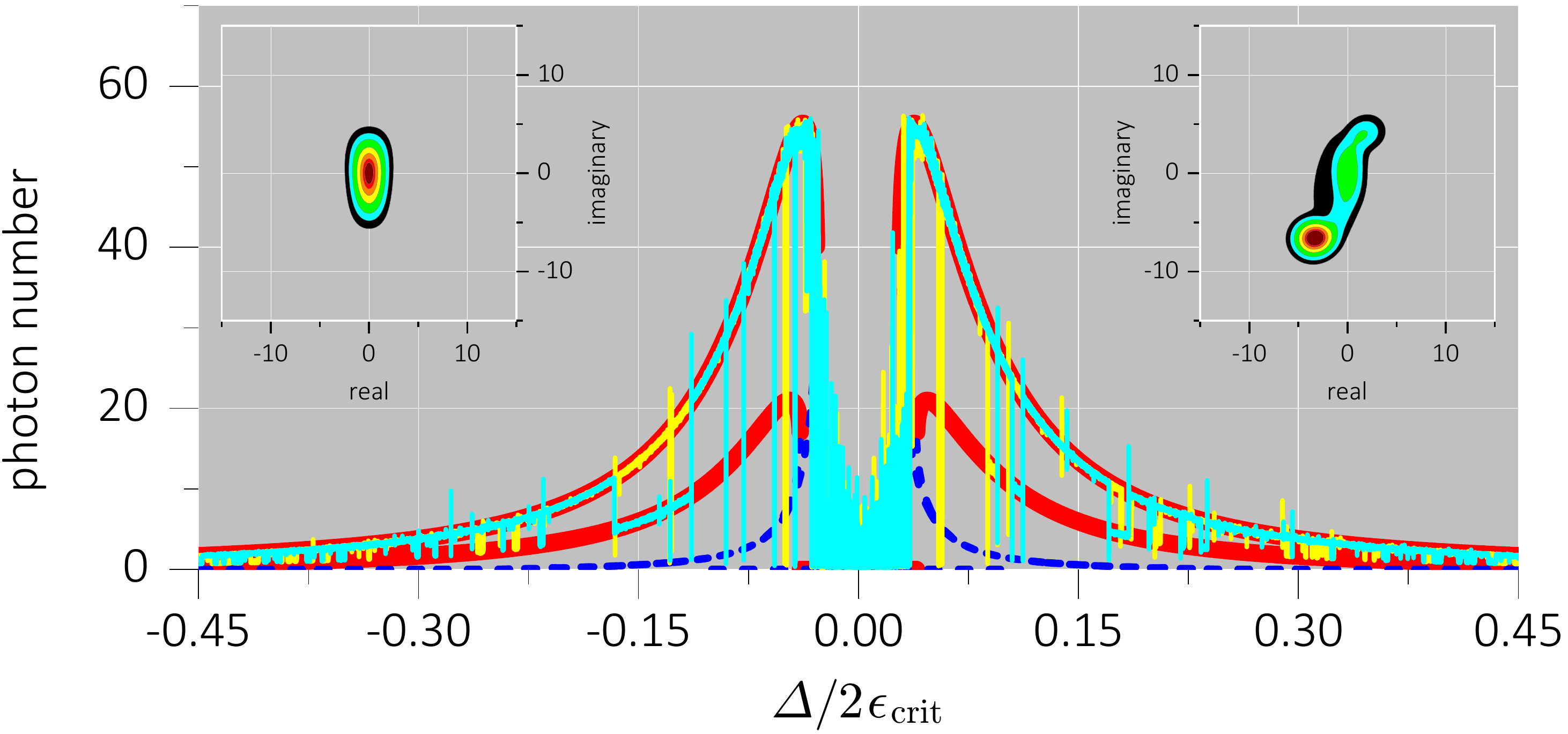}
\end{center}
\caption{Sample quantum trajectories as a function of scanned detuning and steady-state $Q$ functions for $N=1$, $\Delta_0=\Delta$, $\epsilon/\epsilon_{\rm crit}=0.2$, $\kappa/\lambda=0.1$, and $\eta=1$, $0.8$, $0.6$ (top to bottom); in all frames the detuning is scanned from $\Delta/\lambda=-1$ to $+1$ in a time $\kappa T=6\times10^4$. Two sample scans are plotted in each frame (solid yellow and cyan lines) against the background of mean-field steady states (solid red and dashed blue curves). The inset $Q$ functions are for detunings $\Delta/2\epsilon_{\rm crit}=0$ (left) and $\Delta/2\epsilon_{\rm crit}=0.04$, $0.015$, $0.04$ (top to bottom) (right).}
\label{fig:fig9}
\end{figure}

\begin{figure}[htpb!]
\begin{center}
\includegraphics[width=3.4in]{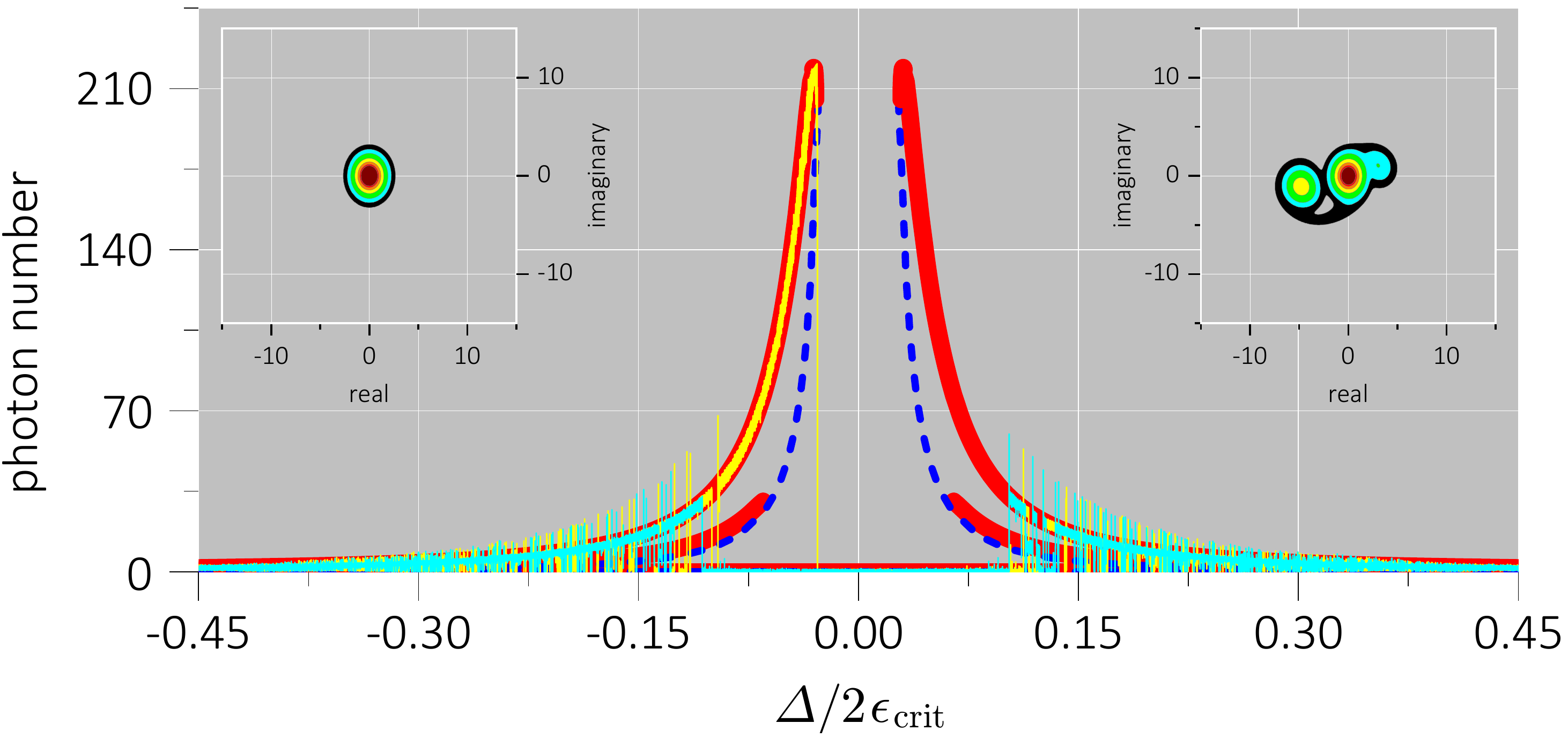}
\end{center}
\caption{As in Fig.~{\ref{fig:fig9}} but for $\kappa/\lambda=0.02$ and $\eta=0.2$, and with the detuning scanned from $\Delta/\lambda=-0.6$ to $+0.6$ in a time $\kappa T=6\times10^4$. The inset $Q$ functions are for detunings $\Delta/2\epsilon_{\rm crit}=0$ (left) and $\Delta/2\epsilon_{\rm crit}=0.12$ (right).}
\label{fig:fig10}
\end{figure}

Amongst the many effects of quantum fluctuations, in the following we target just two: first, mean-field steady states that are stable under Maxwell-Bloch equations are expected to be metastable in the presence of quantum fluctuations; and, second, isolated stable steady states---e.g., the lower state in frames (b) and (c) of Fig.~\ref{fig:fig5} [the minus sign in Eq.~(\ref{eqn:photon_number_eta=1_unscaled})]---might be accessed via quantum fluctuations. These effects are illustrated in Figs.~\ref{fig:fig9} and \ref{fig:fig10}, where we plot quantum trajectories of the photon number expectation while the detuning is slowly swept, from negative to positive. The coupling $\lambda/\kappa=10$ is used in Fig.~\ref{fig:fig9} in order to keep the maximum photon number relatively low, while the larger value in Fig.~\ref{fig:fig10} maps to the mean-field results of Fig.~\ref{fig:fig7}.

Figure \ref{fig:fig9} presents a sequence of plots illustrating the role of quantum fluctuations as we move away from the limit of the coherently driven extension of the Dicke phase transition of Sec.~\ref{sec:Dicke_coherent_drive_eta=1} into the intermediate regime of Sec.~\ref{sec:coherent_drive_intermediate_eta}. Beginning with $\eta=1$, the upper frame shows quantum trajectories tracking the two mean-field curves plotted from Eq.~(\ref{eqn:photon_number_eta=1_unscaled}). Both trajectories (yellow and cyan lines) start on the left by following the higher mean-field branch, but quantum fluctuations allow the isolated [see frames (b) and (c) of Fig.~\ref{fig:fig5}] lower branch to be accessed too. The two branches correspond to fields that are $\pi$ out of phase in the imaginary direction at zero detuning---inset $Q$ function to the left---and rotate to eventually align with the real axis as the detuning is changed---inset $Q$ function to the right.

Similar results are plotted for $\eta=0.8$ and $\eta=0.6$ in the middle and bottom frames, respectively. Once again, mean-field curves are faithfully followed over segments of the path, but the switching between branches is more common. The most prominent feature, however, is the dramatic loss of stability around zero detuning: although the mean-field analysis finds a stable steady state at zero photon number [region $R_2^a$ in frame (a) of Fig.~\ref{fig:fig5}], the full quantum treatment yields fluctuations spanning the two previously stable coherent states; the fluctuations are particularly apparent from the inset $Q$ functions in the middle frame of Fig~{\ref{fig:fig9}}. The spikes that accompany switches between branches are not numerical artifacts; they are decaying oscillations---evidence of a spiraling trajectory for the field amplitude in the approach to the new locally stable state.

Figure \ref{fig:fig10} presents the results of two detuning scans for $\lambda/\kappa=50$ and $\eta=0.2$, corresponding to the parameters of Fig.~\ref{fig:fig7}. In one scan the quantum trajectory follows the highest branch of stable mean-field solutions all the way up to its maximum. Much more commonly, though, the trajectory switches between this branch and the vacuum state in the region of $\Delta/2\epsilon_{\rm crit}=\pm 0.1$, as illustrated by the second scan. In this region the quantum fluctuations show clear evidence of the three coexisting stable mean-field steady states illustrated in frame (e) of Fig.~\ref{fig:fig7} (region $R_6$)---inset $Q$ function to the right.

\section{Conclusions}
\label{sec:conclusions}
\noindent
We have generalized the dissipative extension \cite{dimer_etal_2007} of the Dicke model \cite{dicke_1954} of light interacting with matter in two directions, thus linking the superradiant phase transition of Hepp and Lieb \cite{hepp&lieb_1973a,hepp&lieb_1973b} to the breakdown of blockade \cite{carmichael_2015,fink_etal_2017}. Although the former was originally approached through exact calculations in the thermodynamic limit for $N$ two-state systems in thermal equilibrium, and the latter as a phenomenon of single systems, both might be engineered in many- and one-two-state-system versions, with the same underlying mean-field phenomenology and where the central issue of photon number in the presence of dissipation is governed not by the number of two-state systems only, but also the ratio of coupling strength to photon loss \cite{carmichael_2015}---even one two-state system can control many photons in cavity and circuit QED \cite{armen_etal_2009,fink_etal_2017}.

We adopted a generalization introduced by Hepp and Lieb \cite{hepp&lieb_1973b}, and taken up in a number of recent publications \cite{stepanov_etal_2008,schiro_etal_2012,tomka_etal_2014,xie_etal_2014,tomka_etal_2015,wang_etal_2016,moroz_2016,kirton_etal_2018}, where the interaction Hamiltonian is made from a sum of rotating and counter-rotating terms of variable relative strength; in this way we span the continuum from the Jaynes-Cummings to the quantum Rabi interaction. We also added direct driving of the field mode, since that, not the counter-rotating interaction, creates photons in the breakdown of photon blockade. We analyzed mean-field steady states as a function of adjustable parameters for this extended model and found that a common critical drive strength, $\epsilon_{\rm crit}=\lambda(1+\eta)/2$, links the superradiant phase transition  to the breakdown of photon blockade---$\lambda$ is the coupling strength and $\eta$ the relative strength of counter-rotating to rotating interactions.  More generally, we found that the extended phase diagram moves from a region of pure superradiant character into the region of broken blockade, passing through a phase that although present in the generalized model of Hepp and Lieb \cite{hepp&lieb_1973b} is not identified in that work.

We then carried our analysis beyond mean-field steady states to a fully quantum treatment for the limiting case of one two-state system: we extended a prior calculation of quasi-energies \cite{alsing_etal_1992} to the generalized Hamiltonian---resonant driving of the field mode and no dissipation---and obtained numerical results with both detuning and photon loss  included. The quasi-energy spectrum for one two-state system was shown to be singular at $\epsilon_{\rm crit}$, where it undergoes a transition from discrete to continuous, and numerical simulations broadly support mean-field results, though expanding the view from earlier work \cite{carmichael_2015,shamailov_etal_2010} of multi-photon resonances at weak drive and exhibiting quantum-fluctuation-induced switching amongst locally stable mean-field steady states.

The aim of this study has been to uncover connections between different dissipative quantum phase transition for light and we have left many directions untouched; for example, a broader investigation of a very rich parameter space and the fully quantum treatment. We expect future work on the theoretical side will fill the gaps and hope that experiments in the spirit of Refs.~\cite{baumann_etal_2010,baumann_etal_2011,baden_etal_2014,fink_etal_2017,armen_etal_2009} will prove feasible.

\section*{Acknowledgments}
This work was supported by the Marsden fund of the RSNZ. Quantum trajectory simulations were carried out on the NeSI Pan Cluster at the University of Auckland, supported by the Center for eResearch, University of Auckland.

\end{document}